\documentclass[twocolumn]{article}

\usepackage{authblk}
\usepackage{amsfonts}
\usepackage{amsmath}
\usepackage{amssymb}
\usepackage{xfrac}
\usepackage{graphicx}
\usepackage{booktabs}
\usepackage{array}
\usepackage[backend=biber,style=phys]{biblatex}
\usepackage{comment}
\usepackage[skip=2pt,font=footnotesize]{caption}
\usepackage{titlesec}
\usepackage[flushleft]{threeparttable}
\usepackage{color}
\usepackage[normalem]{ulem}
\usepackage{tablefootnote}
\usepackage{caption}
\usepackage{subcaption}
\usepackage{hyperref}
\usepackage{tikz}

\newcolumntype{L}[1]{>{\raggedright\arraybackslash}m{#1}}
\newcolumntype{M}[1]{>{\centering\arraybackslash}m{#1}}
\newcolumntype{R}[1]{>{\raggedleft\arraybackslash}m{#1}}

\titleformat*{\section}{\normalsize\bfseries\rmfamily}
\titleformat*{\subsection}{\normalsize\bfseries\rmfamily}
\titleformat*{\subsubsection}{\normalsize\bfseries\rmfamily}

\graphicspath{{./Figures_Revised/}, {./Workflows/}}
\DeclareGraphicsExtensions{.jpg,.png,.pdf}
\title{Efficient dataset construction using active learning and uncertainty-aware neural networks for plasma turbulent transport surrogate models}
\author[1]{A. Ho}
\author[2]{L. Zanisi}
\author[3]{B. de Leeuw}
\author[1]{V. Galvan}
\author[1]{P. Rodriguez-Fernandez}
\author[1]{N. T. Howard}
\affil[1]{MIT, PSFC, Cambridge MA 02139, United States}
\affil[2]{UKAEA, CCFE, Culham Science Centre, Abingdon, OX14 3DB, United Kingdom}
\affil[3]{Radboud University, 6525 XZ Nijmegen, Netherlands}

\begin{document}

\maketitle

\begin{abstract}
	This work demonstrates a proof-of-principle for using uncertainty-aware architectures, in combination with active learning techniques and an in-the-loop physics simulation code as a data labeller, to construct efficient datasets for data-driven surrogate model generation. Building off of a previous proof-of-principle successfully demonstrating training set reduction on static pre-labelled datasets, using the ADEPT framework, this strategy was applied again to the plasma turbulent transport problem within tokamak fusion plasmas, specifically the QuaLiKiz quasilinear electrostatic gyrokinetic turbulent transport code. While QuaLiKiz provides relatively fast evaluations, this study specifically targeted small datasets to serve as a proxy for more expensive codes, such as CGYRO or GENE. The newly implemented algorithm uses the SNGP architecture for the classification component of the problem and the BNN-NCP architecture for the regression component, training models for all turbulent modes (ITG, TEM, ETG) and all transport fluxes ($Q_e$, $Q_i$, $\Gamma_e$, $\Gamma_i$, and $\Pi_i$) described by the general QuaLiKiz output. With 45 active learning iterations, moving from a small initial training set of $10^{2}$ to a final set of $10^{4}$, the resulting models reached a $F_1$ classification performance of $\sim$0.8 and a $R^2$ regression performance of $\sim$0.75 on an independent test set across all outputs. This extrapolates to reaching the same performance and efficiency as the previous ADEPT pipeline, although on a problem with 1 extra input dimension. While the improvement rate achieved in this implementation diminishes faster than expected, the overall technique is formulated with components that can be upgraded and generalized to many surrogate modeling applications beyond plasma turbulent transport predictions.
\end{abstract}

\section{Introduction}
\label{sec:Introduction}

Integrated fusion plasma models~\cite{rodriguez-fernandezEnhancingPredictiveCapabilities2024,romanelliJINTRACSystemCodes2014,tardiniFullypredictiveTransportModelling2021} generally couple together multiple independent physics codes describing specialized phenomena in the physical system, an approach which has been demonstrated to be useful for understanding and replicating fusion plasma experiments~\cite{fajardoLModeImpurityTransport2024,howardSimultaneousReproductionExperimental2024,angioniTokamakLModeConfinement2024,kimValidationDTFusion2023,cassonPredictiveMultichannelFluxdriven2020,marinFirstprinciplesbasedMultipleisotopeParticle2019,hoApplicationGaussianProcess2019,rodriguez-fernandezExplainingColdPulseDynamics2018}. This approach simplifies many of the time and length scale resolution problems and enables whole-plasma calculations to be done within a reasonable time frame. Unfortunately, it is still computationally expensive for fully automated reactor design optimization studies or other highly iterative physics applications, such as investigating the plasma dynamics of a physical discharge by numerically evolving the plasma state with fine time resolution~\cite{marinMultipleisotopePelletCycles2021,koechlITERFuellingRequirements2020}, such as sawteeth, pellets, or edge-localized modes. In spite of this, efforts continue to be made in applying these frameworks through the use of additional constraints and simplifications~\cite{rodriguez-fernandezOverviewSPARCPhysics2022,casiraghiFirstPrinciplebasedMultichannel2021,parailSelfConsistentSimulationITER2013}.

Previous efforts have leveraged machine learning techniques, e.g. neural networks (NNs), to bridge the gap in computational requirements needed to enable highly exploratory exercises while retaining the accuracy necessary for a faithful representation of the underlying physics regarding plasma transport~\cite{hoNeuralNetworkSurrogate2021,morosohkNeuralNetworkModel2021,vandeplasscheFastModelingTurbulent2020,meneghiniSelfconsistentCorepedestalTransport2017}, heating and fuelling~\cite{wallaceFastAccuratePredictions2022,morosohkAcceleratedVersionNUBEAM2021}, and magnetic geometry~\cite{joungGSDeepNetMasteringTokamak2023,degraveMagneticControlTokamak2022}. However, this approach relies heavily on the creation and curation of large datasets with sufficient coverage in input space to encapsulate the target application domain and provide adequate resolution of its salient features. When generalizing these approaches to the highest available model fidelities~\cite{citrinFastTransportSimulations2023}, it becomes apparent that the dataset generation process itself represents a substantial computational investment and may not be currently feasible.

This study attempts to address this dataset generation issue by applying a combination of uncertainty-aware machine learning model architectures and active learning techniques. It focuses on the tokamak plasma microturbulence-driven transport problem, specifically using the QuaLiKiz quasilinear gyrokinetic model~\cite{citrinTractableFluxdrivenTemperature2017} as the primary data source, as a demonstration platform for the proposed algorithm. However, there is strong evidence that this workflow can be applied to other regression problems in the fusion domain and perhaps beyond. It builds upon a previously developed framework addressing this particular problem, called ADEPT~\cite{zanisiEfficientTrainingSets2024}, and aims to improve the robustness and utility of the original approach. It also targets the application of ML techniques to smaller dataset sizes in an attempt to create a proxy exercise for more expensive gyrokinetic codes, e.g. CGYRO~\cite{candyHighaccuracyEulerianGyrokinetic2016}, GENE~\cite{jenkoElectronTemperatureGradient2000}, etc..

Active learning (AL)~\cite{renDeepActiveLearning2021,aggarwalActiveLearning2014} is a class of sequential machine learning techniques which leverage the information captured in previous iterations of the learned model to determine the regions in input space where extra data should be acquired to improve said learned model. The algorithm typically proceeds with the following steps:
\begin{enumerate}
    \item a collection of input values with no corresponding outputs, called \emph{candidates}, spanning the domain of interest is constructed;
    \item these candidates are then further evaluated through a customizable function, called the \emph{acquisition function}, which determines a fitness score for each point reflecting their priority for inclusion into the training set;
    \item a decision criteria is applied to the fitness score and the chosen candidates are passed forward to a process, called a \emph{labeller}, which labels them with the ground truth outputs to be used in a supervised machine learning application.
\end{enumerate}
Generally, the candidates are sampled out of an \emph{unlabelled pool} of possible input points~\cite{cohnImprovingGeneralizationActive1994}, defined \emph{a priori} to restrict the input domain within which the algorithm tries to construct a suitable surrogate model. Optional steps involving data filtering catered for the target problem can be placed before or after any of these steps, if deemed necessary.

It is worth mentioning that the use of acquisition functions hint at the similarities between AL techniques and the more well-studied field of Bayesian optimization (BO)~\cite{snoekPracticalBayesianOptimization2012}. Indeed, these two methods only differ in their final objectives~\cite{difioreActiveLearningBayesian2024}, which in turn influences the form of its acquisition function~\cite{wilsonMaximizingacquisitionfunctionsbayesian2018}. Whereas BO aims to acquire the input point which provides the best value of a target quantity, often through the training of surrogate models to represent the problem, AL aims to acquire the best overall surrogate model to represent the target domain. In more specific terms, AL leans more towards \emph{exploration} while BO leans more towards \emph{exploitation} of the surrogate within its problem domains.

Section~\ref{sec:ModelSelection} introduces the uncertainty-aware models used in this study and discusses their advantages and disadvantages, while Section~\ref{subsec:Uncertainties} provides a brief description of the uncertainties being presented by said models, which is important for any future UQ exercises. Section~\ref{sec:MachineLearning} describes the implemented AL algorithm and the optimization principles used within it. Then, Section~\ref{sec:Results} describes the measures taken to ensure both that the developed model is sufficient for the purposes of rapid and accurate computations. Finally, a summary is provided in Section~\ref{sec:Conclusion} and comments are made on any potential future work.

\section{Model selection}
\label{sec:ModelSelection}

Standard feed-forward NNs (FFNNs)~\cite{schmidhuberDeepLearningNeural2014,hintonLearningMultipleLayers2007}, also known as artificial neural networks (ANN) in past literature, have been used as the architecture of choice for developing surrogate models in the field of fusion plasmas, due to their simple structure and ease of implementation. Exceptions can be found in applications containing information-rich locally-correlated data, such as images, in which the family of convolutional neural networks (CNNs) are preferred~\cite{lecunLearningMethodsGeneric2004}. However, the most prevalent AL acquisition functions generally require metrics which compare the quality of information in the existing training set and the information complexity~\cite{bozdoganInformationComplexity2000,traubComplexity1998} represented by the ML model output, which is \emph{not} simply a function of the number of free parameters due to the effects of regularization. In more approachable terms, a measure is needed which can discern whether the training set is sufficiently populated to have confidence in the variations observed in the model output. Unfortunately, information of this nature is not inherently available as an output of FFNN architectures and it gets increasingly more difficult to heuristically estimate as the number of input dimensions increases.

While this does not automatically exclude FFNNs from being used in AL applications, as methods to extract uncertainties from them exist~\cite{galDropoutUncertainty2016}, it does demand more complicated acquisition functions and statistical methods to account for arbitrarily distributed predictive posteriors. Within literature, other methods developed to obtain this information elsewhere include:
\begin{itemize}
	\item comparing the candidate pool directly to existing training set distributions~\cite{lindenbaumSelectiveSamplingNearest2004};
	\item extracting local output gradients with respect to the input variables from the trained NN~\cite{ashDeepbatchactivelearning2020};
	\item extracting uncertainty estimates from the trained NN~\cite{kirschBatchbaldefficientdiversebatch2019,houlsbyBayesianactivelearningclassification2011}.
\end{itemize}
This study focuses on the final option of this list, as it allows the trained models to be leveraged for downstream applications in uncertainty quantification (UQ). Specifically, it will investigate NN architectures which incorporate a direct uncertainty estimate as a part of its prediction outputs~\cite{huMultidimensionaluncertaintyawareevidentialneural2021}, referred to here as \emph{uncertainty-aware} architectures. Previous attempts at obtaining uncertainties from FFNNs used ensembles~\cite{zanisiEfficientTrainingSets2024,hoNeuralNetworkSurrogate2021}, where each ensemble member was trained using a different randomized free parameter initialization and shuffled training set to ensure sufficient variation. It was found that, depending on the degree of regularization, the ensemble can become overconfident when interpolating in sparsely filled regions of the training set~\cite{dawoodAddressingDeepLearning2023}. The uncertainty-aware models aim to circumvent the overconfidence issue present in ensembles of FFNNs while additionally allowing the pipeline to require less computational resources and reducing the memory requirements of any downstream applications. It should be noted that initial attempts to use a regressor version of the evidential neural network (ENN) architecture~\cite{sensoyEvidentialDeepLearning2018}, as it had been successfully applied to AL exercises in literature~\cite{soleimanyEvidentialDeepLearning2021}, did not produce uncertainty estimates with the desired properties of identifying holes within the training set. This motivated a search for alternative uncertainty-aware architectures, which is detailed below.


Similarly to past work on NN surrogates for plasma turbulent transport processes~\cite{hornsbyGaussianProcessRegression2024}, the problem was divided into a classifier for predicting the presence of a particular turbulent mode instability and a regressor for determining the level of transport driven by said instability. Also, the QuaLiKiz outputs were separated such that each network only predicts the flux for one fluid transport channel.

\begin{figure*}[tb]
	\centering
	\begin{tikzpicture}
		\node[rectangle] at (0,0) (world) {\includegraphics[scale=0.8]{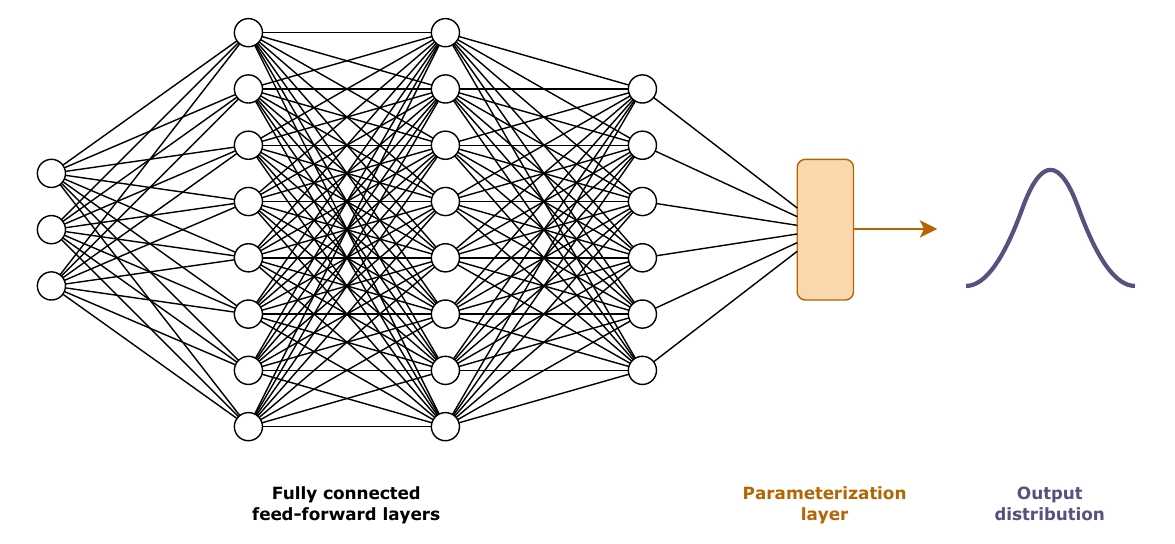}};
		\node[rectangle] at (3.23,1.12) (mu) {$\mathbf{\mu}$};
		\node[rectangle] at (3.23,0.62) (sigma) {$\mathbf{\sigma}$};
		\node[rectangle] at (3.23,0.12) (etc) {$\vdots$};
	\end{tikzpicture}
	\caption{Representative schematic of the Bayesian last layer (BLL) architecture class, which is identical to a standard feed-forward neural network (FFNN) except the last layer being replaced with a parameterization layer that recasts the output values as the parameters needed to describe a desired probability distribution. While the generic concept supports a wide variety of parameterizations, the difficulty lies in implementing robust techniques for training these parameters to return meaningful results. For simplicity, the specific architectures used in this study recast the outputs into Gaussian distributions, fully described by their means, $\mu$, and standard deviations, $\sigma$, which can be constrained via well-established distance functions.}
	\label{fig:BayesianLastLayerSchematic}
\end{figure*}
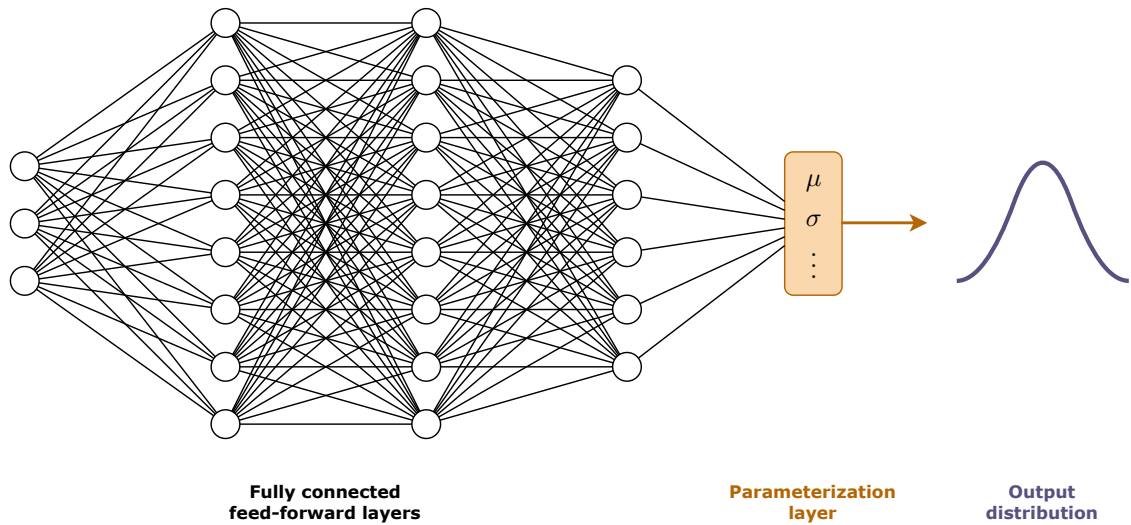

The spectral-normalized Gaussian Process (SNGP) model architecture~\cite{liuSinglemodeldeepuncertainty2023} was selected as the classifier model and the Bayesian neural network with noise contrastive prior (BNN-NCP) model architecture~\cite{hafnerNoisecontrastivepriorsfunctional2019} was selected as the regressor model. Both of these models handle uncertainty estimation by design and effectively fall under the category of Bayesian last layer (BLL) models~\cite{watsonBayesianlastlayer2021,lazaroMarginalizedNeuralNetworkMixtures2010}, which accelerate the computation of uncertainty-aware networks by hybridizing the architecture with FFNN and BNN layers. A representative diagram of this class of architectures is provided in Figure~\ref{fig:BayesianLastLayerSchematic}. The SNGP model was chosen due to its success in binary classification problems in multiple domains~\cite{gasparinDistanceawarecalibrationLanguagemodels2024,siebertUncertaintydeepkernellearning2023} whereas the BNN was chosen based on previous success in applying it to fusion data~\cite{paneraalvarezEuroPEDNNUncertaintyAware2024}.

While the original paper detailing the chosen regressor model categorizes the BNN-NCP model as a traditional Bayesian neural network (BNN), it does not contain the probabilistic tuning parameters, e.g. weights and biases, characteristic of a true Bayesian inferential model~\cite{blundellWeightUncertaintyInNeuralNetwork2015,titteringtonBayesianMethodsNeural2004,lampinenBayesianApproachNeural2001}. Instead, it effectively reparameterizes the final output layer such that each output value is represented by the moments of a Gaussian distribution, namely its mean, $\mu$, and standard deviation, $\sigma$. However, for the purposes of remaining true to the original publication, this paper will continue to refer it as the BNN model.

Likewise, while the chosen SNGP classifier architecture has Gaussian process (GP) in its name, it only borrows some concepts from a true GP~\cite{rasmussenGaussianProcessesMachine2006}. Specifically, it uses a randomly initialized layer to act as a covariance-generating, or kernel-generating, function of a GP with arbitrary features. However, unlike a regular GP, the free parameters of this kernel-generating layer are not modified during the training process. Instead, the trainable parameters are within the layers prior to this last kernel-generating layer, which learns where in this arbitrary feature space to place input points in order to properly capture its relation to other data points. Again, this kernel-generating layer functionally acts as a reparameterization of the final output layer such that each output value is represented as a Gaussian distribution, via its mean, $\mu$, and standard deviation, $\sigma$.

However, note that the SNGP output is actually trained as a \emph{logit}~\cite{hilbeLogisticRegressionModels2009}, or logistic unit, denoted here as $z$, expressed as follows:
\begin{equation}
	\label{eq:LogisticUnit}
	z = \ln\!\left(\frac{c}{1 - c}\right)
\end{equation}
which performs a smooth continuous mapping from $c \in \left(0, 1\right) \rightarrow z \in \left(-\infty, \infty\right)$. Thus, the model output must be further post-processed to obtain the final class prediction. For the binary classification implementation, as used in this study, a shifted sigmoid function is applied to the logit mean prediction, $\mu$, to remap it to $c \in \left(0, 1\right)$ and then rounded to the nearest integer to obtain the predicted class. The value of this shift is also a learned parameter by analyzing the Receiver Operating Characteristic Area Under the Curve (ROC AUC) score of the classifier~\cite{fawcettROCAnalysis2006} within the training loop.

\subsection{Nature of represented uncertainties}
\label{subsec:Uncertainties}

Within the context of numerical modelling and analysis, it is often useful to know the uncertainty associated with the predicted value in addition to the prediction itself. These uncertainties can be used to make decisions whether a given prediction can be taken at face value or whether incorporating information from other sources is necessary. However, including any sort of uncertainty estimate into a model generally elicits a discussion on the type of information encapsulated within said estimate, as these details become critical in avoiding improper interpretations in downstream applications.

Unlike GPs, the underlying mathematics of NNs are not built on probabilistic theory and thus cannot guarantee that any uncertainty estimates derived from them are rigourously defined. Since the uncertainties of the chosen uncertainty-aware models should reflect properties of a collection of data points, as opposed to a property of a specific point, it is not obvious how to capture this information inside a training target for the back-propagation algorithm. Nonetheless, previous efforts to construct NN models with these estimates have proven fruitful~\cite{paneraalvarezEuroPEDNNUncertaintyAware2024}.

The original BNN paper~\cite{hafnerNoisecontrastivepriorsfunctional2019} uses the terms \emph{epistemic} and \emph{aleatoric} to describe the different uncertainties represented by the model. Within Ref.~\cite{zouReviewUncertaintyEstimationMedical2023}, epistemic uncertainty is defined to arise from inconsistencies in the underlying model, data distribution, and / or model structure, which can generally be reduced by additional sampling and aleatoric uncertainty is defined to arise from inherent noise in the data, such as measurement error or ambiguous annotation, which cannot generally be reduced by additional sampling. As the inputs of this training set contain zero uncertainty and their corresponding simulation outputs are deterministic, it becomes confusing when communicating the results of this exercise across disciplines with the original BNN terms, most notably because the ``aleatoric" uncertainty branch of the BNN model no longer represents the randomness inherent in the process of data collection.

For this reason, this paper relabels epistemic as \emph{model} uncertainty and aleatoric as \emph{data} uncertainty. This relabelling is meant to more closely reflect:
\begin{itemize}
	\item the model uncertainty, $\sigma_m$, representing the probability distribution of the prediction that could result from different training attempts with random free parameter initializations;
	\item the data uncertainty, $\sigma_d$, representing the degree of variability in the output data within a localized input space, e.g. due to numerical errors, spatial resolution or convergence criteria.
\end{itemize}
It is then worth noting that the BNN model used in this study provides an explicit hyperparameter calibrating the size of this local input space for the purposes of defining the aleatoric error, called the out-of-distribution (OOD) width. Its value was roughly optimized in the setup of this exercise such that it only captures variability not explained by the given set of input variables. The SNGP model does not differentiate between the model and data uncertainty, so its output can best be interpreted as combining both into its uncertainty estimate.

\section{Active learning pipeline}
\label{sec:MachineLearning}

This section outlines AL pipeline and the various developments made to facilitate its construction for the purposes of this study.

\subsection{Improvements to BNN model}
\label{subsec:UncertaintyAwareModels}

Out-of-distribution (OOD) sampling is invariably required to obtain the extra information necessary to capture local uncertainty information inside a BNN. While the NCP training algorithm avoids the need for an exhaustive and time-consuming Monte-Carlo-based OOD sampling method~\cite{daxbergerLaplaceApproximation2021,nealBNN1992}, an issue regarding a lack of robustness in its convergence arose within the early trials of this pipeline. To identify potential causes for these convergence issues, a deeper investigation was carried out into the recommended loss function used when training the BNN model, expressed as:
\begin{equation}
	\label{eq:BNNLossFunction}
	\small
	L_{\text{BNN}} = L_{\text{NLL}} + L_{\sigma_m} + L_{\sigma_d}
\end{equation}
where
\begin{equation}
	\label{eq:NLLLossTerm}
	\small
	L_{\text{NLL}} = -\log p_l\!\left(y_t\right)
\end{equation}
\begin{equation}
	\label{eq:EpistemicLossTerm}
	\small
	L_{\sigma_m} = D_{\text{KL}}\!\left(p_m\,||\,p_{m,\text{prior}}\right)
\end{equation}
\begin{equation}
	\label{eq:AleatoricLossTerm}
	\small
	L_{\sigma_d} = D_{\text{KL}}\!\left(p_d\,||\,p_{d,\text{prior}}\right)
\end{equation}
and
\begin{equation}
	\label{eq:OriginalNLLProbability}
	\small
	p\!\left(y\right) = \mathcal{N}\!\left(\mu, \sigma_d^2\right)
\end{equation}
\begin{equation}
	\label{eq:EpistemicProbability}
	\small
	p_m = \mathcal{N}\!\left(y_s, \sigma_m^2\right) \; , \quad p_{m,\text{prior}} = \mathcal{N}\!\left(y_t, \sigma_{m,\text{prior}}^2\right)
\end{equation}
\begin{equation}
	\label{eq:AleatoricProbability}
	\small
	p_d = \mathcal{N}\!\left(y_t, \sigma_d^2\right) \; , \quad p_{d,\text{prior}} = \mathcal{N}\!\left(y_t, \sigma_{d,\text{prior}}^2\right)
\end{equation}
where $y_t$ is the ground truth output value from the labeller, $y_s$ is an output value sampled from the predicted model output distribution, and $\sigma_{m,\text{prior}}$ and $\sigma_{d,\text{prior}}$ are the user-defined prior standard deviations for the model and data uncertainty distributions, respectively. The $D_{\text{KL}}$ function in Equations~\eqref{eq:EpistemicLossTerm} and \eqref{eq:AleatoricLossTerm} is known as the Kullback-Leibler (KL) divergence, which acts as a measure of the distance between two probability distributions and is defined as follows:
\begin{equation}
	\label{eq:KLDivergence}
	\small
	D_{\text{KL}}\!\left(p \, || \, q\right) = \int_{-\infty}^{\infty} p\!\left(x\right) \log\!\left(\frac{p\!\left(x\right)}{q\!\left(x\right)}\right) \text{d}x
\end{equation}

By analyzing Equation~\eqref{eq:BNNLossFunction}, these convergence issues were determined to be likely due to use of KL-divergence as the loss term to control the adjustment of these uncertainties within the training routine, Equations~\eqref{eq:EpistemicLossTerm} and \eqref{eq:AleatoricLossTerm}. While the KL-divergence does represent an effective distance between two distributions, it is not symmetric, i.e. $D_{\text{KL}}\!\left(p\,||\,q\right) \ne D_{\text{KL}}\!\left(q\,||\,p\right)$, and it represents a square-distance, i.e. it diverges quickly when the two probability distributions being compared drift further apart. These properties lead to large spontaneous loss gradients which can drive the NN training routine far away from any meaningful local optima, also known as an exploding gradient. This behaviour was commonly observed in the early attempts to train a BNN model on QuaLiKiz data and severely limited its potential application in an AL exercise.

To remedy this, the closed form of the geodesic Fisher-Rao (FR) distance metric for Gaussian distributions~\cite{miyamotoClosedformExpressionsFisher2024} was proposed by this work and can be computed using the expression:
\begin{equation}
	\label{eq:FisherRaoDistanceMetric}
	\small
	D_{\text{FR}}\!\left(p_1,p_2\right) = 2 \sqrt{2} \tanh^{-1} \left(\sqrt{\frac{\left(\mu_1 - \mu_2\right)^2 + \left(\sigma_1 - \sigma_2\right)^2}{\left(\mu_1 - \mu_2\right)^2 + \left(\sigma_1 + \sigma_2\right)^2}}\right)
\end{equation}
where $\mu_1$ and $\mu_2$ represent the means of the Gaussian distributions being compared and $\sigma_1$ and $\sigma_2$ represent the standard deviations of the same distributions. This FR metric then replaces any instance of the KL-divergence within the NN training loss function, which in this case served as the epistemic (model) and aleatoric (data) uncertainty loss terms of the BNN-NCP architecture.

Additionally, the previously used leaky rectified linear unit (LReLU) activation functions were replaced with the Gaussian error linear unit (GELU) due to the increased fitting power provided by reintroducing a larger nonlinear domain motivated by the statistics of stochastic dropout processes~\cite{hendrycksGaussianErrorLinearUnits2016}. Theoretically, this specific activation function improves on the solution provided by LReLU to the zero gradient issue inherent on the negative half of the ReLU activation function. Practically, the inclusion of this activation function in all of the layers within the BNN architecture provided an additional degree of robustness with regards to the loss function choice discussed above.

\begin{figure}
	\centering
	\includegraphics[scale=0.26]{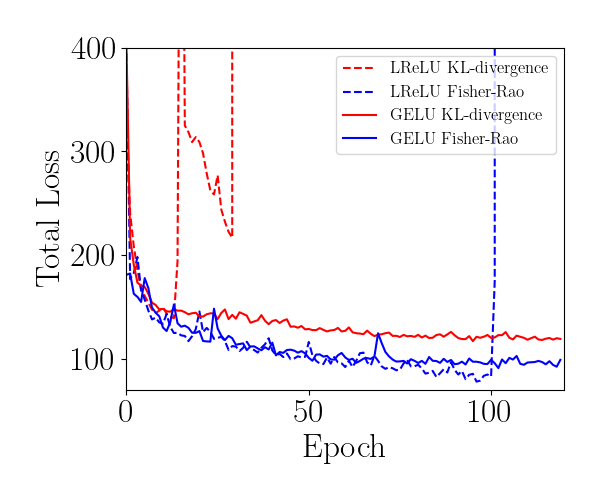}%
	\includegraphics[scale=0.26]{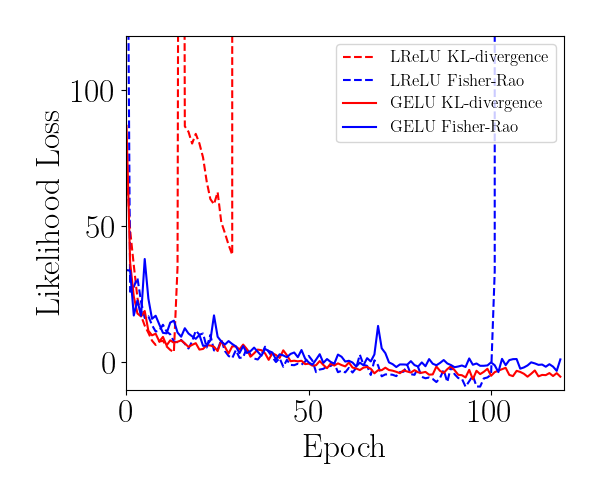}\\
	\includegraphics[scale=0.26]{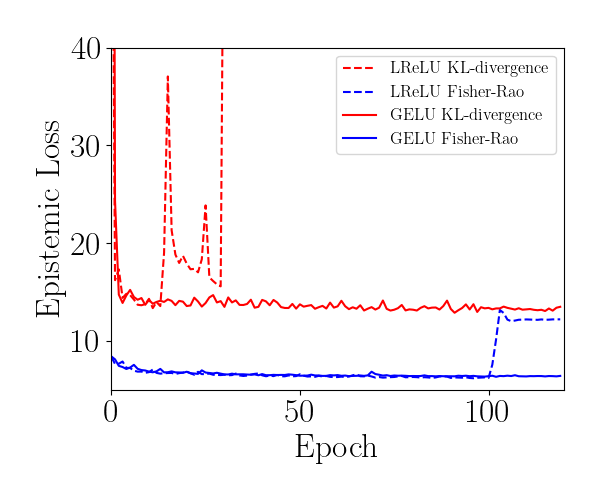}%
	\includegraphics[scale=0.26]{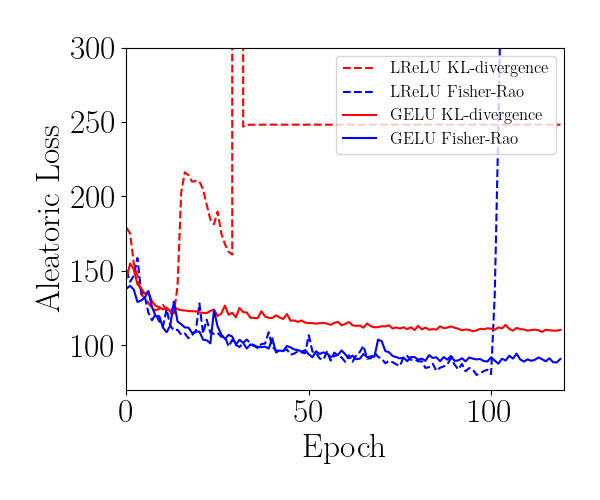}
	\caption{Loss curves for the BNN architecture training on $Q_{i,{\text{ITG}}}$ from an example QuaLiKiz dataset, differentiating the training behaviour using the GELU activation function (solid) and the LReLU activation function (dashed) in addition to using the KL-divergence loss term (red) from the Fisher-Rao metric loss term (blue). The large discontinuities in these plots is an example of the catastrophic impact of a spontaneously large gradient, as the routine generally does not recover from this excursion and settles on a very sub-optimal local solution involving extremely large uncertainty estimates. While this behaviour has also been observed with the Fisher-Rao loss term, it occurs significantly less frequently hence leading to its improved robustness in training.}
	\label{fig:LossComparison}
\end{figure}

Figure~\ref{fig:LossComparison} compares the loss curves between using the GELU and the LReLU activation function and between using the KL-divergence and the FR loss function, while training the BNN model on an example dataset of $\sim$8000 QuaLiKiz points with 16 inputs and 1 output parameter, detailed further in Section~\ref{subsec:PipelineImplementation}. This demonstrates both the prevalence and effect of these spontaneous and large loss gradients, which effectively reduce the robustness of each individual model training attempt. Consequently, switching to the FR loss term resulted in more reliable AL iterations and allowed this study to proceed with the chosen BNN model, although it should be noted that it does not completely eliminate the problem. It should be noted that an additional rudimentary study using gradient clipping~\cite{zhangGradientClipping2020} was performed to tackle the exploding gradient problem but preliminary results showed worse overall fit performance and decreased AL efficiency. This is suspected to occur due to the interaction between the lower effective learning rate due to gradient clipping and the early stopping routine. However, an extensive hyperparameter scan on the clipping threshold was not performed, thus the technique may still be able to resolve any remaining robustness issues within the BNN model but this is left for future work.


Finally, it was found that these exploding gradients can also occur due to the negative-log-likelihood (NLL) loss term, $L_{\text{NLL}}$, given in Equation~\eqref{eq:NLLLossTerm}. Since this term becomes extremely large when $\sigma_d \rightarrow 0$ and $\mu \ne y_t$, it can prematurely stop the training routine when the hyperparameters are set to strongly minimize its data uncertainty prediction. However, the AL algorithm benefits greatly when this data uncertainty is low in regions of sparse training point density. Thus, to minimize the chance of this occurring for the purposes of improving the overall robustness of the AL routine, the probability function used in the loss term was modified to the following:
\begin{equation}
	\label{eq:ModifiedNLLProbability}
	p\!\left(y\right) = \mathcal{N}\!\left(\mu, \sigma_d^2 + \gamma_s \sigma_m^2\right)
\end{equation}
where $\gamma_s$ is an extra hyperparameter for the training routine, with $\gamma_s = 0.1$ in this study. While this modification to the NLL loss term was observed to improve the overall robustness of the BNN architecture to exploding gradients, it is unclear whether this change has additional undesired effects on the final model uncertainty estimate. It is noted that a trial using $p\!\left(y\right) = \mathcal{N}\!\left(\mu, \sigma_d^2 + \epsilon\right)$ where $\epsilon = 10^{-6}$ was also attempted, but it did not significantly reduce the frequency of exploding gradients. A deeper investigation of this impact is warranted but currently left as future work.

As a general point, it was observed that the frequency of these exploding gradients can also be reduced by increasing the strength of the L1 and L2 regularization terms. However, this has a trade-off with the maximum goodness-of-fit metric by encouraging underfitting. For this reason, adjusting these parameters was deemed an unsuitable method for addressing the observed convergence issues.

Based on these findings, this study proceeded with both classifier and regressor uncertainty-aware models incorporating the GELU activation function and the BNN regressor model additionally switching to the FR loss terms. The modifications made to these models caused the hyperparameter choice for training these models to deviate from the recommended values from the previous application to fusion data~\cite{paneraalvarezEuroPEDNNUncertaintyAware2024}. Finally, as detailed in Table~\ref{tbl:RegressorTrainingHyperparameters}, these loss studies were performed using the Adam optimization routine~\cite{kingmaAdamMethodStochastic2015}, one of the dominant routines within ML literature. An exploration into different optimization schemes was outside the scope of this study and it is recommended to check whether other optimization routine, such as L-BFGS~\cite{byrdLBFG1995}, suffers from the same exploding gradient behaviour in future work.

\subsection{Pipeline implementation}
\label{subsec:PipelineImplementation}

\begin{figure}
	\centering
	\includegraphics[scale=0.7]{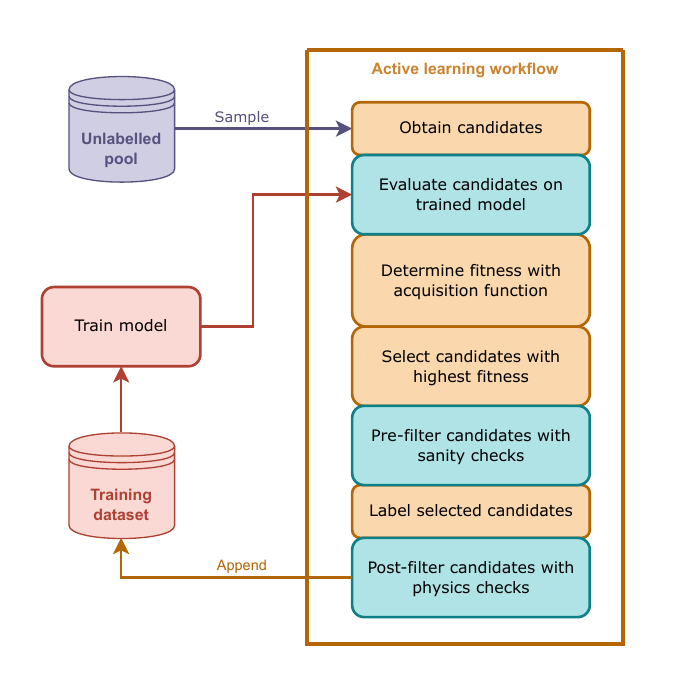}
	\caption{Workflow diagram outlining the general steps taken inside the AL algorithm implemented for this study, separating the mandatory steps for AL (orange blocks) and the optional ones (teal blocks). These optional steps are: pre-constructing an unlabelled pool of 500 million QuaLiKiz input points, evaluating the trained uncertainty-aware NNs as a method of extracting information for the acquisition function, and pre- and post-labeller filtering the selected candidates based on checks specific to the QuaLiKiz code.}
	\label{fig:ActiveLearningPipeline}
\end{figure}

Based on the general AL algorithmic steps described in Section~\ref{sec:Introduction}, a pipeline specific to the QuaLiKiz turbulent transport problem was constructed and can be seen depicted in a flowchart in Figure~\ref{fig:ActiveLearningPipeline}. Firstly, the chosen approach requires the generation of an unlabelled pool which is sufficiently populated to cover the point density needed to resolve the salient features of the problem in the chosen domain. Each point within this pool consists of one value for each of the 16 input variables described in Table~\ref{tbl:NeuralNetworkInputs}, which when combined provide a fully constrained local plasma state for a plasma with 4 species under the electrostatic plasma turbulence assumptions:
\begin{itemize}
	\itemsep 0mm
	\item electrons ($e$);
	\item fuel ions -- D ($i$);
	\item a light impurity -- Be ($z1$);
	\item and a heavy impurity -- Ni ($z2$).
\end{itemize}

\begin{table}[t]
	\centering
	\caption{The 16 dimensionless input parameters chosen for the NNs trained in this study. Although each dimensionless parameter is a function of many physical values, computing them in order from top to bottom in this list allow each dimensionless value to be defined solely on its associated physical value. The gyro-Bohm scaling factor accounts for the true turbulence scaling factors, $L_{\text{ref}}=R_0$ and $B_{\text{ref}}=B_0$, which should be placed at the top of this list. See Ref.~\cite{hoNeuralNetworkSurrogate2021} for more information on calculating these parameters.}
	\begin{tabular}{cc}
		Dimensionless & Approximate Physical \\
		Parameter & Parameter \\
		\midrule
		$\varepsilon$ & $a$ \\
		$x$ & $r$ \\
		$q$ & $q$ \\
		$\hat{s}$ & $\nabla q$ \\
		$\log_{10}\!\left(\nu^*\right)$ & $n_e$ \\
		$\alpha$ & $T_e$ \\
		$R/L_{T_e}$ & $\nabla T_e$ \\
		$R/L_{n_e}$ & $\nabla n_e$ \\
		$T_i/T_e$ & $T_i$ \\
		$R/L_{T_i}$ & $\nabla T_i$ \\
		$n_i/n_e$ & $n_i$ \\
		$R/L_{n_i}$ & $\nabla n_i$ \\
		$Z_{\text{eff}}$ & $n_{z1}$ or $n_{z2}$ \\
		$R/L_{Z_{\text{eff}}}$ & $\nabla n_{z1}$ or $\nabla n_{z2}$ \\
		$M_{\text{tor}}$ & $\Omega_{\text{tor}}$ \\
		$R/L_{u_{\text{tor}}}$ & $\nabla \Omega_{\text{tor}}$ \\
	\end{tabular}
	\label{tbl:NeuralNetworkInputs}
\end{table}


As mentioned before, the turbulent transport prediction was split into a classification and regression problem. Within the electrostatic physics encoded inside QuaLiKiz, there exists three categories of turbulent instabilities: the ion temperature gradient (ITG) mode, the trapped electron mode (TEM), and the electron temperature gradient (ETG) mode. Each of these instability modes are governed by distinct physical drives and are known to have different dependencies on the input variables. However, since the presence of one mode do not exclude the others from being present, mode boundaries can be difficult to quantitatively capture purely from the transport flux and smoothing over mode transitions can greatly impact the regressor performance. For this reason, previous iterations of the QuaLiKiz-neural-network (QLKNN) found it advantageous to also split the network predictions along these mode categories~\cite{zanisiEfficientTrainingSets2024,hoNeuralNetworkSurrogate2021,vandeplasscheFastModelingTurbulent2020}. From here, the classification problem consists of determining whether a mode of that particular category is present and unstable in the local solution or not.

When a particular mode is present and unstable, it drives plasma transport along different channels described by fluid moments of the plasma distribution function:
\begin{itemize}
	\itemsep 0mm
	\item density via particle flux, $\Gamma$;
	\item momentum via momentum flux, $\Pi$;
	\item and energy via heat flux, $Q$.
\end{itemize}
This is further differentiated by the particular particle species since they can have different responses to the same perturbation, although only the electrons, $e$, and main ions, $i$, are investigated in this study. Again, previous QLKNN iterations~\cite{hoNeuralNetworkSurrogate2021} found it advantageous to train an independent model for each of these transport channels to improve its predictive capabilities and minimize training times at the cost of needing to evaluate more models. Naturally, the regression component consists of determining the gyro-Bohm (GB) normalized flux for each channel individually. In order to improve the performance of these networks, the regressor training set only contains points which are unstable for the targeted turbulent mode, which avoids the problem of numerically matching a large domain of zeros~\cite{vandeplasscheFastModelingTurbulent2020}. In order to simplify the software development requirements of the pipeline, this study only trains the models corresponding to one turbulent mode (i.e. ITG, TEM, or ETG) per AL algorithm iteration. This leads to a total of 14 networks to describe the QuaLiKiz output: 3 classifiers (one for each mode) and 11 regressors:
\begin{itemize}
	\itemsep 0mm
	\item ITG: $Q_e$, $Q_i$, $\Gamma_e$, $\Gamma_i$, $\Pi$;
	\item TEM: $Q_e$, $Q_i$, $\Gamma_e$, $\Gamma_i$, $\Pi$;
	\item ETG: $Q_e$
\end{itemize}

Once the NN models were trained, they were evaluated on a random subset of the unlabelled pool in order to obtain an adequate sampling of the NN uncertainties across the target application domain. Since the regressor networks were only trained on unstable data, it is expected to return meaningless but non-zero values in their respective stable domains. To provide a coherent transport flux prediction and associated uncertainty, the regressor outputs were blended with its respective mode classifier output to provide the network prediction according to the following expression:
\begin{equation}
	\label{eq:NetworkBlending}
	y_{\text{reg}} = \mu_{\text{reg}} \, \mu_{\text{cls}} \, \max\!\left(\frac{1 - \sigma_{\text{cls}}}{1 - \sigma_{\text{thr}}}, 1\right)
\end{equation}
where $\max$ function takes the maximum value of the items contained within the bracketed list and $\sigma_{\text{thr}}$ is a user-defined parameter indicating the value of $\sigma_{\text{cls}}$ below which the classifier is deemed to be accurate, chosen as $\sigma_{\text{thr}} = 0.1$ for this study. Equation~\eqref{eq:NetworkBlending} serves a second function in ensuring a gentle transition into the stable region, preventing numerical instabilities corresponding to large discontinuities around the critical threshold in an eventual transport solver implementation~\cite{vandeplasscheFastModelingTurbulent2020}.

Along with their evaluated NN predictions and uncertainties, the selected unlabelled points were then passed through a custom acquisition function to assign a score to each one, detailed in Section~\ref{subsec:PipelineSetup}. A list of potential candidates can be generated by selecting the top $N$ points with regards to the highest score to pass to the labeller, QuaLiKiz. Since there are potentially multiple transport channels for each turbulent mode, this selection procedure effectively turns into a multi-objective optimization problem. This introduces a minor complication in that imbalances may occur in the prioritization of each output during the candidate selection process, potentially neglecting to improve certain outputs completely in a given iteration. Although the AL algorithm should automatically balance this given enough iterations in theory, it is still worth addressing directly as it does affect its overall efficiency. The first attempt to counteract this was by defining extra acquisition function hyperparameters, $\gamma_{\text{reg}}$, detailed further in Section~\ref{subsec:PipelineSetup}, which provided a way to globally scale the scores for each output with respect to the others. While it was found that these static multipliers were effective at preventing one output from overshadowing all the others, it was less effective at preventing one from being overshadowed.

\begin{figure}[h]
	\centering
	\includegraphics[scale=0.35]{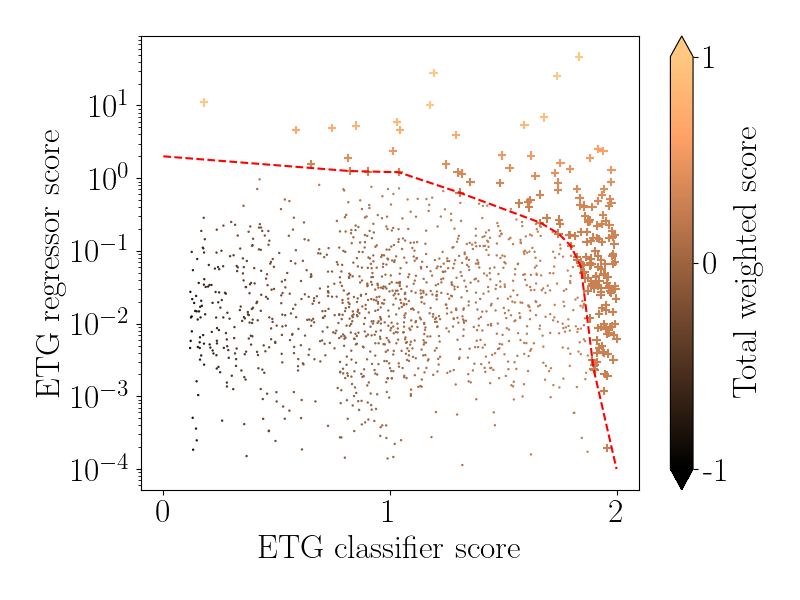}
	\caption{Sample acquisition from an AL iteration targeting the ETG turbulence mode differentiating the chosen candidates (plusses) from the rejected candidates (dots) based on their maximum fitness score after applying the Sobol sampled weights (colorbar). Since the ETG network configuration only has two output parameters informing the acquisition step, specifically the mode classifier and the electron heat flux regressor, the approximate Pareto boundary (red dashed line) of the multi-output acquisition separating the chosen and not chosen candidates can be easily visualized by directly plotting the candidates according to the weighted contribution of each of their outputs.}
	\label{fig:SampleETGParetoAcquisition}
\end{figure}

\begin{figure}[h]
	\centering
	\includegraphics[scale=0.35]{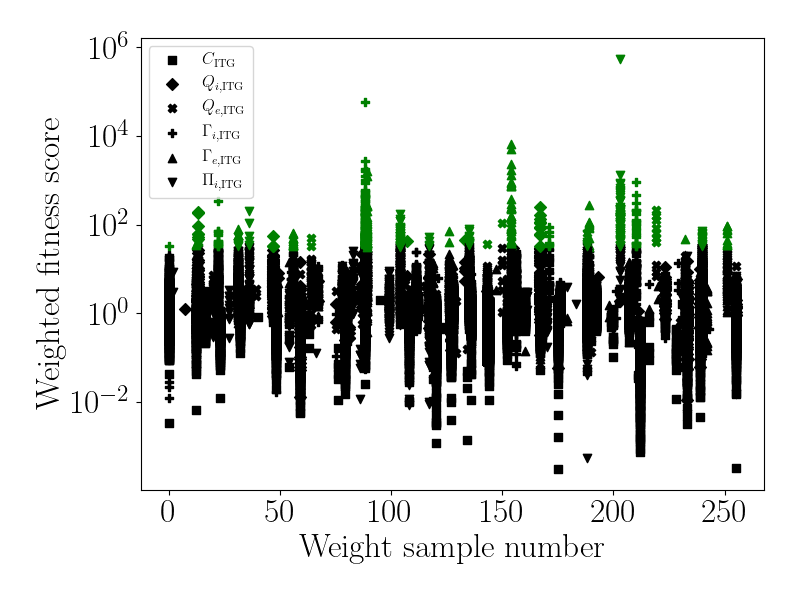}
	\caption{Sample acquisition from an AL iteration targeting the ITG turbulence mode differentiating the chosen candidates (green) from the rejected candidates (black) based on their maximum fitness score after applying the Sobol sampled weights indexed by the horizontal axis. Since the ITG network configuration has more than two output parameters informing the acquisition step, a different visualization was chosen where the dominant contribution to the maximum fitness score is indicated in the legend (symbols). The vertical structure indicates that some weight combinations from the Sobol sampler never result in a maximum fitness score for any candidates while others result in many maxima, but the sufficiency of the Sobol sampling to provide a reasonable distribution over both the weight combinations returning maxima and the dominant contributing output is clearly visible.}
	\label{fig:SampleITGParetoAcquisition}
\end{figure}

Thus, to provide a more generalized solution, a Sobol sampler~\cite{burhenneSobolSequences2011} was used to randomly generate a set of $M$ weight vectors, $w \in \left[0,1\right]$, which are then applied to the acquisition scores from each network for each potential candidate. By selecting the $N$ points with the highest total score using this method, there is greater assurance that the selected candidates have a variety of dominant contributors and thus providing a better resolution of the Pareto front of optimal improvement. Both the score scaling hyperparameters and the Sobol weight generator solutions were kept in the final implementation to ensure robustness, even though it is suspected that a more elegant method that combines them likely exists. Figures~\ref{fig:SampleETGParetoAcquisition} and \ref{fig:SampleITGParetoAcquisition} show an example of this multi-objective acquisition method for the ETG and ITG mode AL configurations, respectively, highlighting its sufficiency in diversifying the selection of optimal candidates.

While the acquisition function was only informed by the uncertainties associated with one turbulent mode, the QuaLiKiz executions continued to simultaneous produce the necessary output for all modes provided that they are unstable for the given local plasma inputs. For example, this means that running an iteration focused on exploring ITG modes can still produce training points containing TEM instabilities. This is not expected to be a problem for the AL efficiency, as the points are still being added to improve an output prediction and can even benefit the algorithm by allowing a secondary form of input space exploration.

Once the selected candidates are labelled via executing the QuaLiKiz code on their respective local plasma inputs, a number of additional physics-based data filters, described in Table~\ref{tbl:PhysicsFilters}, were applied to them as a post-processing step. As found in Ref.~\cite{hoNeuralNetworkSurrogate2021}, this reduces the amount of unexplained variance in the dataset, which in turn both improves the convergence robustness of the NN training routine and reduces the degree of undesired uncertainty expansion from the BNNs attempting to replicate the source of this noise. The labelled candidates which pass this filtering step are then appended to the training set to be used in the next AL algorithm iteration. Since the fraction of the labelled candidates which will be filtered out in post-processing cannot be known \emph{a priori}, it is often the case that this pipeline does not add exactly the number of points requested.

\begin{table*}[t]
	\centering
	\caption{Description of the physics consistency filters applied to the QuaLiKiz outputs for the purposes of this study. The last four filters are useful since QuaLiKiz performs separate integrations to compute the total fluxes, $\left\lbrace \Gamma, \, Q \right\rbrace$, compared to the diffusion and convection coefficients, $\left\lbrace D, \, \chi \right\rbrace$ and $\left\lbrace V_n, \, V_T \right\rbrace$, respectively, effectively providing a useful check for the consistency of the gyrokinetic eigenfunctions and linear spectra. More information on how these filters impact the QuaLiKiz dataset can be found in Ref.~\cite{hoNeuralNetworkSurrogate2021}.}
	\begin{tabular}{lcc}
		Filter description & Expression & Threshold \\
		\midrule
		Abnormally small particle flux & $\left|\Gamma_{\text{GB}}\right| = 0 \; \cup \; \left|\Gamma_{\text{GB}}\right| \ge 10^{-4}$ & -- \\
		Negative heat flux & $Q_{e,i,\text{GB}} \ge 0$ & -- \\
		Ambipolar particle flux & $\left|\sum_s \Gamma_{s,\text{GB}} \, Z_s\right| \le \epsilon \left|\Gamma_{e,\text{GB}}\right|$ & $\epsilon = 0.1$ \\
		Electron particle flux equivalence & $\left|\Gamma_{e,\text{GB}} - \varepsilon D_{e,\text{GB}} \left(R/L_{n_e}\right) - V_{n,e,\text{GB}}\right| \le \epsilon \left|\Gamma_{e,\text{GB}}\right|$ & $\epsilon = 0.05$ \\
		Ion particle flux equivalence & $\left|\Gamma_{i,\text{GB}} - \varepsilon D_{i,\text{GB}} \left(R/L_{n_i}\right) - V_{n,i,\text{GB}}\right| \le \epsilon \left|\Gamma_{i,\text{GB}}\right|$ & $\epsilon = 0.1$ \\
		Electron heat flux equivalence & $\left|Q_{e,\text{GB}} - \varepsilon \chi_{e,\text{GB}} \left(R/L_{T_e}\right) - V_{T,e,\text{GB}}\right| \le \epsilon \left|Q_{e,\text{GB}}\right|$ & $\epsilon = 0.05$ \\
		Ion heat flux equivalence & $\left|Q_{i,\text{GB}} - \varepsilon \chi_{i,\text{GB}} \left(R/L_{T_i}\right) - V_{T,i,\text{GB}}\right| \le \epsilon \left|Q_{i,\text{GB}}\right|$ & $\epsilon = 0.1$ \\
	\end{tabular}
	\label{tbl:PhysicsFilters}
\end{table*}

Since the applied filter preferentially removes unstable data points, the number of stable points included in the final selection of candidates are adjusted to ensure that a large class imbalance is not introduced as more iterations are performed, which would have a negative effect the performance of the trained classifiers. This down-selection of stable points was done by ranking them according to their classifier acquisition score and choosing points with the highest score until the number of stable points balanced the number of unstable points. While this method was proposed to also improve the resolution of the stability threshold, it can also bias the classifier contribution to the candidate fitness to explore edges of the known stable region not related to the stability boundary. While this exploration is not inherently detrimental to the classifier performance, the relative area of the exploration boundary compared to the stability boundary makes any unnecessary exploration undesirable under the restriction of efficiency. Nevertheless, the detailed development of an optimal acquisition function and subsequent filtering is beyond the scope of this study but is recommended for future work.

Overall, this filtering step effectively turns the dataset construction component of the AL algorithm into a semi-stochastic process in terms of data points added per iteration. This could become problematic if a given iteration adds an overwhelmingly large number of points to the training set, resulting in a significant distribution shift that ultimately lowers the computational efficiency of the generated dataset. In an attempt to correct this for the purposes for this study, the number of potential candidates passed to the labeller in a given iteration, $N$, was manually adjusted per iteration. Note that this exercise aimed to add $\sim$200 point per iteration, $\sim$100 unstable and $\sim$100 stable, without any change in schedule. This choice was made purely to simplify the number of changes introduced compared to the previous study~\cite{zanisiEfficientTrainingSets2024}, although it is suspected that an optimal schedule likely also exists.

\subsection{Pipeline execution setup}
\label{subsec:PipelineSetup}

This section details the various components required to execute the AL algorithm and how they were chosen for this particular study. In order to provide boundaries for the exploration-focused AL algorithm, all datasets were generated by sampling a multivariate Gaussian distribution over the 16 input variables and additionally bounded with upper and lower limits on specific variables. The mean and covariance matrix determining this multivariate Gaussian distribution was constructed based on the simplified statistics gathered from a previous QuaLiKiz dataset built around the parameter space of the JET tokamak~\cite{hoNeuralNetworkSurrogate2021}, then expanded along certain inputs in an attempt to cover other tokamaks. Further information concerning this method and the pool generation can be found in Appendix~\ref{app:MultivariateSampling}.


\textbf{Unlabelled pool:} A set of $5\times10^8$ QuaLiKiz input points were generated from the multivariate distribution.

\textbf{Seed set:} A set of 100 QuaLiKiz input points were generated from the multivariate distribution, independently from the unlabelled pool, and consequently labelled using QuaLiKiz to form the initial training set, or seed set. In principle, a factor 100 more QuaLiKiz evaluations were made in order to generate this set, both due to the physics filters applied to the QuaLiKiz outputs and that it was crucial that the seed set contains an approximate balance between stable / unstable points and across all of the various instability modes. This balance ensures that the initial trained model contains some information about the salient features in output space. This is only to accelerate progress in the first iterations as it is expected that the AL algorithm will eventually stumble upon those features through the exploratory process. While the bias towards exploration is higher in the chosen acquisition function, described below, there is equally no guarantee that discovering regions with new salient features will happen in an efficient manner.

\textbf{Validation set:} A set of 1000 QuaLiKiz input points were generated from the multivariate distribution, independently from both the unlabelled pool and seed set, and consequently labelled using QuaLiKiz to form the validation set. While the final dataset was examined to ensure points exist which are stable and unstable for each turbulent mode, it was not modified to ensure balance between them. Based on previous attempts, it was found that maintaining a constant validation set was required to observe the improvements provided by the AL algorithm. It is suspected that the potentially large distribution shifts introduced to the training set by the AL algorithm cause a self-sampled validation set to effective become a moving target for the early stopping trigger of successive iterations, giving mixed results about both the trained model performance and AL efficiency.

\textbf{Test set:} A set of 1000 QuaLiKiz input points were sampled from the multivariate distribution, independently from the unlabelled pool, seed set and validation set, and consequently labelled by QuaLiKiz to form the test set. This method guarantees that it closely resembles the unlabelled pool due to the considerations around distribution mismatches explored in Appendix~\ref{app:DistributionMismatch}.

\textbf{Model training hyperparameters:} Tables~\ref{tbl:ClassifierTrainingHyperparameters} and \ref{tbl:RegressorTrainingHyperparameters} show the hyperparameters chosen for training the classifier networks and the regressor networks, respectively. Note that the same training hyperparameters were used for all modes and all transport fluxes. While it is expected that different target fluxes could benefit from different training hyperparameter configurations, it is unknown how much of this benefit comes from the fixed and information-dense nature of typical training sets. Since the AL algorithm starts from a very sparse training set which shifts as it progresses, it was deemed useful to standardize these hyperparameters across all training executions to provide a more stable foundation for interpreting any results. More information regarding the meanings of these hyperparameters can be found in Ref.~\cite{liuSimpleprincipleduncertaintyestimation2020} and \cite{paneraalvarezEuroPEDNNUncertaintyAware2024} for the SNGP and BNN architectures, respectively.

\begin{table}[t]
	\centering
	\caption{List of critical NN training hyperparameters for the SNGP network architecture used for the classification component of the turbulent transport flux prediction and the corresponding values used within this study.}
	\begin{tabular}{cc}
		Hyperparameter & Value \\
		\midrule
		Batch size & 64 \\
		Early stopping & 500 \\
		Neurons & [5000, 5000] \\
		Number of classes & 1 \\
		Entropy weight & 10 \\
		Optimizer & Adam \\
		Learning rate & ${10}^{-3}$ \\
		First moment decay & 0.9 \\
		Second moment decay & 0.999 \\
		Learning decay rate & 0.99 \\
		Learning decay epoch & 20 \\
	\end{tabular}
	\label{tbl:ClassifierTrainingHyperparameters}
\end{table}

\begin{table}[t]
	\centering
	\caption{List of critical NN training hyperparameters for the BNN network architecture used for the regression component of the turbulent transport flux prediction and the corresponding values used within this study.}
	\begin{tabular}{cc}
		Hyperparameter & Value \\
		\midrule
		Batch size & 64 \\
		Early stopping & 500 \\
		Neurons & [3000, 3000, 3000] \\
		Number of outputs & 1 \\
		L1 regularization weight & 0.2 \\
		L2 regularization weight & 0.8 \\
		OOD width & 0.5 \\
		Epistemic prior & ${10}^{-3}$ \\
		Aleatoric prior & ${10}^{-3}$ \\
		NLL weight & 1 \\
		Epistemic weight & 0.1 \\
		Aleatoric weight & 0.1 \\
		Regularization weight & ${10}^{-3}$ \\
		Optimizer & Adam \\
		Learning rate & ${10}^{-3}$ \\
		First moment decay & 0.9 \\
		Second moment decay & 0.999 \\
		Learning decay rate & 0.99 \\
		Learning decay epoch & 5 \\
	\end{tabular}
	\label{tbl:RegressorTrainingHyperparameters}
\end{table}

\textbf{Acquisition function:} Since the chosen classifier architecture operates using logits, it provides an uncertainty based on the probability represented by the logit output that is naturally bounded within $\sigma_{\text{cls}} \in \left[0,1\right]$. This uncertainty can then be used to define a simple classifier acquisition function for this study expressed as:
\begin{equation}
	\label{eq:ClassifierAcquisition}
	f_{\text{cls}} = \gamma_{\text{cls}} \, \sigma_{\text{cls}}
\end{equation}
where $\gamma_{\text{cls}}$ is a user-defined hyperparameter controlling the relative contribution of the classifier to the acquisition score. However, the regressor uncertainties, $\sigma_{\text{reg}}$, can span multiple orders of magnitude and thus selecting points based on the maximum of an acquisition function which relies solely on absolute uncertainty, $\sigma$, will generally bias towards regions of higher normalized flux. Due to this, it is useful to provide a counterbalance in the acquisition function that preferentially selects lower fluxes, especially as it is known that downstream applications of these networks often require better resolution near of the critical turbulent thresholds where the normalized fluxes are smaller. This counterbalance was determined to be the relative uncertainty, $\sigma/\mu$, leading to a regressor acquisition function of the form:
\begin{equation}
	\label{eq:RegressorAcquisition}
	\small
	f_{\text{reg}} = \begin{cases}
		\mu_{\text{cls}} \, \gamma_{\text{reg}} \left(f_{\text{abs}} + \gamma_{\text{rel}} \, f_{\text{rel}}\right), & \text{if $\left|\mu_{\text{reg}}\right| \le 600$}\\
		0, & \text{otherwise}
	\end{cases}
\end{equation}
where $\mu_{\text{cls}}$ is the output of the corresponding mode classifier (0 or 1), $\gamma_{\text{reg}}$ is a user-defined hyperparameter which controls the overall scale of the specific regressor acquisition function term, $\gamma_{\text{rel}}$ is a user-defined hyperparameter which controls the contribution of the relative uncertainty term and
\begin{equation}
	\label{eq:UncertaintyAwareAcquisition}
	\begin{gathered}
		f_{\text{abs}} = \sigma_{\text{reg},m} + \gamma_{\text{abs},d} \, \sigma_{\text{reg},d} \\
		f_{\text{rel}} = \frac{\sigma_{\text{reg},m}}{\mu_{\text{reg}}} + \gamma_{\text{rel},d} \, \frac{\sigma_{\text{reg},d}}{\mu_{\text{reg}}}
	\end{gathered}
\end{equation}
where $\gamma_{\text{abs},d}$ and $\gamma_{\text{rel},d}$ are user-defined hyperparameters which control the relative impact of the data uncertainty compared to the model uncertainty represented by the specific BNN model chosen as the regressor architecture for this study. Setting $\gamma_{\text{abs},d}$ and $\gamma_{\text{rel},d}$ to be negative allows the acquisition function to discount regions where a large uncertainty is due to an inherent scatter of the output variable locally within the training data as opposed to a complete lack of data. Note that a flux cap of 600 in GB units is applied to the regressor acquisition function, Equation~\eqref{eq:RegressorAcquisition}, to pre-emptively minimize QuaLiKiz evaluations which might be filtered out by the large flux physics filter anyway. Then, using Equations~\eqref{eq:ClassifierAcquisition} and \eqref{eq:RegressorAcquisition}, the final AL acquisition function for a given iteration, trained to target only one turbulent mode at a time, was simply implemented in this study as the weighted sum for each network expressed as:
\begin{equation}
	\label{eq:ModeAcquisition}
	f_j = w_j f_{\text{cls}, j} + \sum_k w_k f_{\text{reg}, k, j}
\end{equation}
where the subscript $j$ represents the turbulent mode (i.e. ITG, TEM, or ETG), the subscript $k$ represents the transport flux channels (i.e. $Q_e$, $Q_i$, $\Gamma_e$, $\Gamma_i$, or $\Pi_i$) driven by that mode, and $w$ represents the weights generated by the Sobol sampler to discover the Pareto front of optimal improvement.

\textbf{Acquisition function hyperparameters:} An educated guess for the hyperparameters of the AL acquisition functions were determined by training a model on the validation set and evaluating the acquisition functions using the outputs of those models on another 1000 input points independently sampled from the multivariate function. The chosen hyperparameter values were then selected such that they would bring the 90th percentile of these acquisition function values for each objective to $f \simeq 1$. This acts to remove potential biases that natural orders of magnitude have on the contributions of each model on the final acquisition function, given by Equation~\eqref{eq:ModeAcquisition}, and also allows each element of the weight vector can be sampled from $w \in \left[0, 1\right]$ for ease of implementation. The final selection of acquisition hyperparameters chosen for this study can be found in Table~\ref{tbl:AcquisitionHyperparameters}, though it should be noted that these have only been selected based on broadly generalized criteria. While this was deemed suitable for the purposes of this study, a more rigourous hyperparameter tuning exercise is recommended as future work, especially since the values chosen to be suitable in the early iterations were observed to be less effective in later ones.

\begin{table}[h]
	\centering
	\caption{The acquisition scoring hyperparameters associated with this AL implementation and the values chosen for them within this study. While all of hyperparameters are explicitly listed here as implemented, note that the chosen values for the various $\gamma_{\text{rel}}$, $\gamma_{\text{abs},d}$, and $\gamma_{\text{rel},d}$ are equal across all the transport channels and all turbulence modes. While it was not extensively studied in this work, it is strongly suspected they can be compressed into a single $\gamma_{\text{rel}}$, $\gamma_{\text{abs},d}$, and $\gamma_{\text{rel},d}$, regardless of the number of outputs, as they likely depend more on the BNN architecture than the specific application.}
	\begin{tabular}{cccc}
		& \multicolumn{3}{c}{Values} \\
		Hyperparameter & ITG & TEM & ETG \\
		\midrule
		$\gamma_{\text{cls}}$ & 2 & 2 & 2 \\
		$\gamma_{\text{reg},Q_e}$ & ${10}^{-5}$ & ${10}^{-5}$ & ${10}^{-5}$ \\
		$\gamma_{\text{rel},Q_e}$ & ${10}^{4}$ & ${10}^{4}$ & ${10}^{4}$ \\
		$\gamma_{\text{abs},d,Q_e}$ & -0.5 & -0.5 & -0.5 \\
		$\gamma_{\text{rel},d,Q_e}$ & -0.5 & -0.5 & -0.5 \\
		$\gamma_{\text{reg},Q_i}$ & ${10}^{-5}$ & ${10}^{-5}$ & -- \\
		$\gamma_{\text{rel},Q_i}$ & ${10}^{4}$ & ${10}^{4}$ & -- \\
		$\gamma_{\text{abs},d,Q_i}$ & -0.5 & -0.5 & -- \\
		$\gamma_{\text{rel},d,Q_i}$ & -0.5 & -0.5 & -- \\
		$\gamma_{\text{reg},\Gamma_e}$ & $5 \times {10}^{-5}$ & $5 \times {10}^{-5}$ & -- \\
		$\gamma_{\text{rel},\Gamma_e}$ & ${10}^{4}$ & ${10}^{4}$ & -- \\
		$\gamma_{\text{abs},d,\Gamma_e}$ & -0.5 & -0.5 & -- \\
		$\gamma_{\text{rel},d,\Gamma_e}$ & -0.5 & -0.5 & -- \\
		$\gamma_{\text{reg},\Gamma_i}$ & $5 \times {10}^{-5}$ & $5 \times {10}^{-5}$ & -- \\
		$\gamma_{\text{rel},\Gamma_i}$ & ${10}^{4}$ & ${10}^{4}$ & -- \\
		$\gamma_{\text{abs},d,\Gamma_i}$ & -0.5 & -0.5 & -- \\
		$\gamma_{\text{rel},d,\Gamma_i}$ & -0.5 & -0.5 & -- \\
		$\gamma_{\text{reg},\Pi_i}$ & ${10}^{-4}$ & ${10}^{-4}$ & -- \\
		$\gamma_{\text{rel},\Pi_i}$ & ${10}^{4}$ & ${10}^{4}$ & -- \\
		$\gamma_{\text{abs},d,\Pi_i}$ & -0.5 & -0.5 & -- \\
		$\gamma_{\text{rel},d,\Pi_i}$ & -0.5 & -0.5 & -- \\
	\end{tabular}
	\label{tbl:AcquisitionHyperparameters}
\end{table}

\textbf{Computing system:} Both the SNGP and BNN models were re-implemented in TensorFlow 2.15.1 (Python 3.11) to allow customization of the training loop, loss functions and data logging\footnote[1]{\label{foot:GitHubPackage}The customized model architectures, loss functions, and training routines are available open-source on GitHub at \url{https://github.com/aaronkho/low-cost-bnn}}. At the upper end of the dataset size ($\sim$10000 points), these SNGP model took between 1--4 hours of walltime to train and the BNN model took between 0.5--1.5 hours of walltime to train, all when parallelized over 4 NVIDIA A100 GPUs.

\section{Active learning results}
\label{sec:Results}

\subsection{Trained network performance}
\label{subsec:NetworkPerformance}

With the seed set, the validation set, and the unlabelled pool defined above, a total of 45 iterations were carried out using the pipeline described above, bringing the constructed training set to a total of $\sim10^{4}$ points. The target turbulent mode for each consecutive iteration was selected by iterating on a repeated list containing ITG, TEM, and ETG, in that order. The candidate buffer size was adjusted between 20 -- 200 every iteration in order to target an addition of $\sim$100 -- 300 new data points per iteration. These adjustments were necessary as the AL acquisition function contains no way to predict where the QuaLiKiz code output does not meet the physics criteria within the post-processing filter. It was observed that the buffer required for the ETG acquisition iterations was $\sim$3 times larger than for the ITG and TEM acquisition iterations, meaning that it is harder for QuaLiKiz to sufficiently resolve points containing ETG fluxes with the strict physics-based filters applied. The number of AL iterations performed in this study was determined purely by time constraints and it is expected that more iterations can be carried out to continue expanding the dataset and further improving the trained models.

\begin{figure*}[tb]
	\centering
	\includegraphics[scale=0.35]{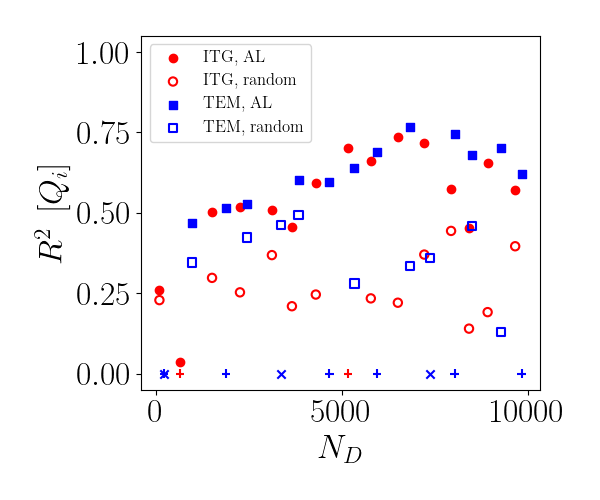}%
	\includegraphics[scale=0.35]{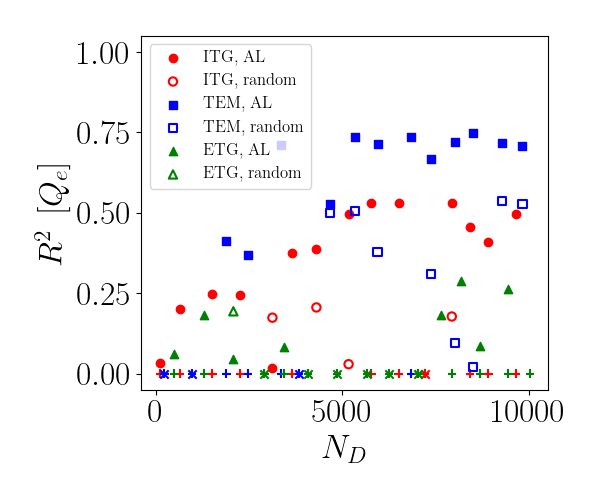}\\
	\includegraphics[scale=0.35]{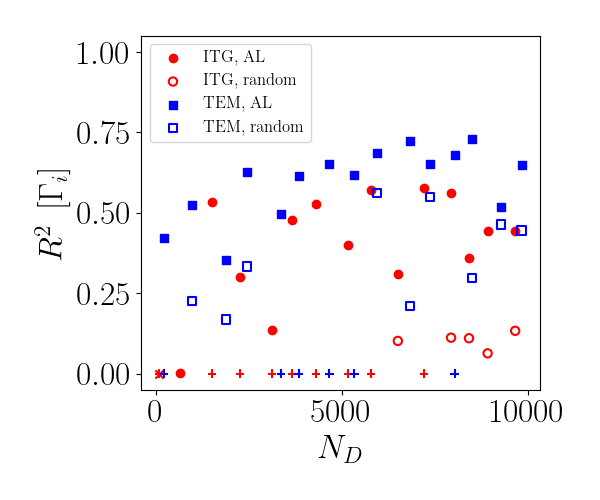}%
	\includegraphics[scale=0.35]{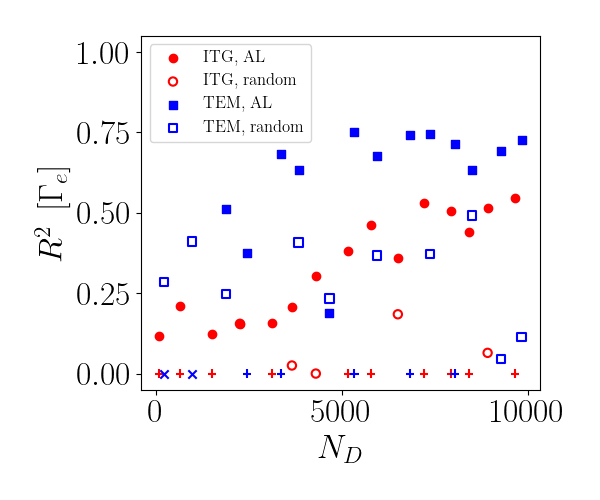}%
	\includegraphics[scale=0.35]{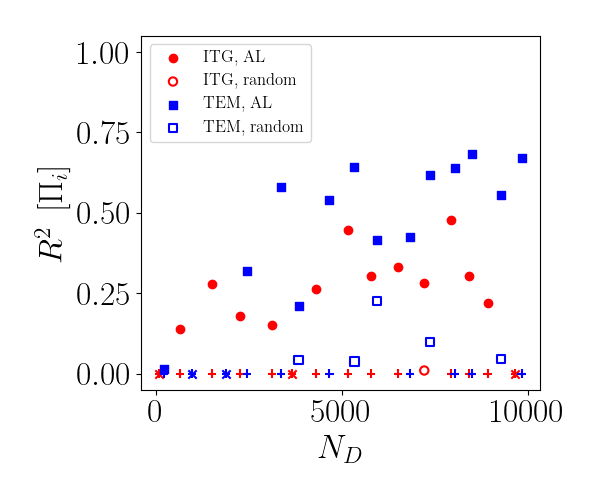}
	\caption{Regressor NN model performances, as measured by the $R^2$ metric, as a function of the training set size using the proposed AL pipeline and acquisition function (solid) and using random sampling (hollow) for the ion heat flux, $Q_i$ (top left), electron heat flux, $Q_e$ (top right), ion particle flux, $\Gamma_i$ (bottom left), electron particle flux, $\Gamma_e$ (bottom center), and ion momentum flux, $\Pi_i$ (bottom right). As these fluxes can be driven by multiple turbulent modes, results from the ITG (red), TEM (blue), and ETG (green) networks are shown simultaneously. Due to the rotating list of target turbulent modes per AL iteration, the points corresponding to each mode are staggered in their location on the horizontal axis. Models returning $R^2 \le 0$ for the AL pipeline (crosses) and the random sampling (plusses) have been set to zero for ease of interpreting the positive results.}
	\label{fig:FluxRegressorR2}
\end{figure*}

Figure~\ref{fig:FluxRegressorR2} shows the performance of the various turbulent mode regressor networks predicting the test set as a function of the training set size, organized based on the transport channels being predicted. As shown, the $R^2$ performance metric of the regressor networks on this test set generally improves with the number of data points. For comparison, the performance of a set of regressor networks trained using a completely random acquisition with the same number of added points from the unlabelled pool is also shown. This demonstrates that the AL algorithm is able to efficiently identify regions in the input space which require more labelled data to obtain a better overall model. While this property is already established from the previous ADEPT study~\cite{zanisiEfficientTrainingSets2024}, the novelty shown here is that this configuration of the AL algorithm combined with these specific network architectures and acquisition functions can adequately perform this role and that this approach extends well to the multi-objective AL application attempted here.

By comparing the regressor performance gains across the output variables, it appears that the momentum flux, $\Pi_i$, networks were not sufficiently prioritized in the later iterations of this study. This shifting bias is a known issue within multi-objective optimization in general. It is suspected that fine-tuning of the acquisition hyperparameters outlined in Table~\ref{tbl:AcquisitionHyperparameters} would help circumvent this effect. Thus, while the chosen hyperparameters for this study were determined by a rudimentary statistical analysis, a production-oriented application of this method would benefit from a more rigourous tuning effort, which was not performed here.

\begin{figure}[h]
	\centering
	\includegraphics[scale=0.35]{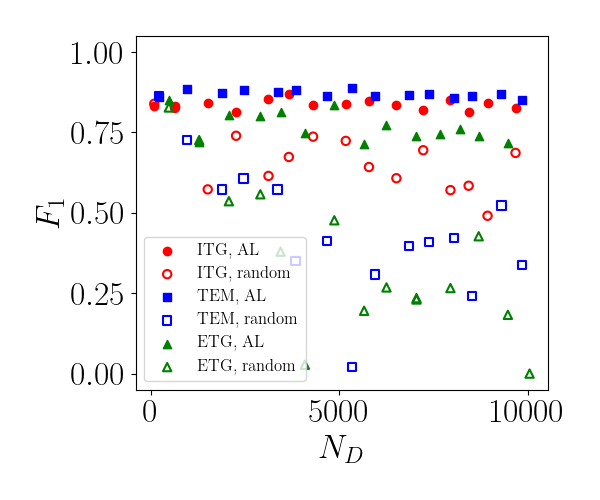}
	\caption{Classifier NN model performance, as measured by the $F_1$ metric, as a function of the training set size using the proposed AL pipeline and acquisition function (solid) and using random sampling (hollow). Results from the ITG (red), TEM (blue), and ETG (green) networks are shown simultaneously. Due to the rotating list of target turbulent modes per AL iteration, the points corresponding to each mode are staggered in their location on the horizontal axis.}
	\label{fig:ModeClassifierF1}
\end{figure}

Similarly, Figure~\ref{fig:ModeClassifierF1} shows the performance of the various classifier networks as a function of the training set size. Also similarly to Figure~\ref{fig:FluxRegressorR2}, the performance of a set of classifier networks trained using a completely random acquisition with the same number of added points from the unlabelled pool is also shown. Unfortunately, there is almost no improvement of the classifier as a function of training set size although this is still notably better than using the random acquisition. Naturally, this outcome can be attributed to the ability of the sequential algorithm to enforce a balance between the number of stable and unstable points added to the training set. While the algorithm takes advantage of the uncertainty-aware nature of the chosen classifier model to preferentially select stable points near its decision boundary, the lack of overall improvement using the AL algorithm may indicate that these decision boundaries are not synonymous to the turbulence instability boundary. This is possible since the SNGP prediction uncertainty can also be large at the edges of the dataset, meaning that the chosen acquisition function may actually bias towards exploration depending on the relative ``surface area'' of the instability boundary compared to the entire stable dataset boundary. Regardless, since the trained classifiers generally were able to reach $F_1 \simeq 0.85$ for the ITG and TEM networks on the test for this study and $F_1 \simeq 0.75$ for the ETG networks, they were deemed acceptable enough to demonstrate the benefits of using uncertainty-aware classifier models. A more detailed investigation into improving the classifier acquisition function or developing a better metric to demonstrate its performance is recommended as future work.

\begin{figure*}
	\centering
	\includegraphics[scale=0.35]{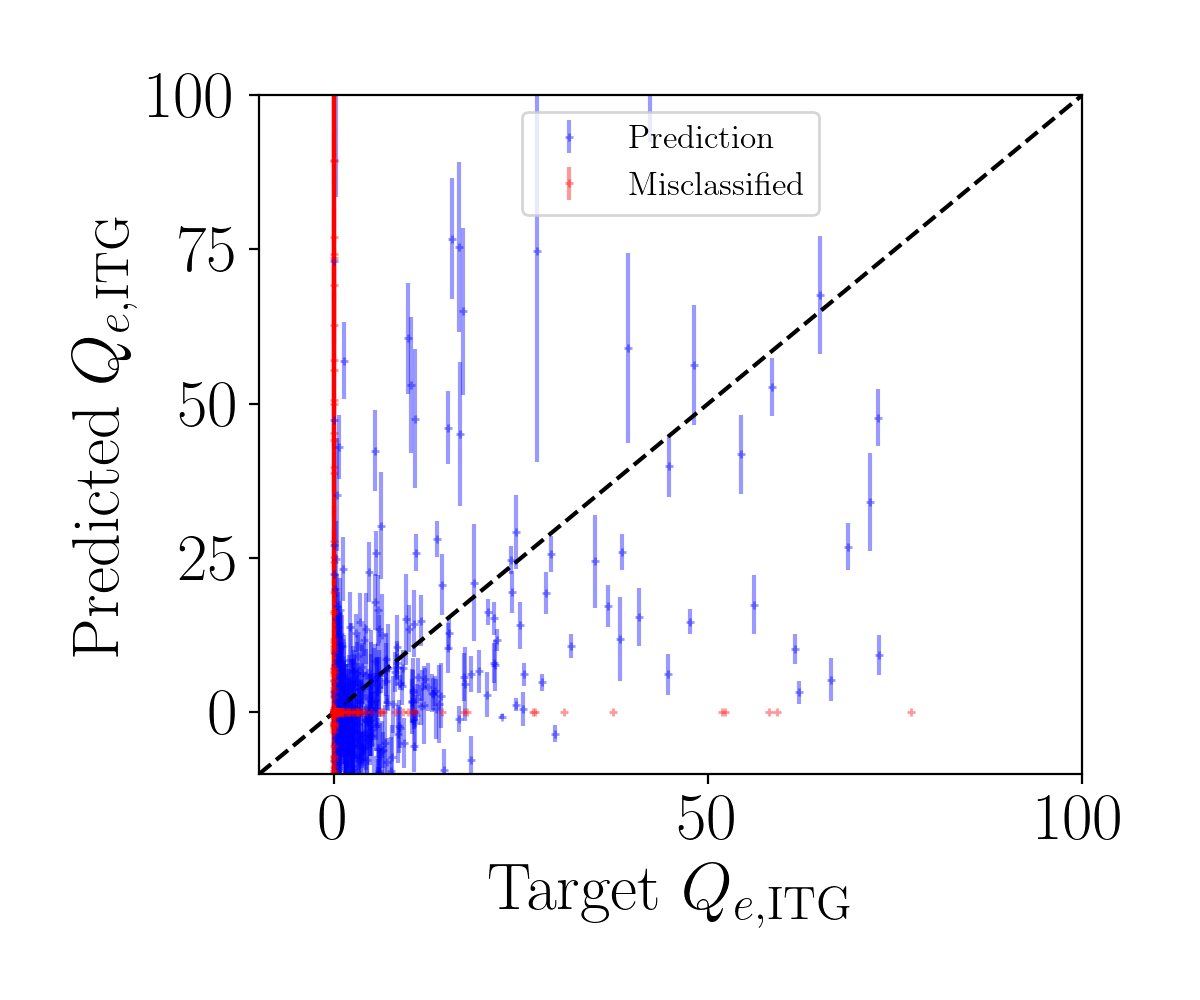}%
	\includegraphics[scale=0.35]{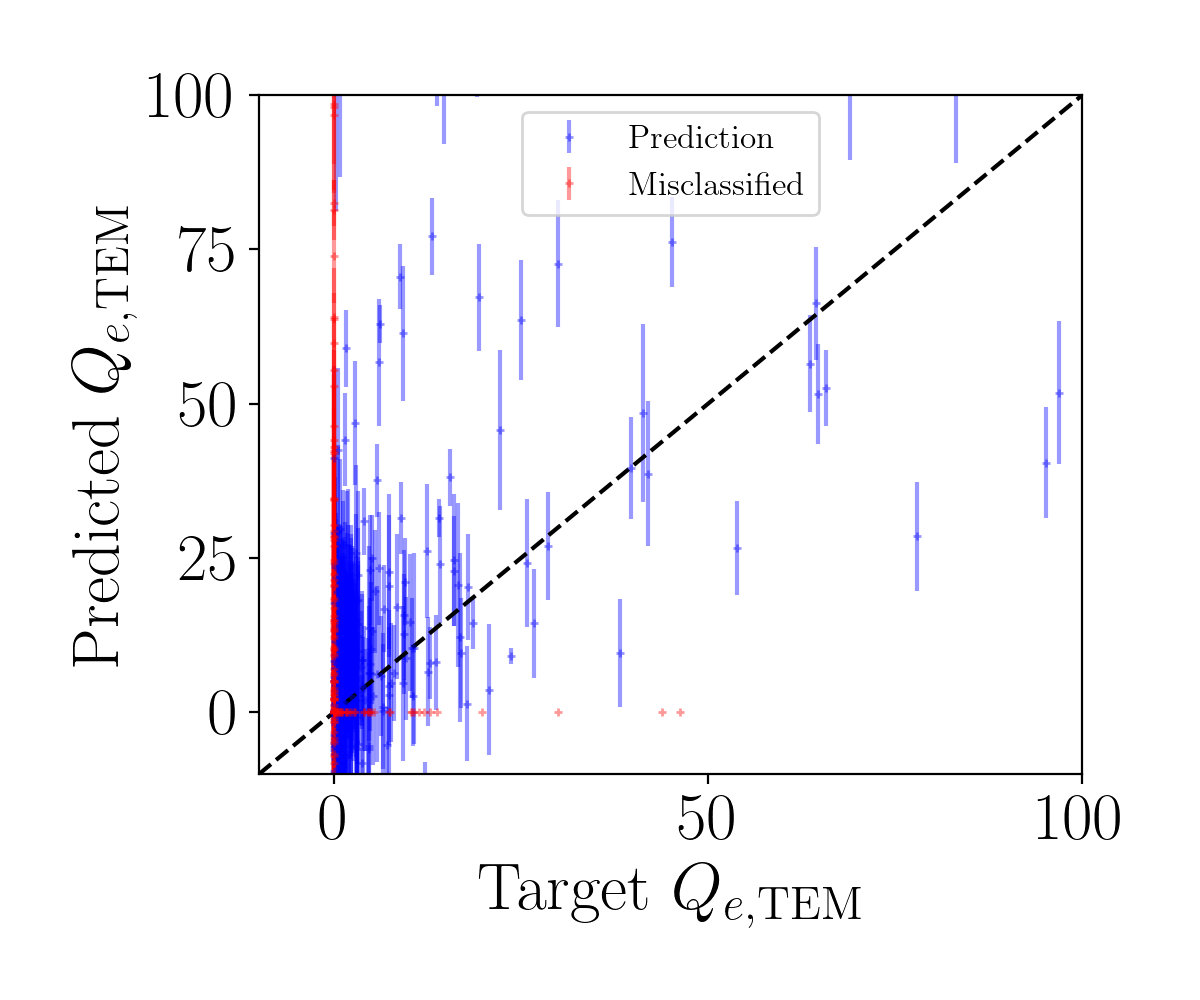}%
	\includegraphics[scale=0.35]{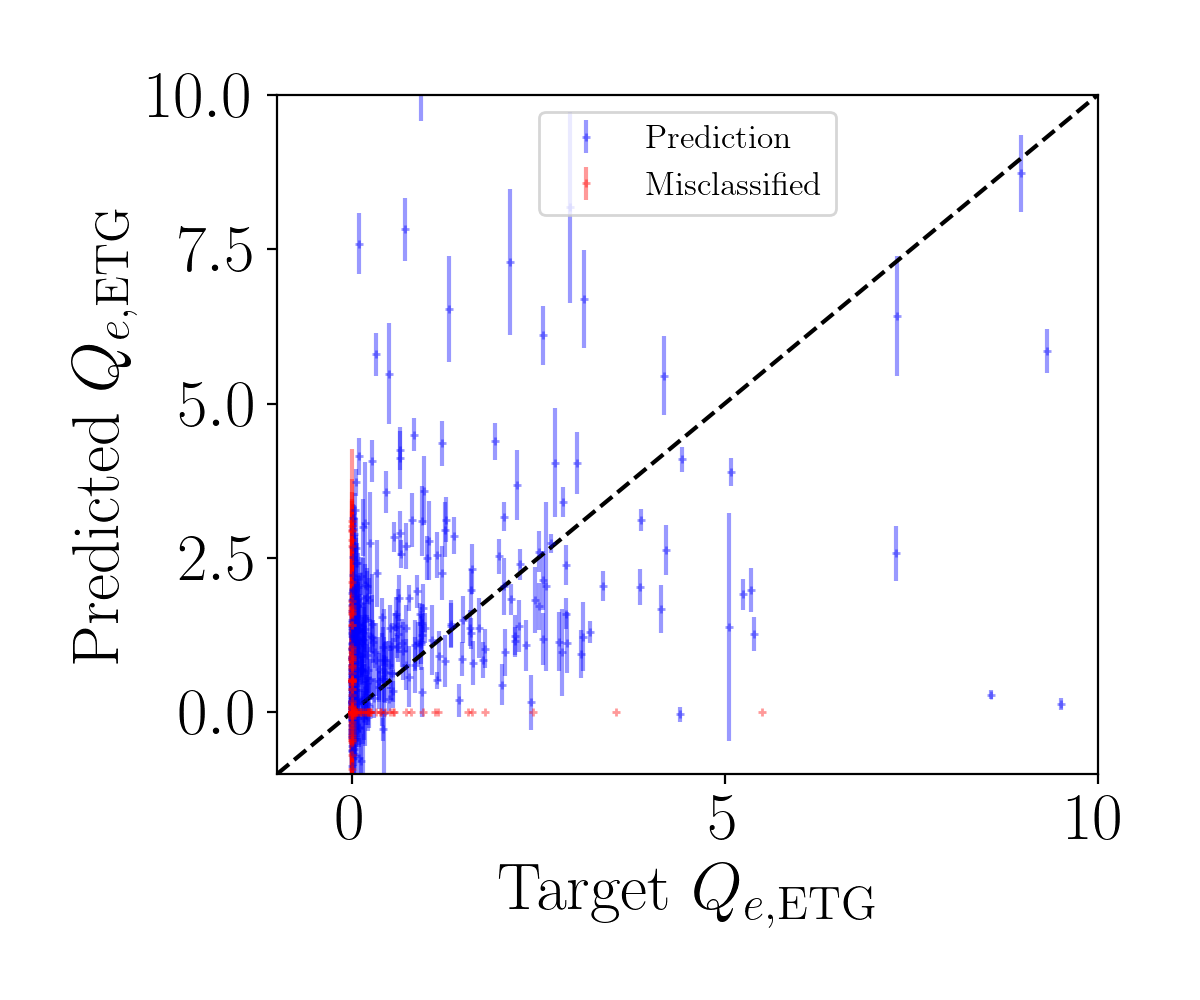}\\
	\includegraphics[scale=0.35]{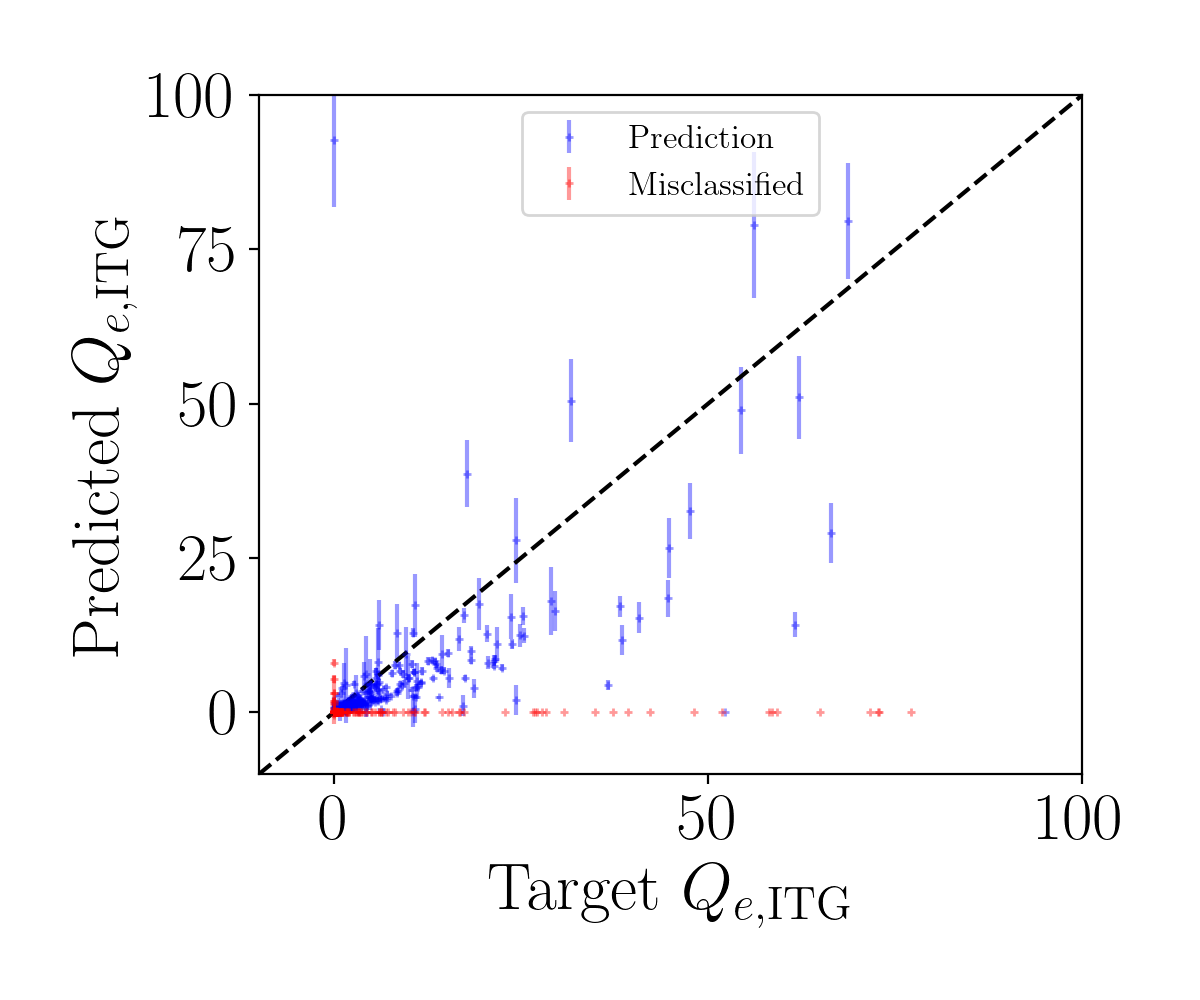}%
	\includegraphics[scale=0.35]{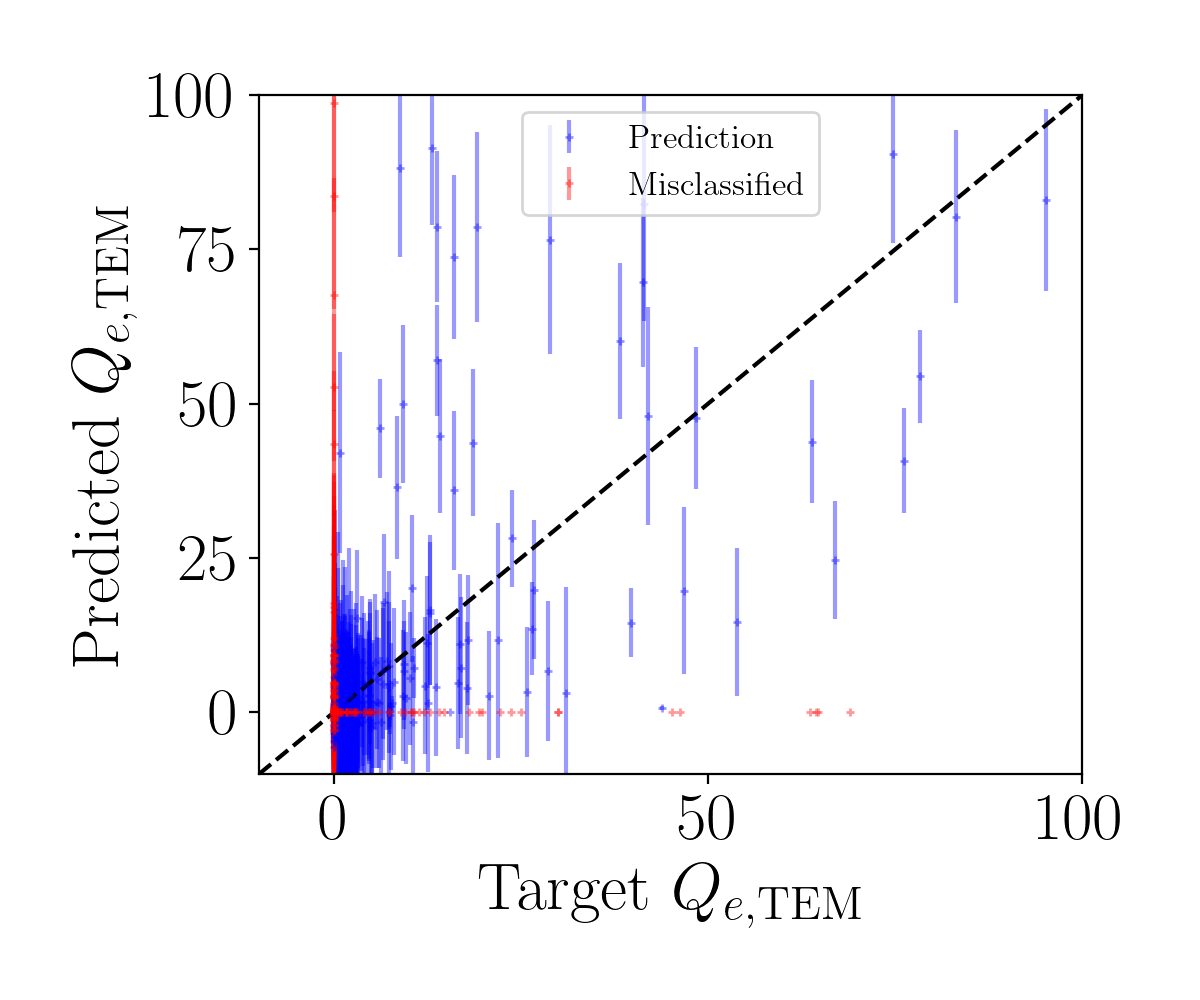}%
	\includegraphics[scale=0.35]{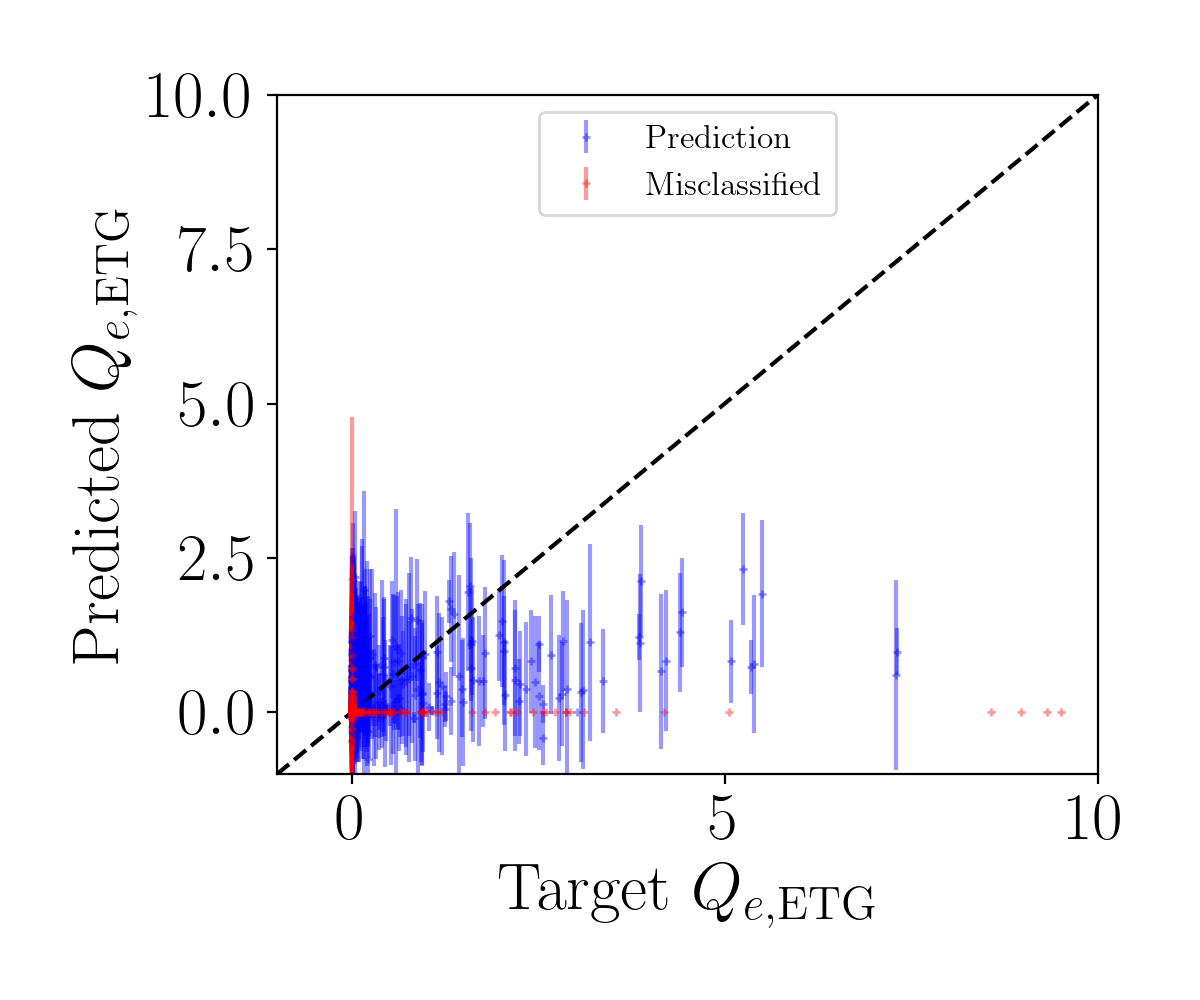}\\
	\includegraphics[scale=0.35]{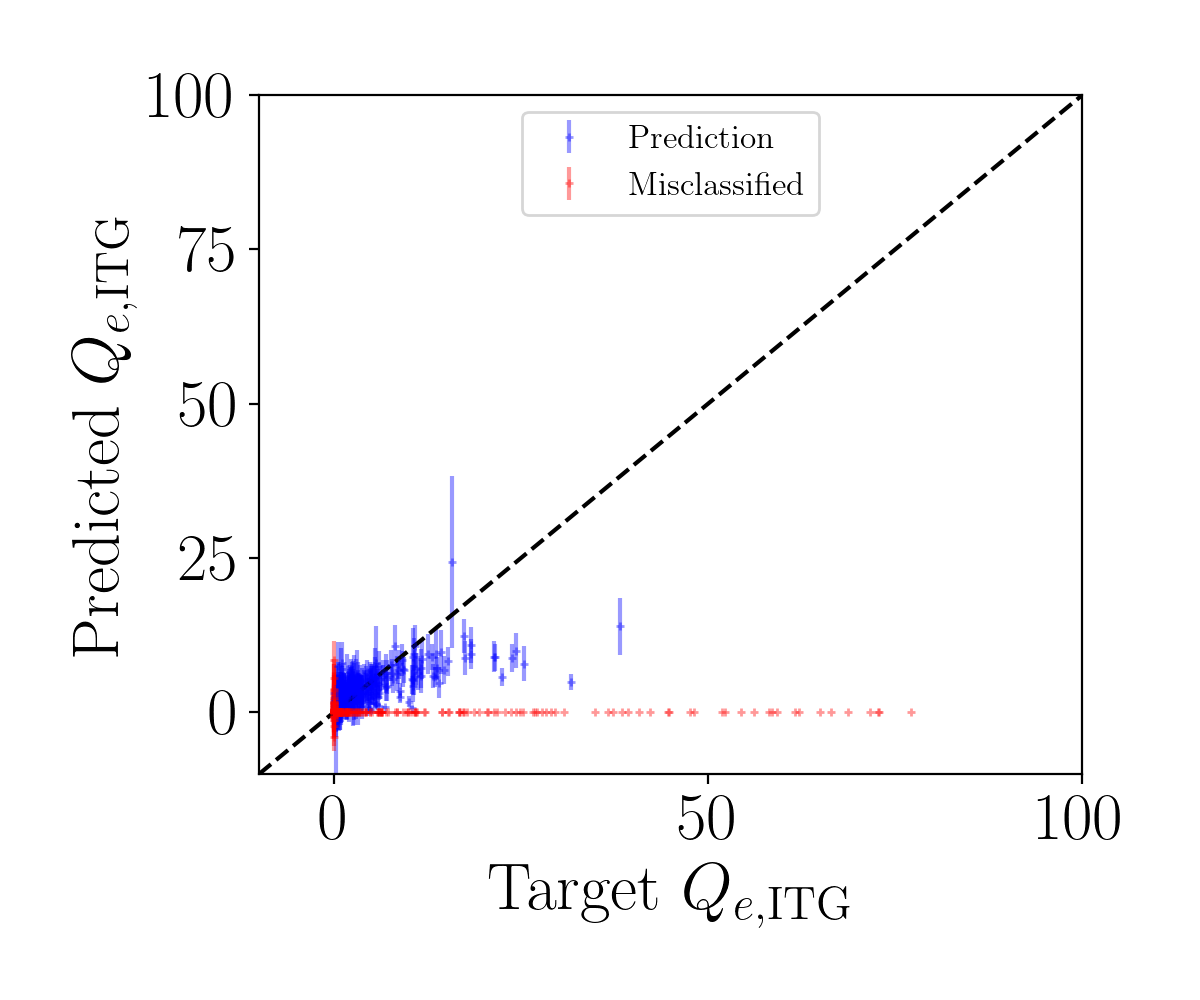}%
	\includegraphics[scale=0.35]{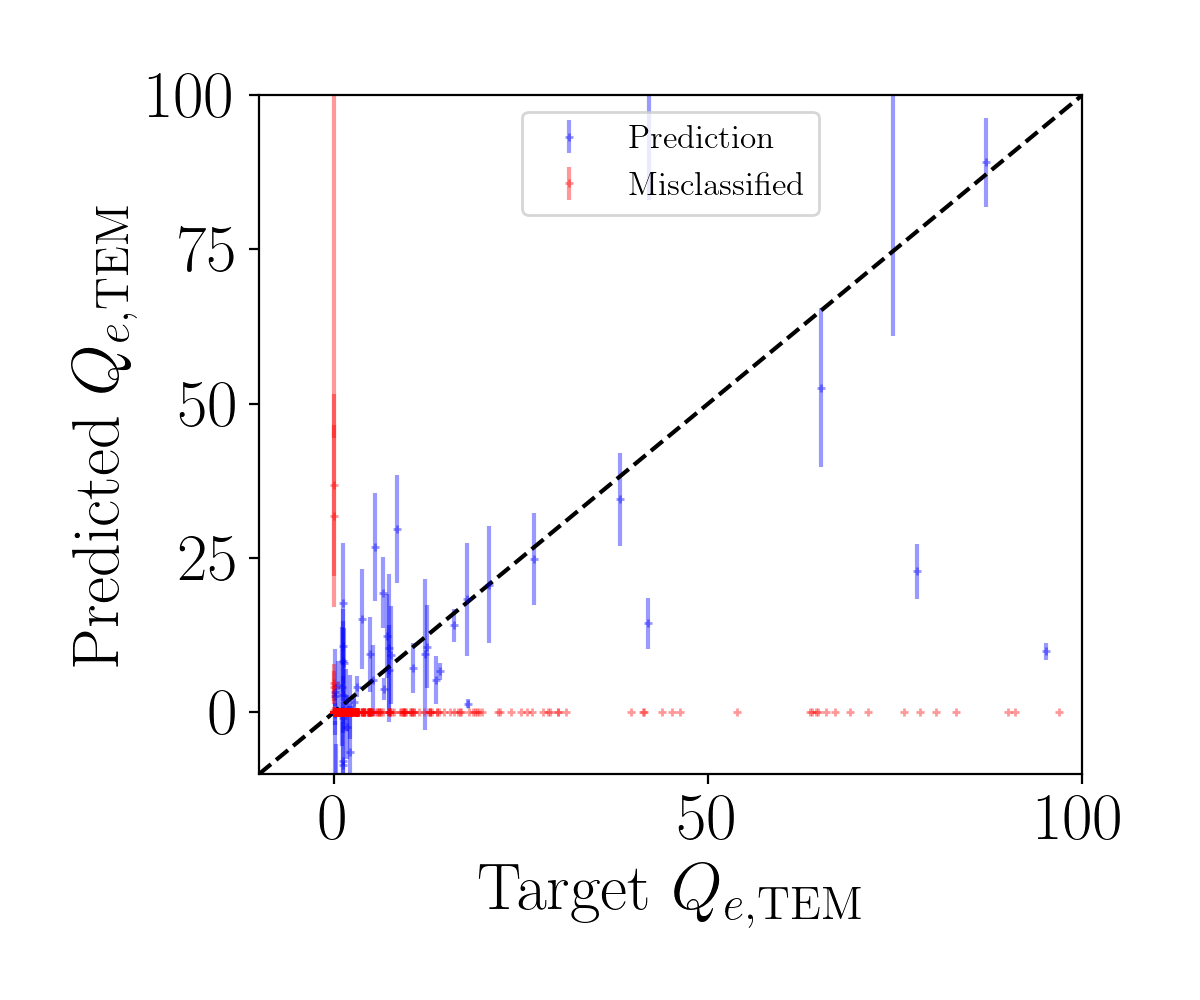}%
	\includegraphics[scale=0.35]{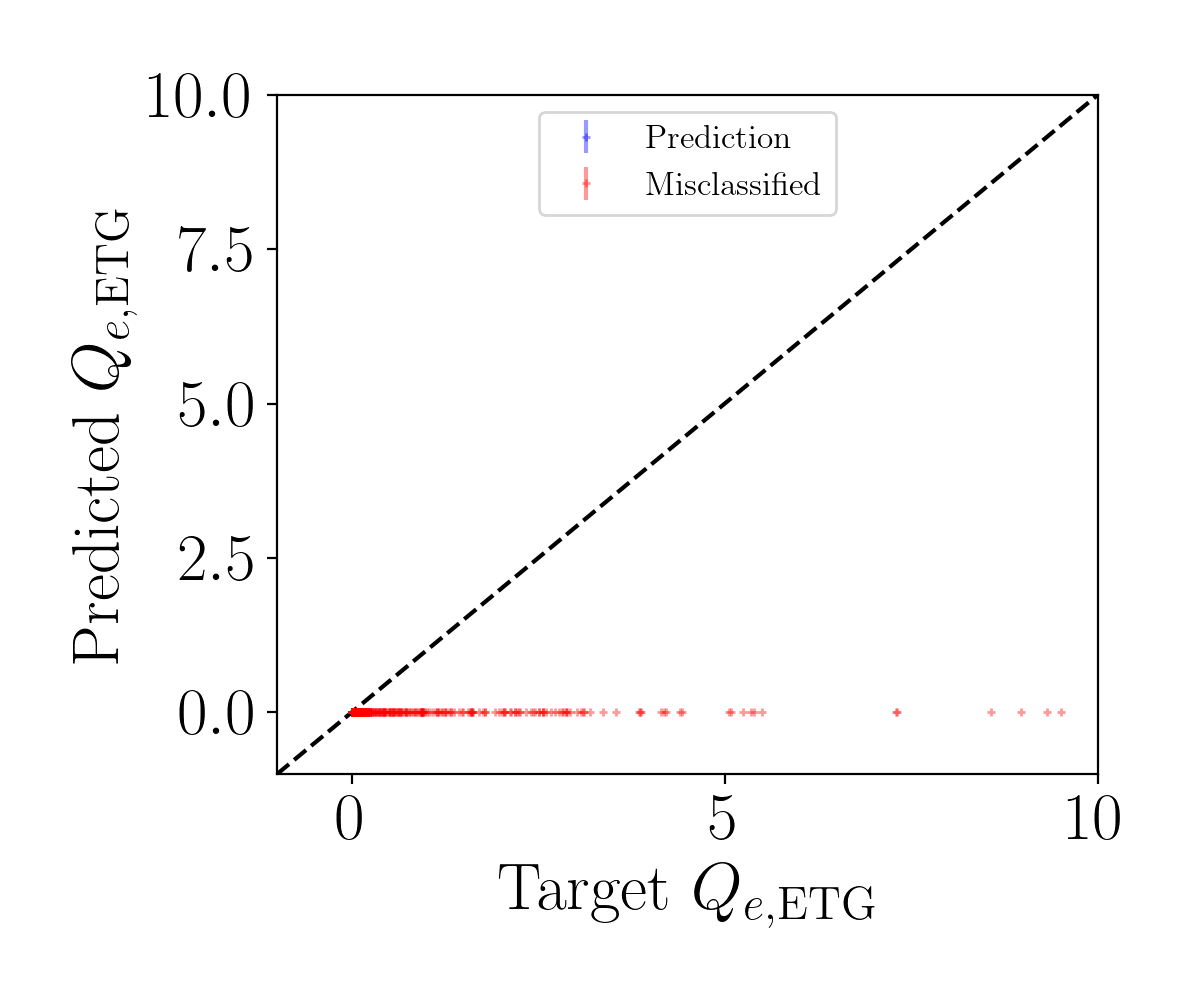}
	\caption{Comparison of the predicted electron heat flux, $Q_e$, against the target flux as evaluated by QuaLiKiz for the ITG (left), TEM (center) and ETG (right) turbulent modes at the initial (top) and the final iteration using the proposed AL pipeline and acquisition function (middle) and using random sampling (bottom), with any misclassified points depicted in red points. These demonstrate the general improvement of the models due to the informed selection of new training points from the AL algorithm, though it is noted that the performance at this volume of data is not yet sufficient for the models to be used in downstream applications, best exemplified by the performance of the $Q_{e,\text{TEM}}$ network.}
	\label{fig:ElectronHeatFluxPerformance}
\end{figure*}

Figure~\ref{fig:ElectronHeatFluxPerformance} compares the predicted electron heat flux, $Q_e$, computed from the requisite classifier and regressor network pairs using Equation~\eqref{eq:NetworkBlending}, against the real target flux from QuaLiKiz from the independent test set, for all three turbulent modes and for the first and last AL iterations executed in this study. These plots give more insight into the performance gain in the networks via the AL algorithm, as reflected by the $R^2$ performance metric previously shown, and the remaining output variable networks can be found in Appendix~\ref{app:NetworkPerformance}. However, as shown by Figure~\ref{fig:FluxRegressorR2}, this performance gain is not so evident in the $Q_{e,\text{ETG}}$ network which may indicate either that the ETG turbulence domain in input space is inherently more vast, requiring even more data to gain any form of predictive power, or that the applied physics filters preferentially reject points with $Q_{e_\text{ETG}} \ne 0$, hinting at the difficulties of QuaLiKiz to return numerically consistent outputs when electron scale turbulence is present and appreciable. Given the relatively poor statistics of ETG presence within QuaLiKiz predictions in JET experimental data space~\cite{hoNeuralNetworkSurrogate2021} and the general dominance of ITG modes in reactor-relevant tokamak plasma regimes~\cite{hollandReactorRelevantITG2023}, this is not foreseen to be problematic for extension into higher-fidelity turbulence applications. Additionally, for the purposes of this AL algorithm investigation, it does not detract from the utility of using uncertainty-aware networks on relatively small volumes of data.


\subsection{Dataset construction}
\label{subsec:DatasetConstruction}

The interest in using AL techniques is not just to produce a final surrogate model but also to generate a data efficient description of the input space necessary to describe the physics problem within the chosen input parameter space. In order to determine whether the implemented AL algorithm demonstrated the latter capability, the KL-divergence was used to evaluate the differences between the 1D parameter distributions of the constructed dataset, $D$, and the unlabelled pool, $P$. Due to the extremely large number of points within the unlabelled pool, it was used as a proxy for the clipped multivariate Gaussian distribution described in Appendix~\ref{app:MultivariateSampling}, removing to need to integrate such a complicated function to compute the continuous KL-divergence. It should be noted that, while this metric was deemed unsuitable to act as a loss function term for NN training, the KL-divergence is still adequate at quantifying the distance between two arbitrary distributions, where a value of 0 means the two distributions are identical and large positive values indicate that they are very different.

\begin{figure*}
	\centering
	\includegraphics[scale=0.35]{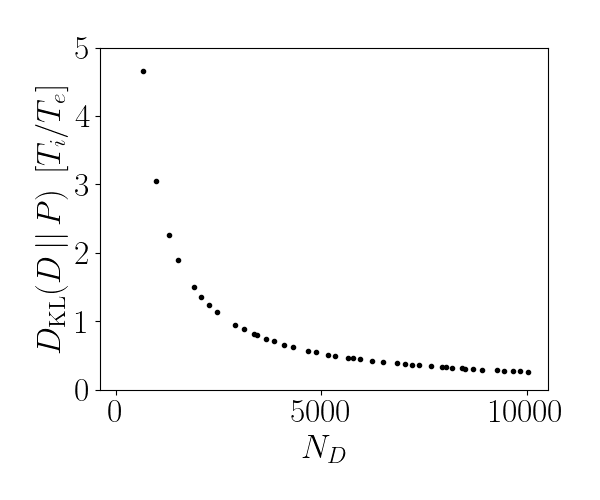}%
	\includegraphics[scale=0.35]{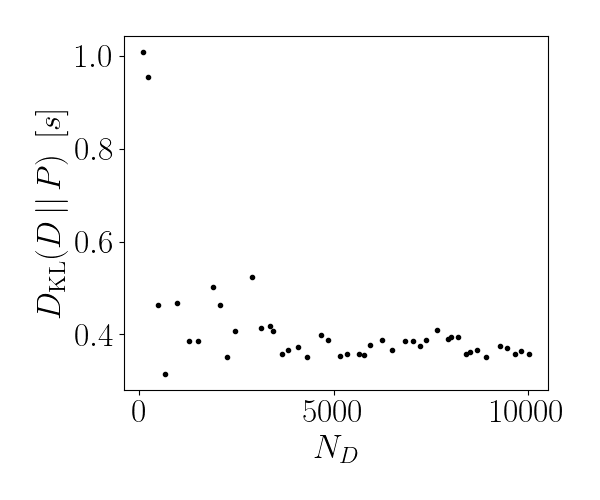}%
	\includegraphics[scale=0.35]{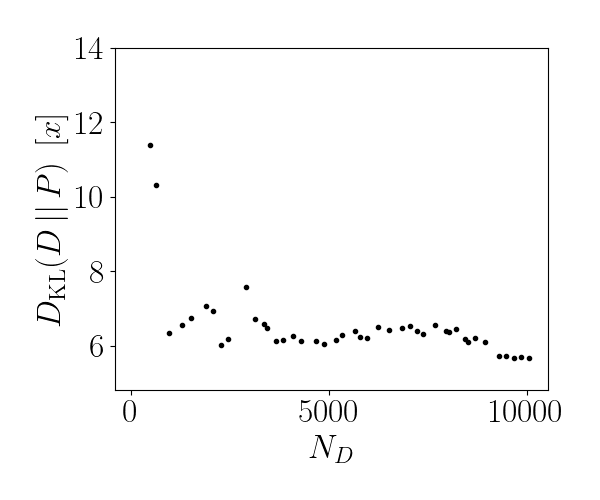}
	\caption{KL-divergence metric, $D_{\text{KL}}$, as a function of the dataset size, $N_D$, of the 1D parameter distributions of the ion temperature ratio, $T_i / T_e$ (left), magnetic shear, $s$ (center), and the radial variable, $x$ (right), computed between the constructed dataset, $D$, and the unlabelled pool, $P$, for which the latter acts as a proxy for the underlying data point sampling distribution due to the sheer number of points within. These parameters were chosen to show a variable which converges smoothly towards identity with the sampling distribution (left), a variable which converges towards a distribution close to identity but different enough to warrant investigation (center), and a variable which converges very far from the sampling distribution (right). The fact that there are input distributions which deviate strongly from identity with the sampling distribution indicates both that there is a more efficient distribution to sample from than the chosen sampling distribution and that the AL algorithm can help identify it.}
	\label{fig:ConstructedDatasetKLDivergence}
\end{figure*}

Figure~\ref{fig:ConstructedDatasetKLDivergence} shows the KL-divergence between the constructed dataset distribution, $D$, and the unlabelled pool distribution, $P$, as a function of the constructed dataset size, $N_D$, for three selected input parameters, the ion temperature ratio, $T_i / T_e$, the magnetic shear, $s$, and the radial variable, $x$. These show that the dataset distribution appears to converge and, for some input parameters (e.g. the radial variable, $x$), towards a distribution which is significantly different than that of the underlying unlabelled pool. In the case of these particular variables, this makes sense as the plasma turbulence is observed to be much stronger in the outer radii due to the increasing drive from the ballooning effect, explaining the difference in the distribution of $x$. The magnetic shear, $s$, is often correlated to the radial variable due to the general shape of natural current profiles, hence explaining why it also converges to a non-zero value. However, while the ion temperature ratio, $T_i / T_e$, is relevant in determining the dominant turbulent mode, it acts as a fairly independent variable in terms of determining the presence of instabilities.

This result, combined with the fact that the models show statistical improvement with the number of iterations, implies that the method and acquisition function chosen do provide some degree of useful information for guiding the dataset towards an efficient distribution of points. To improve readability, only the KL-divergence behaviour of these three input parameters are shown here while the remaining parameters provided in Appendix~\ref{app:DatasetDistributionComparison}.

\section{Conclusion}
\label{sec:Conclusion}

Overall, this study demonstrated that applying AL principles with a broadly defined unlabelled pool and including the simulator in-the-loop is both feasible for improving a surrogate model towards providing generalized predictions over a wide domain and capable of efficiently constructing a ML training set for said domain. This was observed through an exercise executing the proposed AL algorithm over 45 iterations, building a training set up from $10^{2}$ to $\sim10^{4}$ points and reaching a $F_1$ classification performance of $\sim$0.8 and a $R^2$ regression performance of $\sim$0.75 on an independent test set across all outputs. Although it is suspected that the performance still needs to be improved before the generated NN models can be used in downstream applications, it should be strongly stressed that the observed level of performance from the AL algorithm is notably better than randomly sampling an equivalently sized dataset from the unlabelled pool, most especially given the small dataset sizes used in this investigation. This underlines the efficient dataset construction capabilities of the AL algorithm, even when the unlabelled pool is not limited to the operational domain of a pre-existing machine as was investigated in the previous ADEPT study. This feature becomes increasingly beneficial as the labeller becomes more computationally expensive to evaluate.

Additionally, the use of uncertainty aware models enabled a similar performance as observed with the ensemble of FFNNs used in the previous ADEPT study, with the added benefit of only needing to train, store and load one model per output variable. The applied multi-objective optimization technique for simultaneously acquiring new candidates for multiple outputs was shown to be an effective solution to the acquisition phase, although additional hyperparameter tuning is recommended as inherent bias towards some outputs over others was observed in this study. In fact, based on the observations found in this study, it is suspected that an adaptive hyperparameter tuning strategy may be the most effective way to handle the underlying data distribution shift as more iterations are performed. By extrapolating the observed performance trajectory, this AL methodology appears to approach the previous ADEPT performance with a dataset of the same order of magnitude in spite of including an additional input dimension, even after applying a rudimentary accounting for the diminishing returns with each successive iteration.

That said, there are specific technical issues that should be addressed before this technique can be recommended for application in a production setting. In spite of the improved robustness implemented into the specific BNN regressor model chosen for this study, the regressor network training still occasionally fell into undesired regions of the optimization space prior to achieving a reasonable solution, although infrequently. Also, the chosen BNN architecture appears to systematically underpredict outputs with a large absolute value even though the network returns a decent performance metric, and it is uncertain whether this is a fundamental issue of the architecture itself or an expected behaviour reflecting of the sparse statistics of regions returning those large absolute values. Thus, while the proposed methodology extrapolates to a comparable performance to dataset size ratio as the previous ADEPT study, it is suspected that this could approach a ceiling before reaching an equal performance based on the behaviours seen in the trained models even on a static dataset. To compound this, the blending of classifier and regressor network outputs may mean that any performance degradation in the trained classifier network for a given iteration may provide a negative feedback mechanism to the regressor performance in the corresponding iteration, which in turn impacts the points acquired for the next iteration and may spiral out of control if applied blindly. These observations imply that there are still more efficiency gains possible by further development of the AL methodology, both in the investigation of more efficient acquisition functions and in improving the robustness of the underlying BNN model convergence. In the case that these obstacles with the uncertainty-aware networks cannot be reasonably and robustly overcome, an alternative could be to create a hybrid workflow which initially uses these uncertainty-aware networks to build a skeleton dataset and swaps over to the deep ensembles used in the ADEPT study after a threshold model performance for dataset refinement.

For completeness of discussion, an attempt was also made in this study to train a single BNN network to match all the transport fluxes for a given mode simultaneously. This was to investigate whether sharing a common hidden layer amongst all the outputs of a given turbulent mode would allow the NN to more accurately reflect the natural covariance between said outputs. Unfortunately, this approach yielded networks that tended to preferentially perform better on some outputs and catastrophically on the others. While the single output method still shows signs of this general issue as a result of the multi-objective acquisition, it manifests more as a minor barrier to improvement for select output variables rather than a complete lack of predictive power (e.g. negative $R^2$ score). As this hypothesis was not exhaustively tested, further work is still recommended into the construction of loss terms which more strongly penalize performance imbalances between multiple outputs of a single uncertainty-aware NN model.

Finally, in a manner similar to previous NN applications to fusion plasma physics, a deep knowledge of the physical system being emulated is a crucial component in determining an optimal model configuration. These careful model design considerations also serve to improve the robustness and sampling efficiency when applying an AL algorithm to the same problem. While the described classifier / regressor configuration and physics-based labelled data filters are specific to the turbulent transport problem being emulated in this study, the implemented multi-objective optimization scheme, the chosen uncertainty-aware models, and concepts within the proposed uncertainty-based acquisition function are expected to generalize well to problems in many other domains.


\section{Acknowledgements}
\label{sec:Acknowledgements}

This work was funded by Commonwealth Fusion Systems under RPP020 and US Department of Energy under Grant DE-SC0024399.

\printbibliography

\appendix

\section{Development of the multivariate Gaussian sampling distribution}
\label{app:MultivariateSampling}

\begin{figure*}[tb]
	\centering
	\includegraphics[scale=0.55]{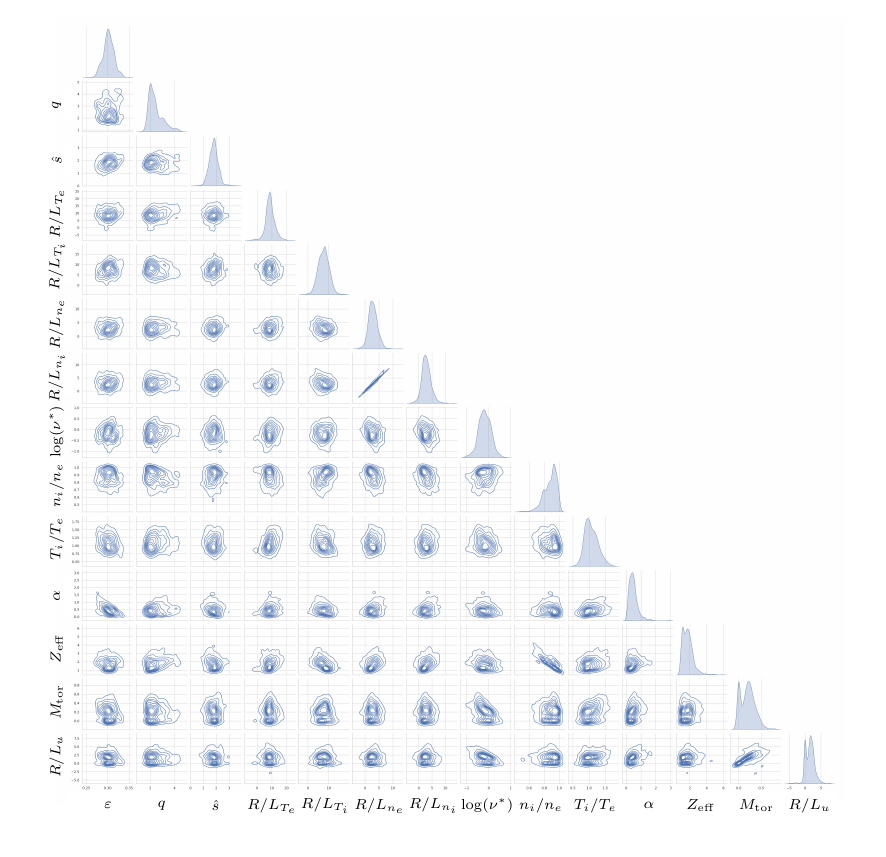}%
	\includegraphics[scale=0.55]{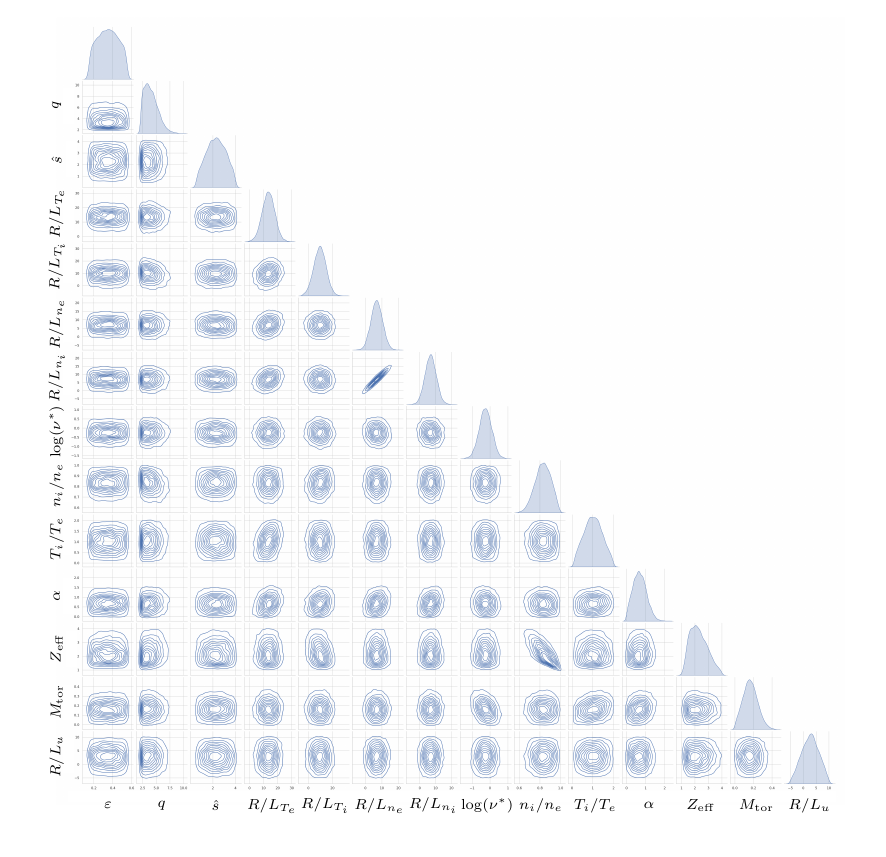}
	\caption{Joint distribution plots for the 16 QuaLiKiz input parameters, except $x$ and $R/L_{Z_\text{eff}}$, extracted from the JET-based dataset at $x\simeq0.8$ -- $\sim1250$ samples (left) and from the constructed multivariate Gaussian distribution at $x\simeq0.8$ -- $\sim10000$ samples (right), demonstrating the capability of the generated covariance functions to approximate the experimentally-based dataset distribution. The radial variable, $x$, was excluded due to the desire to enforce explicit radial dependency and the effective charge gradient variable, $R/L_{Z_\text{eff}}$, was excluded since the original dataset was constructed assuming $R/L_{Z_\text{eff}} = 0$.}
	\label{fig:DatasetCorrelation}
\end{figure*}

In order to build upon the initial results from the ADEPT study~\cite{zanisiEfficientTrainingSets2024}, this study aimed to move away from the method of directly sampling experimental plasma profiles to obtain the unlabelled pool. However, using a uniform sampling across a fixed range would result in a high degree of oversampling in the corners of the parameter space, especially due to the dimensionality of this electrostatic turbulence problem~\cite{vandeplasscheFastModelingTurbulent2020}. Thus, an alternative method was devised where a multivariate Gaussian distribution was constructed based on the general correlations found in a previously released JET-based dataset~\cite{hoNeuralNetworkSurrogate2021}. 

A multivariate Gaussian distribution is completely defined by a mean vector, $\mathbf{M} \equiv \left\{ \mu \right\}$, with length equal to the number of dimensions, $D$, and a covariance matrix, $\mathbf{\Sigma} \equiv \left\{ \Sigma_{i,j} \right\}$, of size $D \times D$. This approach naturally avoids oversampling the edges and the corners of the input parameter space due to the diminishing probability of the Gaussian tails. Additionally, this method allows correlations between strongly coupled parameters to be retained via the off-diagonal elements of the covariance matrix, according to the relation:
\begin{equation}
	\label{eq:CovarianceElement}
	\Sigma_{i,j} \equiv \Sigma_{j,i} = \sigma_i \sigma_j \rho_{i,j}
\end{equation}
where $\rho_{i,j}$ is the Pearson correlation coefficient of dimensions $i$ and $j$. Note that this also applies where $i=j$, such that the diagonal of the covariance matrix represents the variance in that dimension, i.e. $\Sigma_{i,i} = \sigma_i^2$ as $\rho_{i,i} \equiv 1$.

The left half of Figure~\ref{fig:DatasetCorrelation} shows the degree of correlation present in the QuaLiKiz input parameter distributions extracted from the experimentally-based dataset at a specific radius, $x=0.8$. An initial analysis of the data across all radii revealed that many of the parameter pairs had weak correlations, $\left|\Sigma_{i,j}\right| = \left|\Sigma_{j,i}\right| \le 0.2$, and the variances and some of the stronger correlations vary significantly with the radial position within the plasma. Based on this result, it was decided to maximize the number of zeros in the off-diagonal elements and describe the remaining ones as functions of the QuaLiKiz radial variable, $x$, effectively removing that parameter from the multivariate Gaussian distribution. In order to construct these functions, the experimentally-based dataset was segmented into radial bin with a width $\delta x = 0.1$ and the covariance statistics of each radially-resolved dataset was computed. From these statistics, the means, $\mu_i$, standard deviations, $\sigma_i$, and strongest correlation parameters, $\rho_{i,j}$, were extracted and manually fitted to function of $x$ with various forms. These can be found in Equations~\eqref{eq:MeanFunctions}, \eqref{eq:StandardDeviationFunctions}, and \eqref{eq:PearsonCorrelationFunctions}, respectively, and have only been designed to work within the range $x \in \left[0,1\right]$. Any parameter pairs without a corresponding correlation function in Equation~\eqref{eq:PearsonCorrelationFunctions} were set with $\rho_{i,j} = 0$. Note that some weakly correlating parameter pairs were specified as non-zero in this approximation as the distribution generation script returned singular covariance matrices at certain values of $x$ without them.

\begin{equation}
	\label{eq:MeanFunctions}
	\scriptsize
	\begin{aligned}
		\mu_\varepsilon = \; & 0.35\\
		\mu_q = \; & 5.5 x^3 - 2.84 x^2 + 0.877 x + 1.35\\
		\mu_{\hat{s}} = \; & 6 x^3 - 2.28 x^2 + 0.79 x\\
		\mu_\alpha = \; & 0.115 e^{8.5 \left(x - 0.7\right)} + 0.3 x + 0.1\\
		\mu_{\log \nu^*} = \; & 4.83 x^2 - 4.32 x + 0.125\\
		\mu_{R/L_{T_e}} = \; & 0.83 e^{5.03 x} - \left(3 x\right)^4\\
		\mu_{R/L_{n_e}} = \; & 0.045 e^{6.1 x} + x^{0.2}\\
		\mu_{T_i/T_e} = \; & 1\\
		\mu_{R/L_{T_i}} = \; & 1.07 e^{4.65 x} - \left(3 x\right)^4\\
		\mu_{n_i/n_e} = \; & 0.88\\
		\mu_{R/L_{n_i}} = \; & 0.045 e^{6.1 x} + x^{0.2}\\
		\mu_{Z_{\text{eff}}} = \; & 1.7\\
		\mu_{R/L_{Z_{\text{eff}}}} = \; & 0\\
		\mu_{M_{\text{tor}}} = \; & {-0.122} x^2 - 0.385 x + 0.545\\
		\mu_{R/L_{u_{\text{tor}}}} = \; & 2.22 x^2 + 0.89 x + 0.6
	\end{aligned}
\end{equation}

\begin{equation}
	\label{eq:StandardDeviationFunctions}
	\scriptsize
	\begin{aligned}
		\sigma_\varepsilon = \; & 0.2\\
		\sigma_q = \; & 0.24 e^{8.64 \left(x - 0.7\right)} + 1.2\\
		\sigma_{\hat{s}} = \; & 2.24 x^2 - 0.62 x + 0.16\\
		\sigma_\alpha = \; & 0.069 e^{11.2 \left(x - 0.7\right)} + 0.175\\
		\sigma_{\log \nu^*} = \; & 0.35\\
		\sigma_{R/L_{T_e}} = \; & 0.39 e^{16.5 \left(x - 0.7\right)} - 2.2 x + 0.7\\
		\sigma_{R/L_{n_e}} = \; & 0.55 e^{14.58 \left(x - 0.7\right)} + x^{0.2} + 0.225\\
		\sigma_{T_i/T_e} = \; & 0.5\\
		\sigma_{R/L_{T_i}} = \; & 1.1 e^{8.5 \left(x - 0.7\right)} + 2.4 x + 1\\
		\sigma_{n_i/n_e} = \; & 0.1\\
		\sigma_{R/L_{n_i}} = \; & 0.55 e^{14.58 \left(x - 0.7\right)} + x^{0.2} + 0.225\\
		\mu_{Z_{\text{eff}}} = \; & 1\\
		\mu_{R/L_{Z_{\text{eff}}}} = \; & 0.03\\
		\mu_{M_{\text{tor}}} = \; & 0.309 x^3 - 0.497 x^2 + 0.064 x + 0.193\\
		\mu_{R/L_{u_{\text{tor}}}} = \; & 22.6 x^3 - 23.4 x^2 + 7.64 x + 1
	\end{aligned}
\end{equation}

\begin{equation}
	\label{eq:PearsonCorrelationFunctions}
	\scriptsize
	\begin{aligned}
		\rho_{\varepsilon, \hat{s}} = \; & {-5.45} x^3 + 6.83 x^2 - 2.59 x + 0.65\\
		\rho_{\varepsilon, \alpha} = \; & {-0.372} x - 0.2\\
		\rho_{\varepsilon, R/L_{n_e}} = \; & {-1.29} x^2 + 0.68 x + 0.15\\
		\rho_{\varepsilon, \log \nu^*} = \; & 0.48 x - 0.48\\
		\rho_{q, \hat{s}} = \; & 0.975 x - 0.7\\
		\rho_{q, \alpha} = \; & {-0.675} x + 0.7\\
		\rho_{q, R/L_{T_i}} = \; & 4.88 x^3 - 6.59 x^2 + 1.13 x + 0.55\\
		\rho_{q, n_i/n_e} = \; & 0.375\\
		\rho_{q, Z_{\text{eff}}} = \; & 0.3\\
		\rho_{\hat{s}, \alpha} = \; & 4.7 x^3 - 4.585 x^2 + 1.28 x - 0.5\\
		\rho_{\alpha, \log \nu^*} = \; & {-0.1}\\
		\rho_{\alpha, R/L_{T_e}} = \; & {-3.23} x^3 + 6.04 x^2 - 3.07 x + 0.5\\
		\rho_{\alpha, R/L_{n_e}} = \; & {-5} x^4 + 9.37 x^3 - 5.69 x^2 + 1.23 x + 0.1\\
		\rho_{\alpha, T_i/T_e} = \; & {-9.4} x^4 + 22.2 x^3 - 16.35 x^2 +\\ & 3.47 x + 0.25\\
		\rho_{\alpha, R/L_{T_i}} = \; & 2.63 x^3 - 4.74 x^2 + 1.69 x + 0.48\\
		\rho_{\alpha, n_i/n_e} = \; & {-0.1}\\
		\rho_{\alpha, R/L_{n_i}} = \; & {-5} x^4 + 9.37 x^3 - 5.69 x^2 + 1.23 x + 0.1\\
		\rho_{\alpha, Z_{\text{eff}}} = \; & {-8.37} x^4 + 18.9 x^3 - 13.45 x^2 +\\
		& 2.96 x + 0.08\\
		\rho_{R/L_{T_i}, Z_{\text{eff}}} = \; & 5.3 x^3 - 6.5 x^2 + 0.66 x + 0.55\\
		\rho_{R/L_{T_i}, n_i/n_e} = \; & {-3.75} x^3 + 4.4 x^2 - 0.275 x - 0.4\\
		\rho_{\log \nu^*, R/L_{n_i}} = \; & {-7.15} x^4 + 16.55 x^3 - 9.45 x^2 +\\
		& 0.72 x - 0.15\\
		\rho_{\log \nu^*, R/L_{n_e}} = \; & {-7.15} x^4 + 16.55 x^3 - 9.45 x^2 +\\
		& 0.72 x - 0.15\\
		\rho_{R/L_{n_e}, R/L_{n_i}} = \; & 0.95\\
		\rho_{R/L_{T_e}, R/L_{n_e}} = \; & 0.2\\
		\rho_{R/L_{T_e}, R/L_{T_i}} = \; & 0.15\\
		\rho_{R/L_{T_e}, T_i/T_e} = \; & {-3.13} x^3 + 4.4 x^2 - 0.76 x - 0.29\\
		\rho_{R/L_{T_e}, R/L_{n_i}} = \; & 0.25\\
		\rho_{n_i/n_e, Z_{\text{eff}}} = \; & {-0.8}\\
		\rho_{M_{\text{tor}}, \alpha} = \; & {-0.8} x^2 + 0.56 x + 0.27\\
		\rho_{M_{\text{tor}}, \log \nu^*} = \; & 1.37 x^2 - 1.27 x - 0.2\\
		\rho_{M_{\text{tor}}, T_i/T_e} = \; & 6.23 x^4 - 11.26 x^3 + 4.27 x^2 +\\
		& 0.6 x + 0.24
	\end{aligned}
\end{equation}

In addition to sampling from the distribution, some rejection limits were defined in order to avoid both taking samples too far into the Gaussian tails as well as removing non-physical parameter combinations. The lower and upper bounds are also a function of the radial variable, $x$, and are given in Equations~\eqref{eq:SamplingLowerBoundFunctions} and \eqref{eq:SamplingUpperBoundFunctions}, respectively. The first physical condition is that all the normalized plasma densities computed from the sampled parameters must be positive, expressed as:
\begin{equation}
	\label{eq:NegativeDensityCondition}
	\small
	\frac{n_i}{n_e} > 0 \quad;\quad \frac{n_{z1}}{n_e} > 0 \quad;\quad \frac{n_{z2}}{n_e} > 0
\end{equation}
where
\begin{equation}
	\label{eq:ImpurityDensityEquations}
	\small
	\begin{gathered}
		\frac{n_{z1}}{n_e} = \frac{Z_{z2} - Z_{\text{eff}} - \frac{n_i}{n_e} Z_i \left(Z_{z2} - Z_i\right)}{Z_{z1} \left(Z_{z2} - Z_{z1}\right)} \\
		\frac{n_{z2}}{n_e} = 1 - \frac{n_i}{n_e} Z_i - \frac{n_{z1}}{n_e} Z_{z1}
	\end{gathered}
\end{equation}
and $Z_i = 1$, $Z_{z1} = 4$ and $Z_{z2} = 28$ based on the ions chosen for this study (D, Be, Ni). This condition is necessary as some combinations of $n_i/n_e$ and $Z_{\text{eff}}$ for this impurity mix can result in negative densities, which is evidently not physical. The second physical condition is that the sampled normalized pressure gradient must have the same sign as the one computed from the sampled normalized logarithmic kinetic gradients, expressed as:
\begin{equation}
	\label{eq:PressureGradientSignCondition}
	\frac{\alpha}{\tilde{\alpha}} > 0
\end{equation}
where
\begin{equation}
	\label{eq:PressureGradientEquations}
	\small
	\begin{gathered}
		\tilde{\alpha} = \sum_s \frac{n_s}{n_e} \frac{T_s}{T_e} \left(\frac{R}{L_{n_s}} + \frac{R}{L_{T_s}}\right) \\
		\frac{R}{L_{n_{z1}}} = \frac{\frac{R}{L_{n_e}} \left(Z_{z2} - Z_{\text{eff}}\right) - \frac{R}{L_{Z_{\text{eff}}}} Z_{\text{eff}} - \frac{n_i}{n_e} \frac{R}{L_{n_i}} Z_i \left(Z_{z2} - Z_i\right)}{\frac{n_{z1}}{n_e} Z_{z1} \left(Z_{z2} - Z_{z1}\right)} \\
		\frac{R}{L_{n_{z2}}} = \frac{R}{L_{n_e}} - \frac{n_i}{n_e} \frac{R}{L_{n_i}} Z_i - \frac{n_{z1}}{n_e} \frac{R}{L_{n_{z1}}} Z_{z1}
	\end{gathered}
\end{equation}
where $R/L_{T_i} = R/L_{T_{z1}} = R/L_{T_{z2}}$. This is necessary as some combinations of $R/L_{n_e}$, $R/L_{n_i}$, and $R/L_{\text{zeff}}$ for this impurity mix yield a total gradient which is the opposite sign compared to $\alpha$ resulting in a negative value for $T_e$, which is also evidently not physical. While QuaLiKiz would return an error when inputs are used which violate these physical conditions, the way this error is thrown in QuaLiKiz makes its execution difficult to automate for the purposes of the AL algorithm. Thus, excluding these points directly from the input point sampling function was the most sensible way to proceed with the study.

\begin{equation}
	\label{eq:SamplingLowerBoundFunctions}
	\scriptsize
	\begin{aligned}
		L_{\varepsilon} = \; & 0.15\\
		L_{x} = \; & 0\\
		L_{q} = \; & 2.7217 x^2 - 0.2368 x + 0.6324\\
		L_{\hat{s}} = \; & 2 \tanh\!\left(5x - 2.5\right) - 1.25\\
		L_{\log \nu^*} = \; & 3.9946 x^2 - 3.7183 x - 1.2334\\
		L_{\alpha} = \; & 0\\
		L_{R/L_{T_e}} = \; & {-168.67} x^4 + 222.87 x^3 - 87.067 x^2 +\\
		& 14.851 x - 1.9215\\
		L_{R/L_{n_e}} = \; & {-2.5} e^{3.7 x} + 7 x^2\\
		L_{T_i/T_e} = \; & 0.1\\
		L_{R/L_{T_i}} = \; & {-110.81} x^4 + 137.54 x^3 - 55.644 x^2 +\\
		& 9.6366 x - 2.1418\\
		L_{n_i/n_e} = \; & 0.6\\
		L_{R/L_{n_i}} = \; & {-2.5} e^{3.7 x} + 7 x^2\\
		L_{Z_{\text{eff}}} = \; & 1\\
		L_{R/L_{Z_{\text{eff}}}} = \; & {-0.2}\\
		L_{M_{\text{tor}}} = \; & 0\\
		L_{R/L_{u_{\text{tor}}}} = \; & {-74.7} x^3 + 83.8 x^2 - 27.3 x - 1.9
	\end{aligned}
\end{equation}

\begin{equation}
	\label{eq:SamplingUpperBoundFunctions}
	\scriptsize
	\begin{aligned}
		U_{\varepsilon} = \; & 0.55\\
		U_{x} = \; & 1\\
		U_{q} = \; & 10\\
		U_{\hat{s}} = \; & 7.856 x^2 + 0.45\\
		U_{\log \nu^*} = \; & 5.6662 x^2 - 4.9129 x + 1.4784\\
		U_{\alpha} = \; & 0.8957 e^{6.6715 \left(x - 0.7\right)} + 0.8\\
		U_{R/L_{T_e}} = \; & 15.12 e^{7.6656 \left(x - 0.7\right)} + 8\\
		U_{R/L_{n_e}} = \; & 2.55 e^{5.07 x} - 216 x^3\\
		U_{T_i/T_e} = \; & 2\\
		U_{R/L_{T_i}} = \; & 17.213 e^{6.4379 \left(x - 0.7\right)} + 10\\
		U_{n_i/n_e} = \; & 1\\
		U_{R/L_{n_i}} = \; & 2.55 e^{5.07 x} - 216 x^3\\
		U_{Z_{\text{eff}}} = \; & 4\\
		U_{R/L_{Z_{\text{eff}}}} = \; & 0.2\\
		U_{M_{\text{tor}}} = \; & {-1.14} x + 1.45\\
		U_{R/L_{u_{\text{tor}}}} = \; & 93.7 x^3 - 104.1 x^2 + 36.84 x + 2.5
	\end{aligned}
\end{equation}


The right half of Figure~\ref{fig:DatasetCorrelation} shows the degree of correlation present in the QuaLiKiz input parameter distributions sampled from the multivariate Gaussian distribution at a specific radius, $x=0.8$, showing that the dominant correlations are retained compared to the experimentally-sampled distributions in the top half. Notable intentional deviations can be seen in the variance of the inverse aspect ratio variable, $\varepsilon$, to allow multi-machine extensions of the resulting trained NNs, and in the correlation between the rotation velocity parameter, $M_{\text{tor}}$, and the rotation velocity gradient, $R/L_{u_{\text{tor}}}$, which was present in the experimental data due to the torque injected by the neutral beam heating system in JET and may be too limiting for a multi-machine exploration. Every point for the unlabelled pool was then generated by first uniformly sampling the radial variable with $x\in\left[0,1\right]$, then sampling the remaining parameters using the multivariate Gaussian distribution generated using these above functions and the sampled $x$ value, iterating the process at a fixed $x$ until a sample is generated that does not violate the aforementioned limits and filters.



\section{Effect of distribution mismatch between unlabelled pool and target application}
\label{app:DistributionMismatch}

\begin{figure}
	\centering
	\includegraphics[scale=0.27]{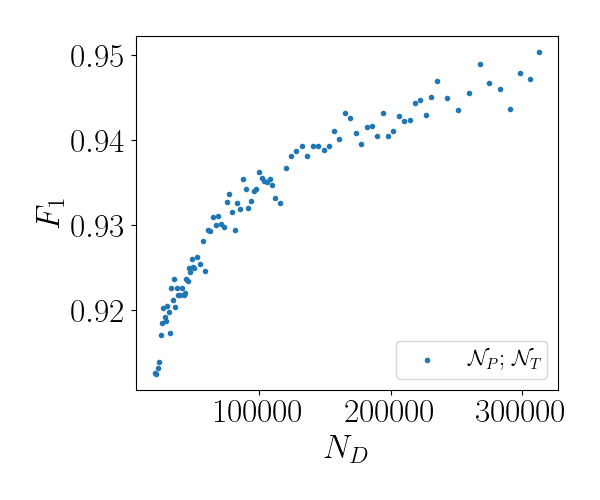}%
	\includegraphics[scale=0.27]{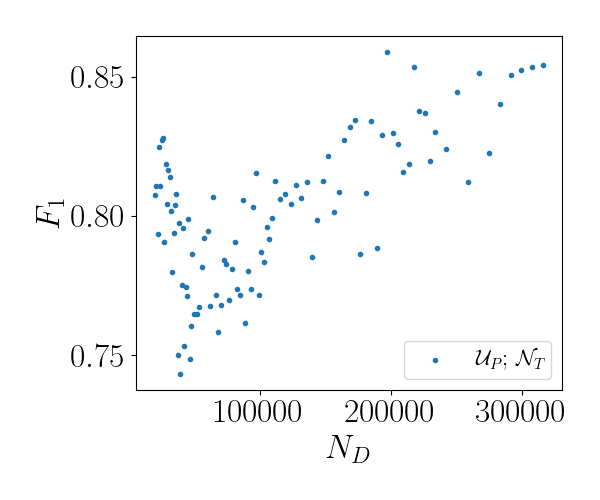}\\
	\includegraphics[scale=0.27]{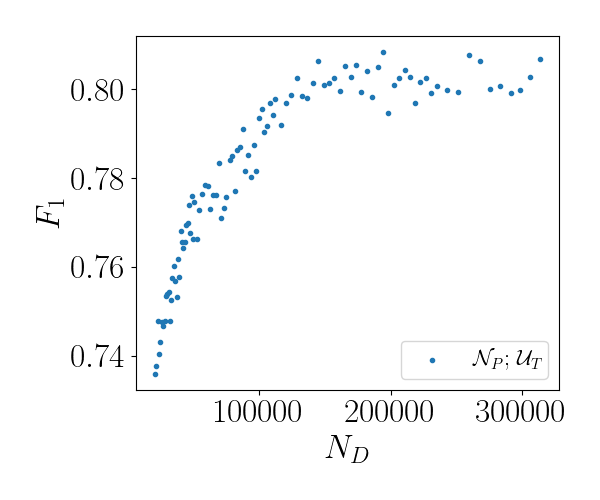}%
	\includegraphics[scale=0.27]{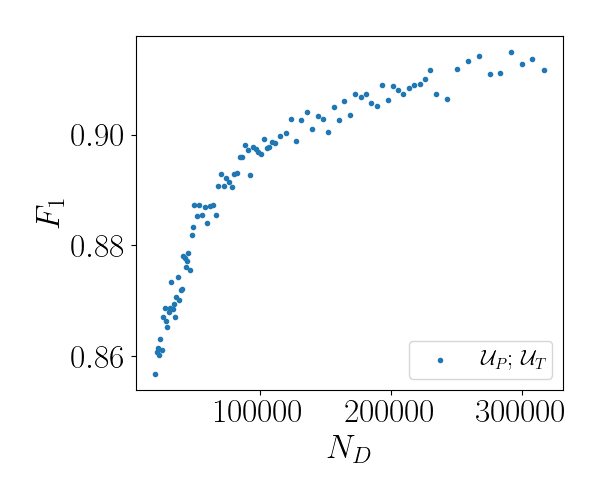}
	\caption{Performance of an ensemble of FFNN classifiers trained on ITG data selected from the AL algorithm from a multivariate Gaussian distributed (left column) and a multivariate uniform distributed (right column) unlabelled pool. To emulate different application domains, the performance is determined on a test set drawn from the multivariate Gaussian distribution (top row) and from the multivariate uniform distribution (bottom row).}
	\label{fig:DistributionMismatchClassifier}
\end{figure}

\begin{figure}
	\centering
	\includegraphics[scale=0.27]{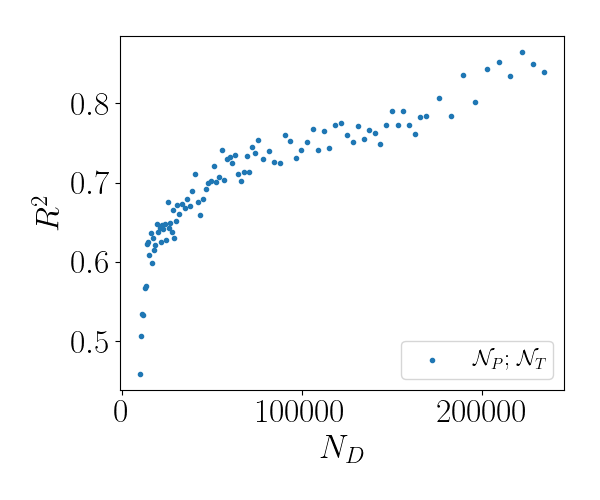}%
	\includegraphics[scale=0.27]{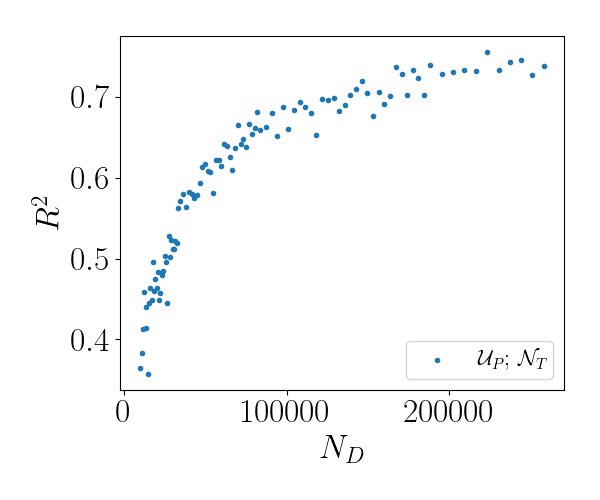}\\
	\includegraphics[scale=0.27]{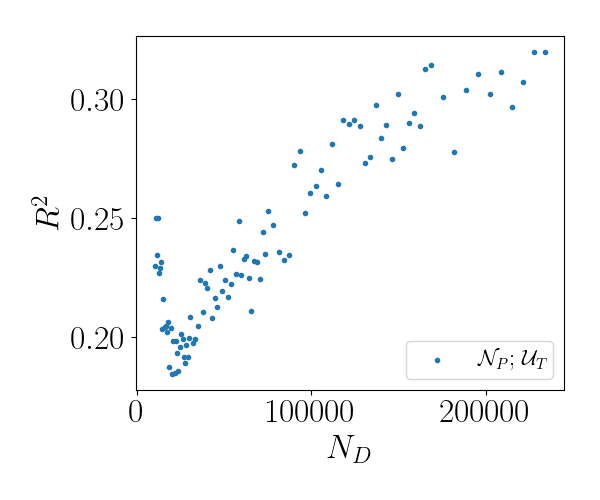}%
	\includegraphics[scale=0.27]{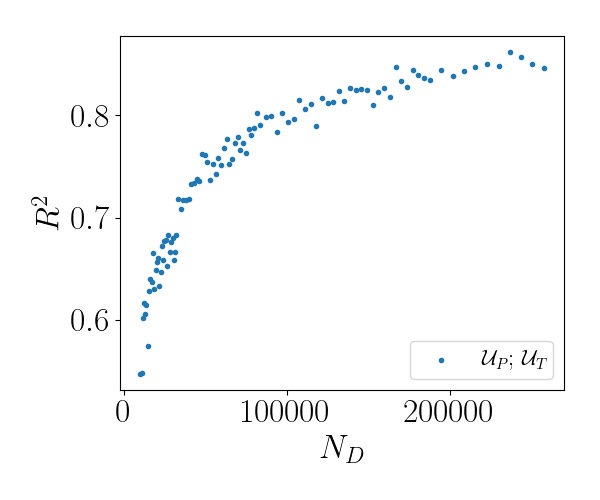}
	\caption{Performance of an ensemble of FFNN regressors trained on ITG ion heat flux data selected from the AL algorithm from a multivariate Gaussian distributed (left column) and a multivariate uniform distributed (right column) unlabelled pool. To emulate different application domains, the performance is determined on a test set drawn from the multivariate Gaussian distribution (top row) and from the multivariate uniform distribution (bottom row).}
	\label{fig:DistributionMismatchRegressor}
\end{figure}

A brief study was performed to further investigate the impact of a distribution mismatch between the unlabelled pool, $P$, and the test dataset, $T$. For simplicity, this was performed using an ensemble of FFNNs and only on the ion heat flux, $Q_i$, transport channel driven of ITG turbulence mode. Figures~\ref{fig:DistributionMismatchClassifier} and \ref{fig:DistributionMismatchRegressor} show the impact when the target application, emulated by the test set, is distributed differently from the unlabelled pool while remaining within the same absolute range. Note that this does not necessarily mean that the AL-constructed training set has the same distribution as the unlabelled pool, since the AL algorithm chooses new points per iteration based on the acquisition function instead of purely random sampling.

From these figures, there is a clear loss of absolute performance in both the classifiers and regressors when there is a strong mismatch between the unlabelled pool and application distributions. For the classifiers, the absolute drop is relatively small and the increasing trend indicate that a performance equivalent to an identical distribution case can be recovered by a rough doubling of the training set size. However, for the regressors, it is clear that there is one mismatch combination that leads to a catastrophic performance loss, which is when the unlabelled pool is multivariate Gaussian distributed and the application is more uniformly distributed. This is likely because the multivariate uniform test set contains a non-negligible number of points far in the tails of the multivariate Gaussian distribution, leading to a poor resolution in those input regions just due to the sparsity of the unlabelled pool and not due to the failure of the acquisition function to identify those regions. Since it remains advantageous to avoid uniform distributions in high-dimensional space, due to its tendency to immensely oversample the corners of the domain, this can then be circumvented by ensuring that the high probability region of the multivariate Gaussian covers the most likely application domain. The advice encapsulated in this exercise was accounted for in this study by severely oversampling the unlabelled pool to $5 \times 10^8$ points, which was ultimately limited by the amount of memory available in the compute nodes.

\section{Additional network performance plots}
\label{app:NetworkPerformance}

\begin{figure}
	\centering
	\includegraphics[scale=0.27]{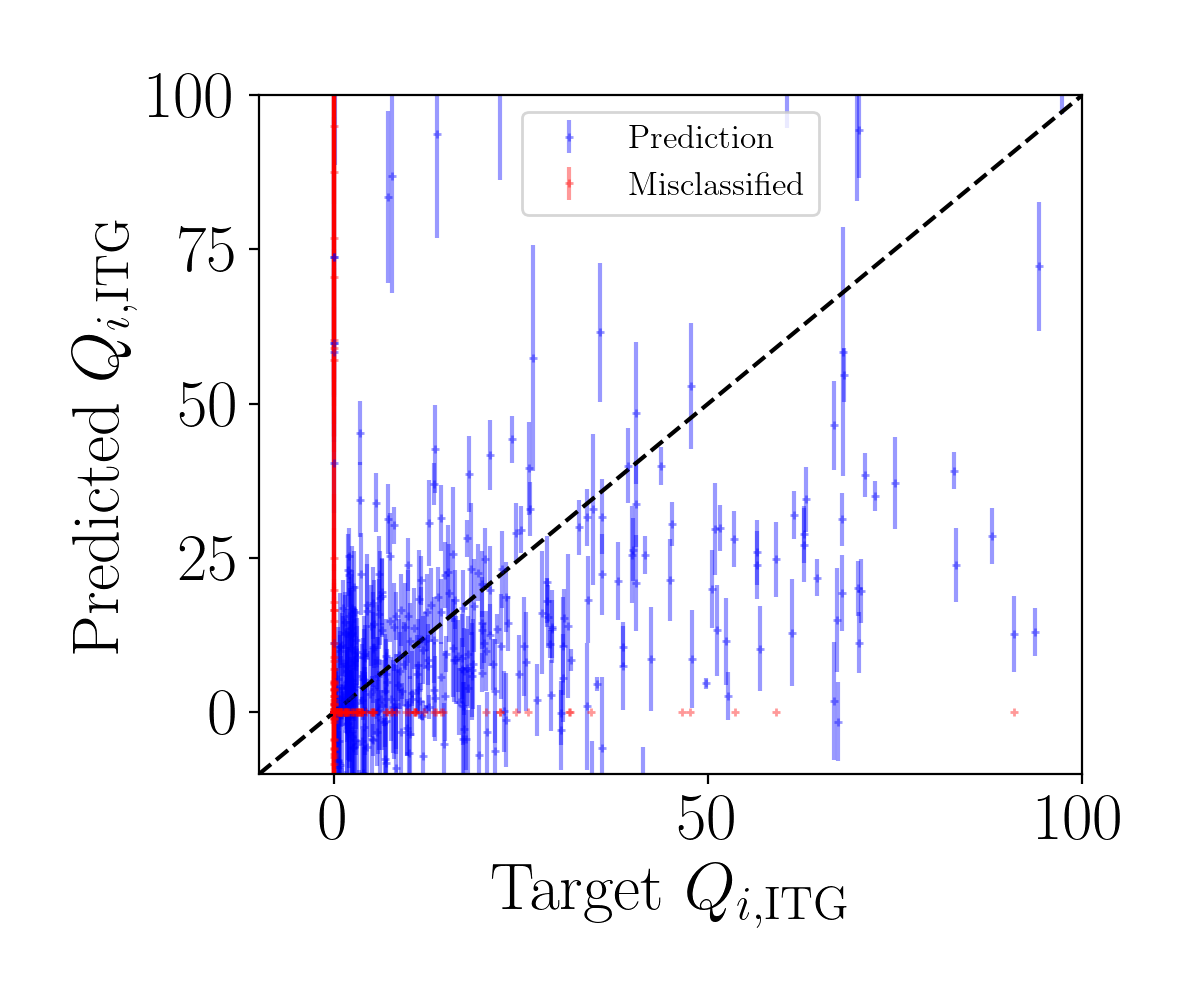}%
	\includegraphics[scale=0.27]{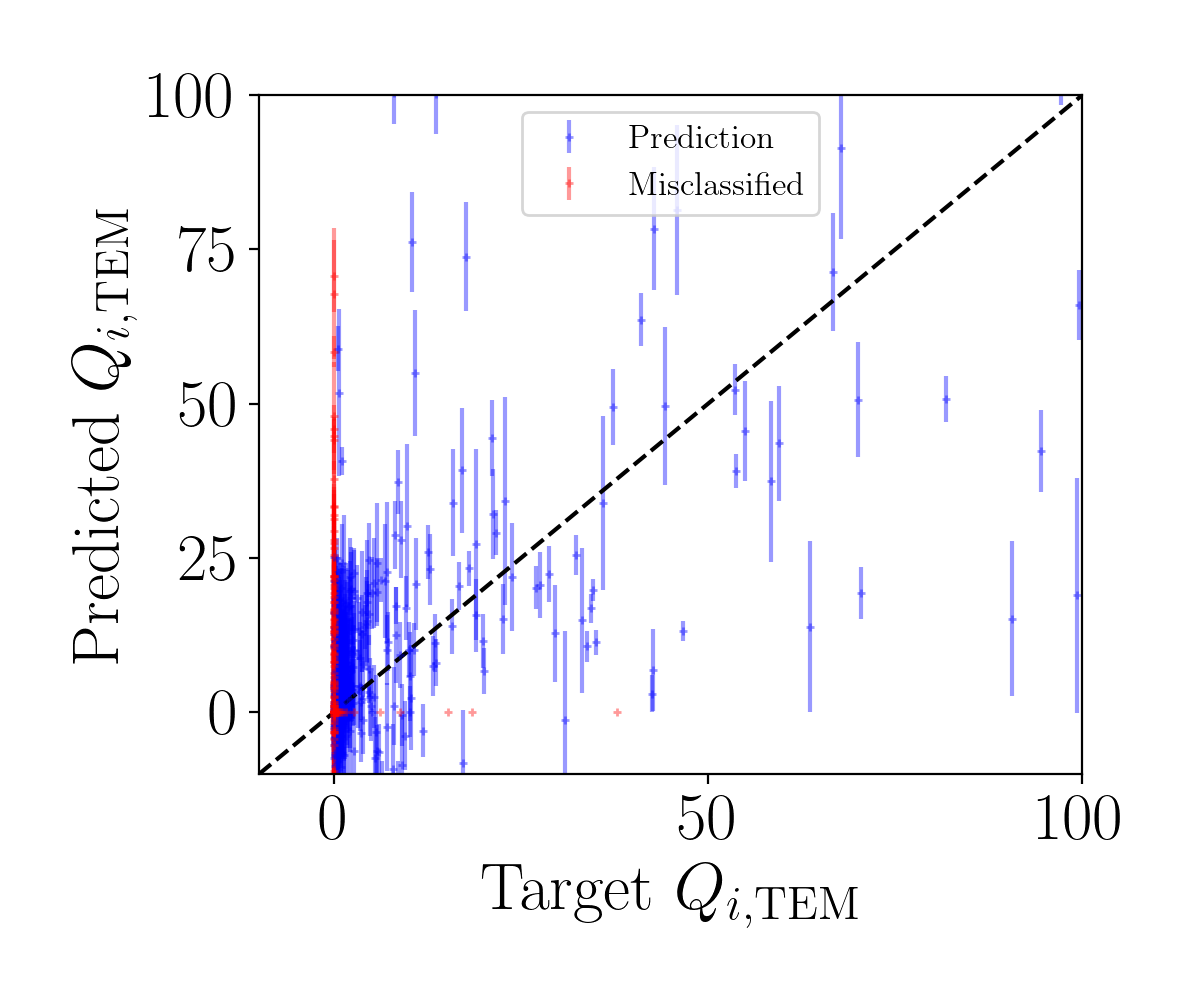}\\
	\includegraphics[scale=0.27]{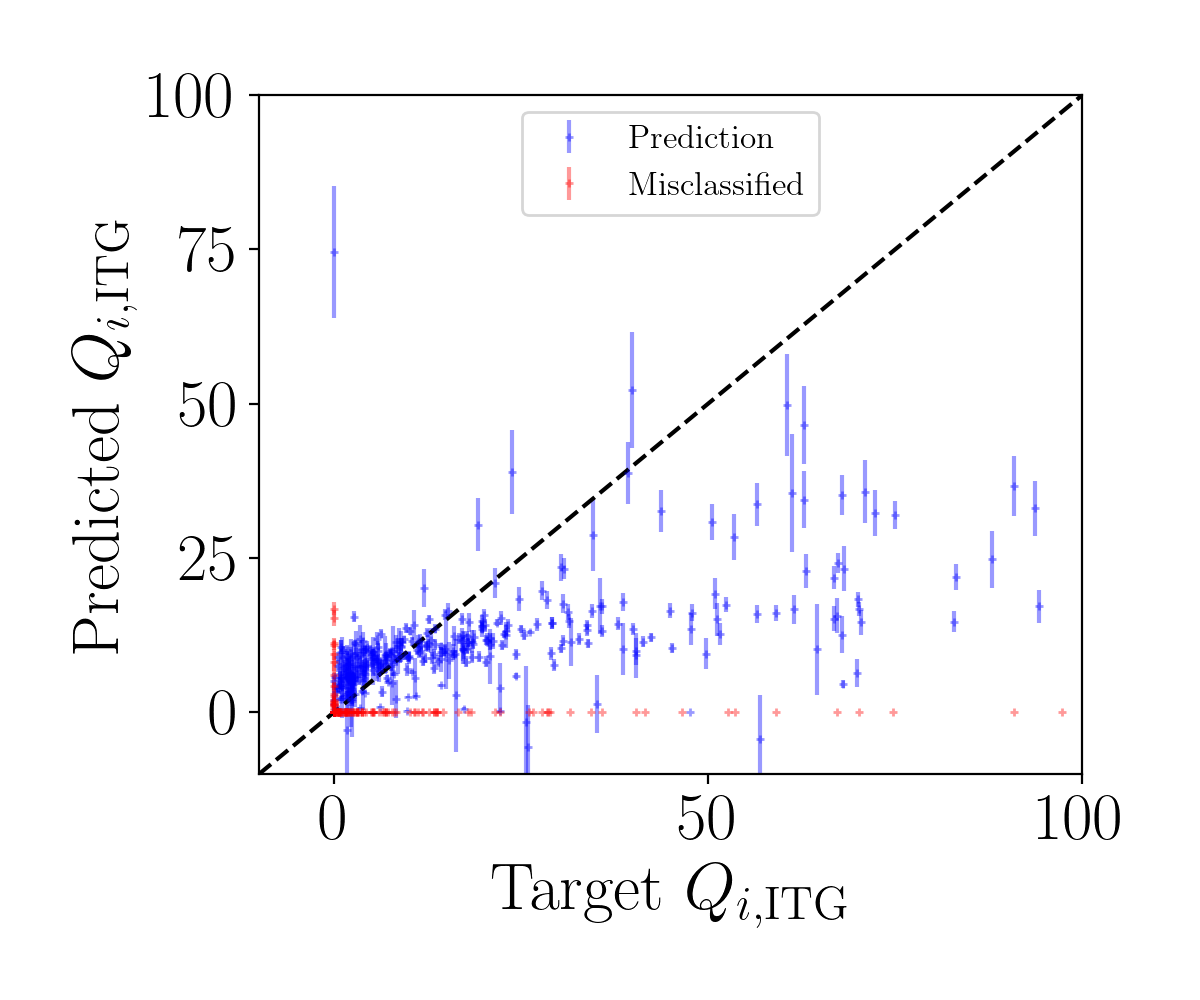}%
	\includegraphics[scale=0.27]{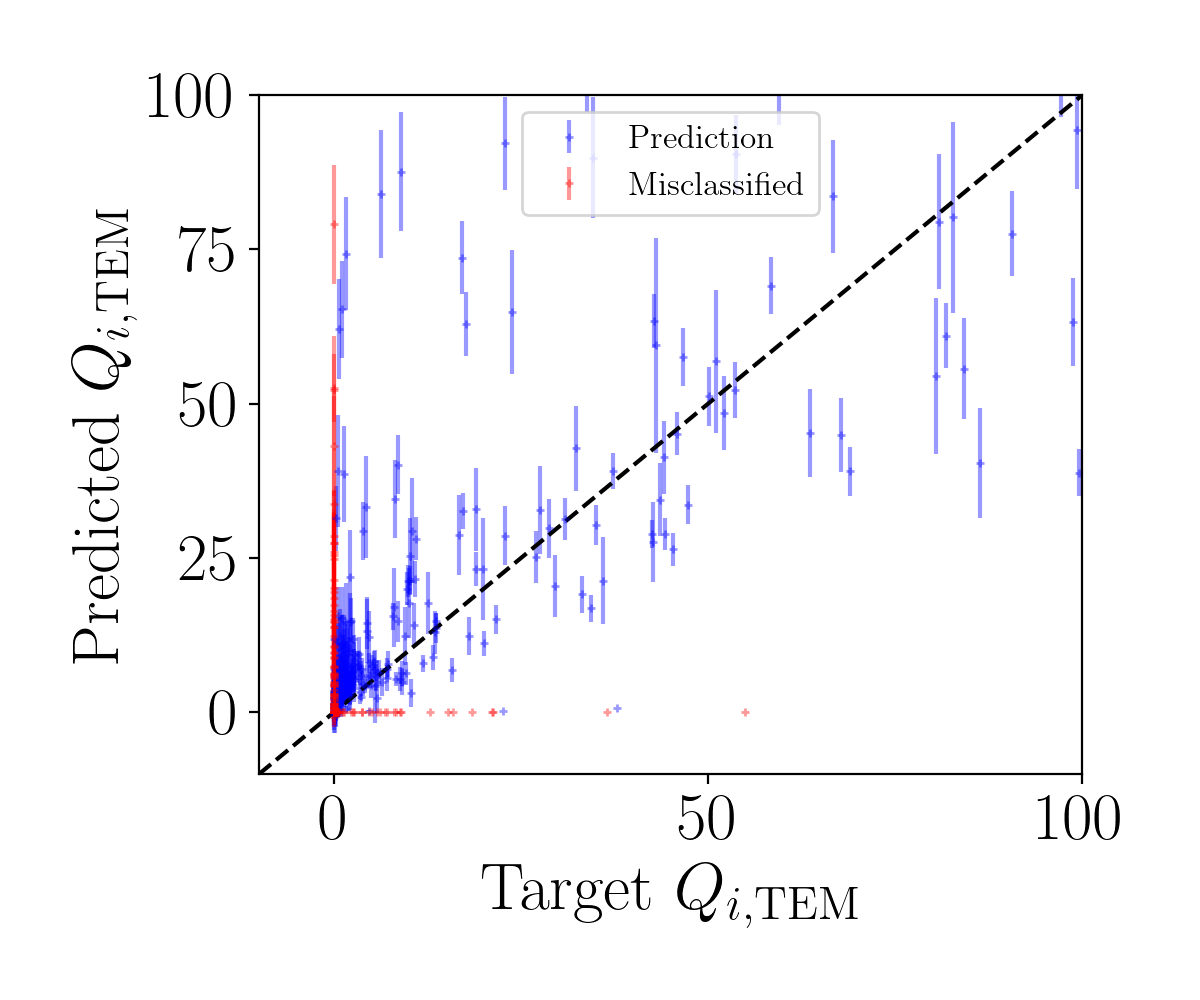}\\
	\includegraphics[scale=0.27]{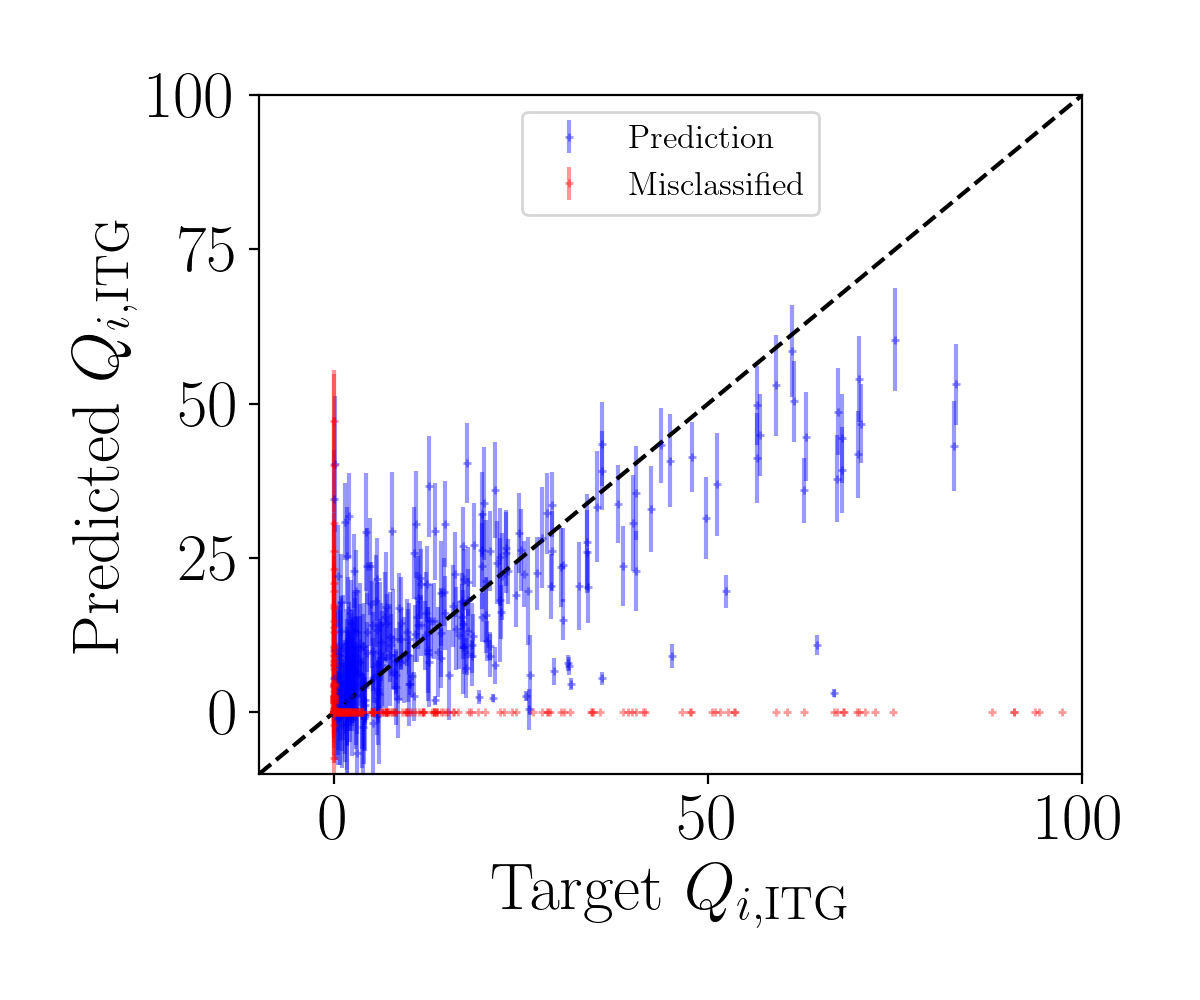}%
	\includegraphics[scale=0.27]{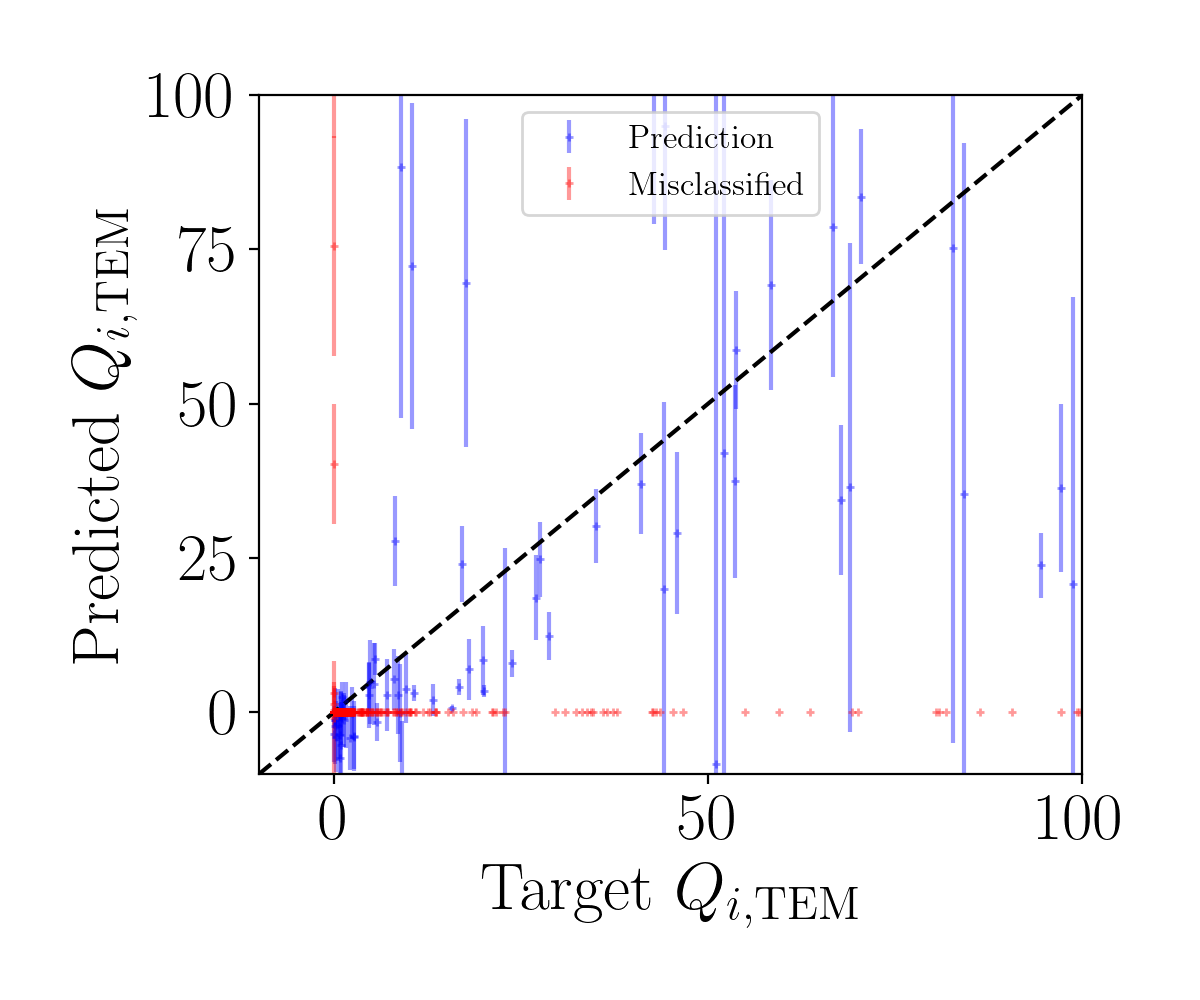}
	\caption{Comparison of the predicted ion heat flux, $Q_i$, against the target flux as evaluated by QuaLiKiz for the ITG (left) and TEM (right) turbulent modes at the initial (top) and the final iteration using the proposed AL pipeline and acquisition function (middle) and using random sampling (bottom).}
	\label{fig:IonHeatFluxPerformance}
\end{figure}

\begin{figure}
	\centering
	\includegraphics[scale=0.27]{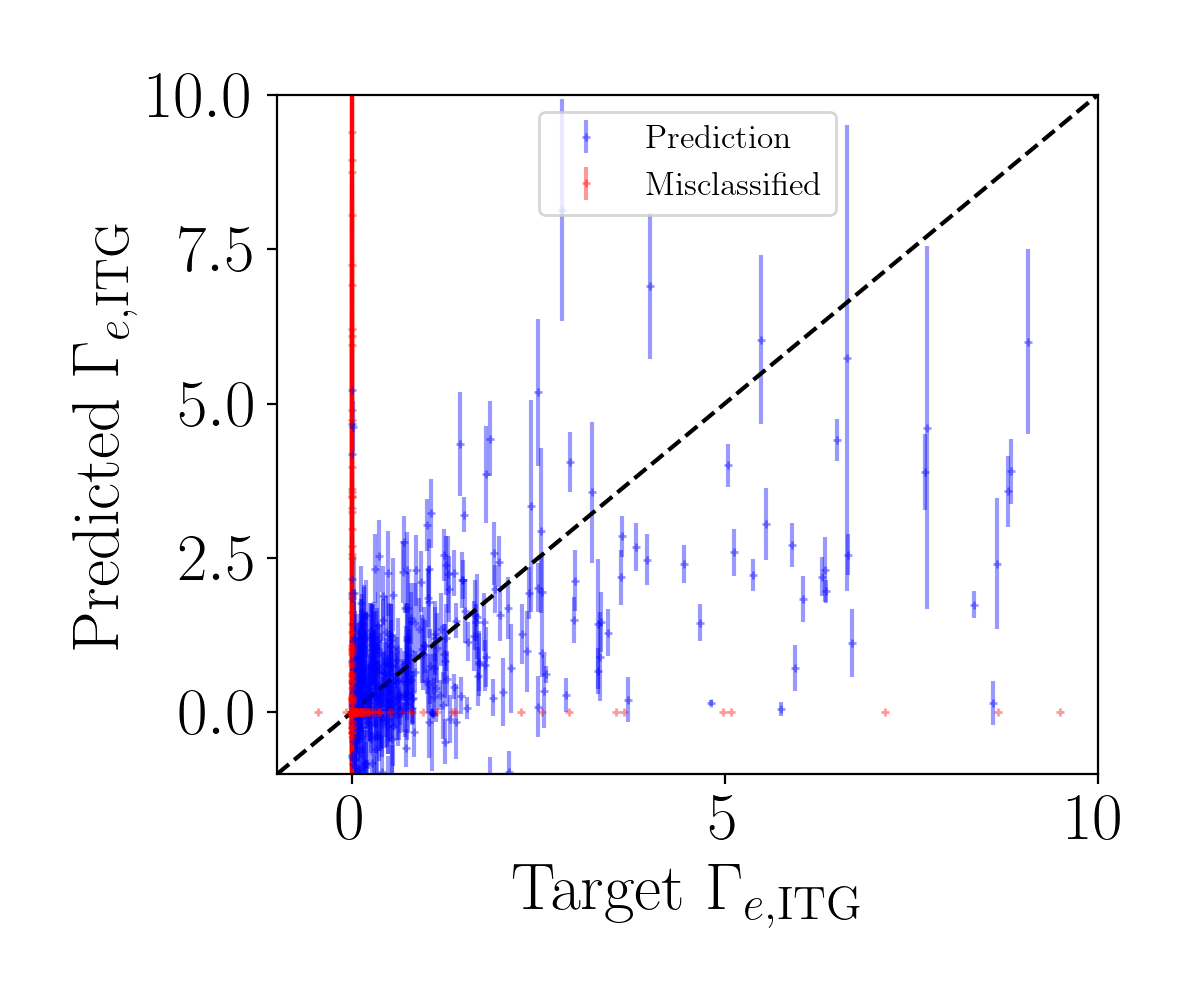}%
	\includegraphics[scale=0.27]{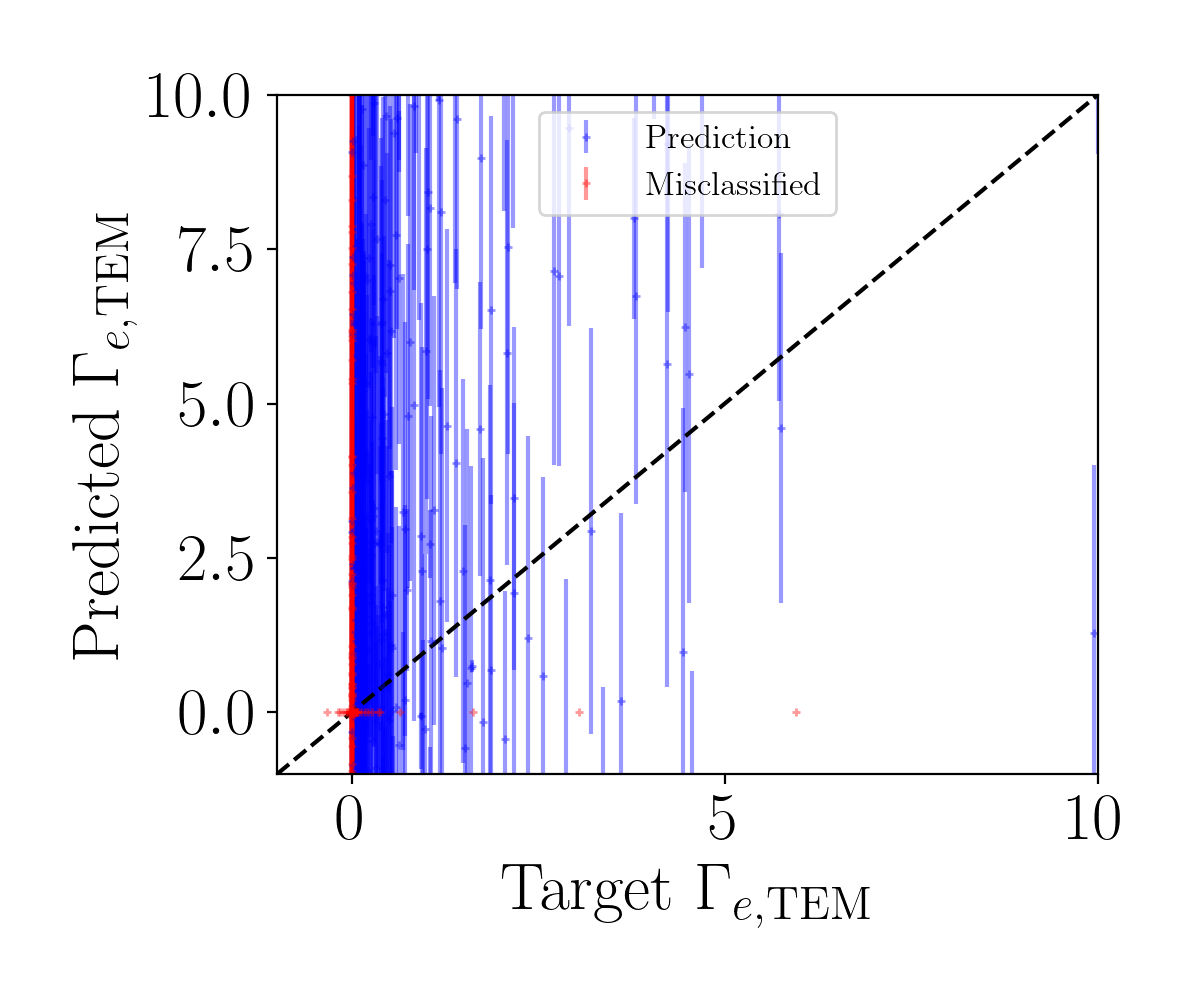}\\
	\includegraphics[scale=0.27]{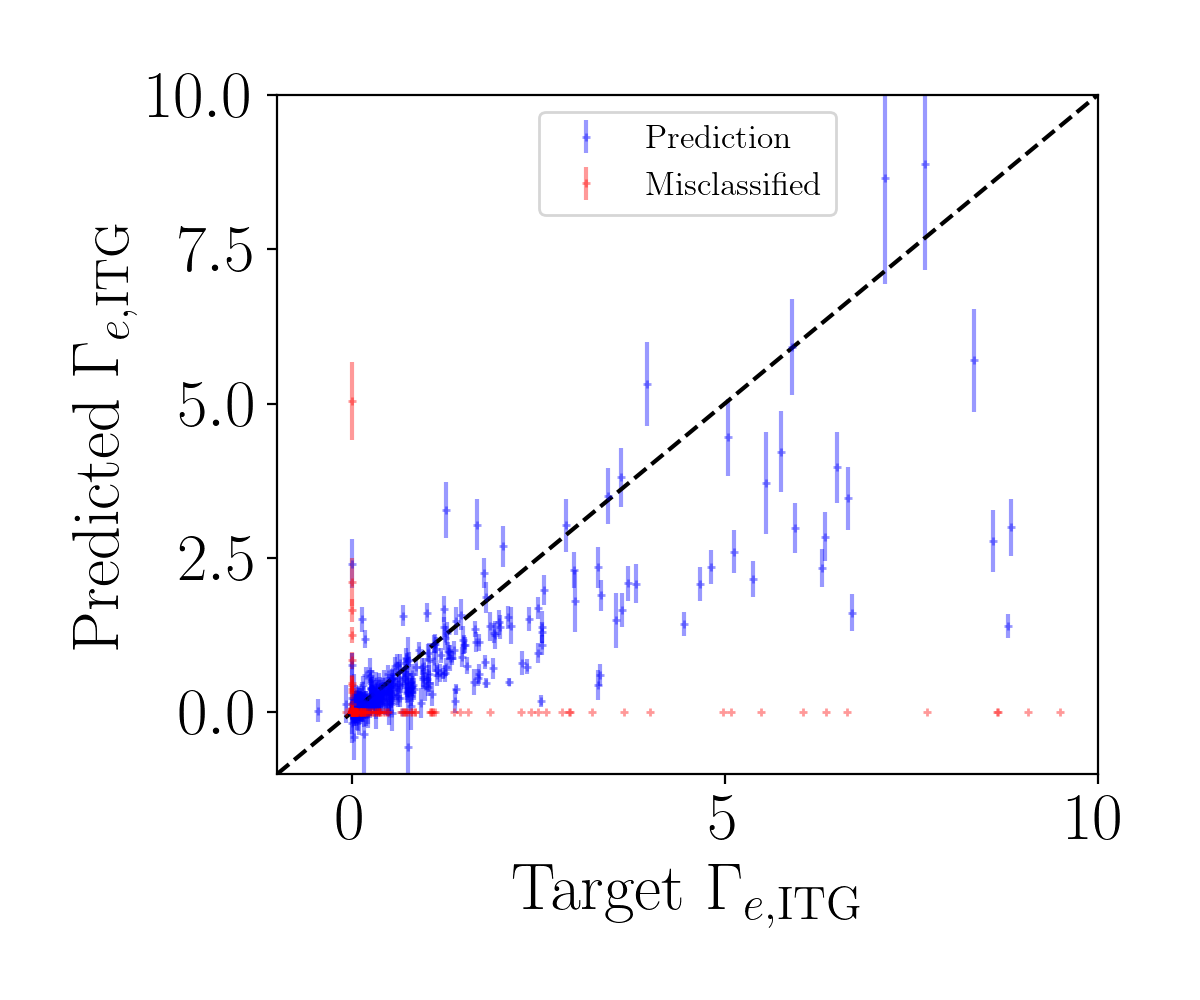}%
	\includegraphics[scale=0.27]{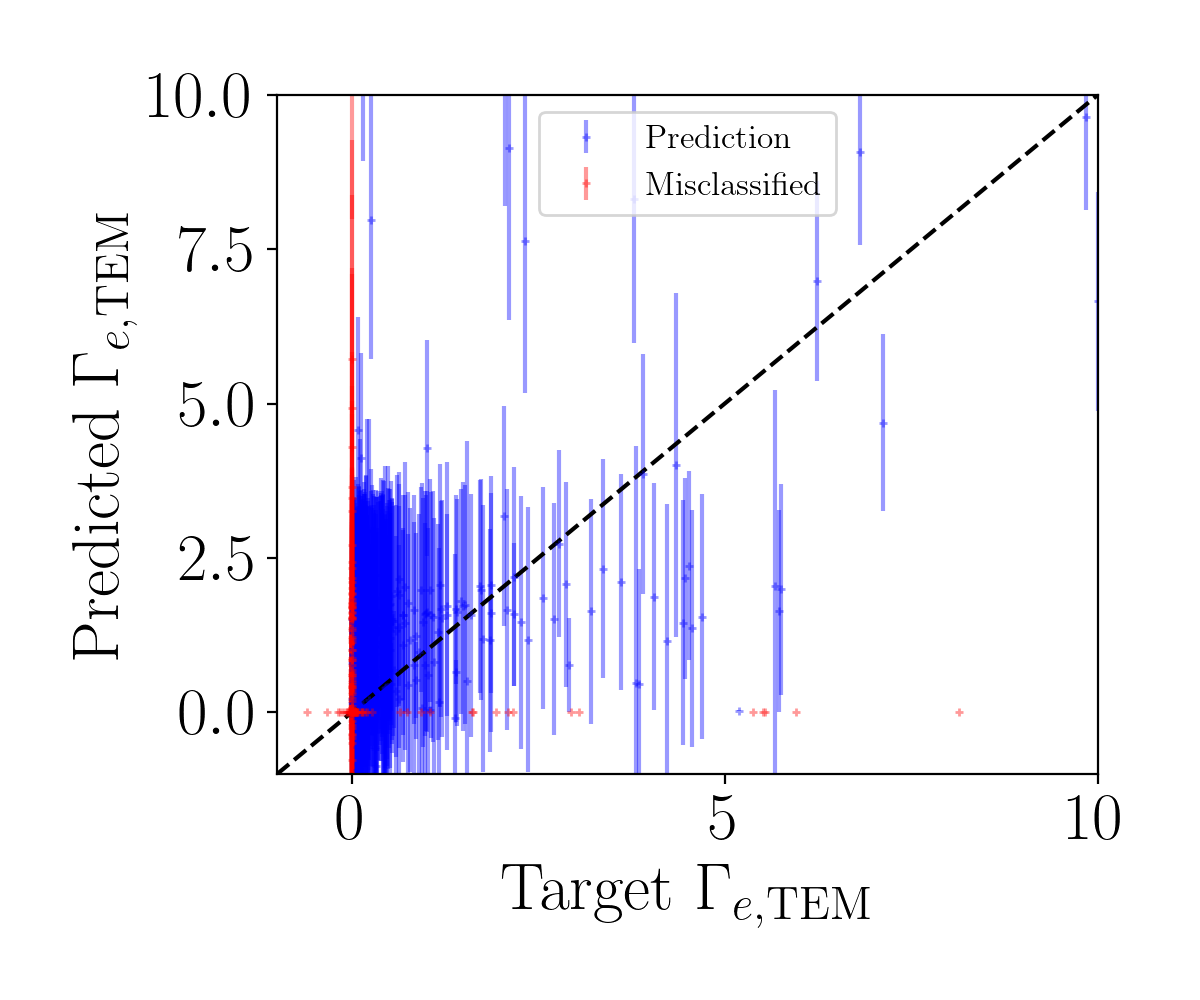}\\
	\includegraphics[scale=0.27]{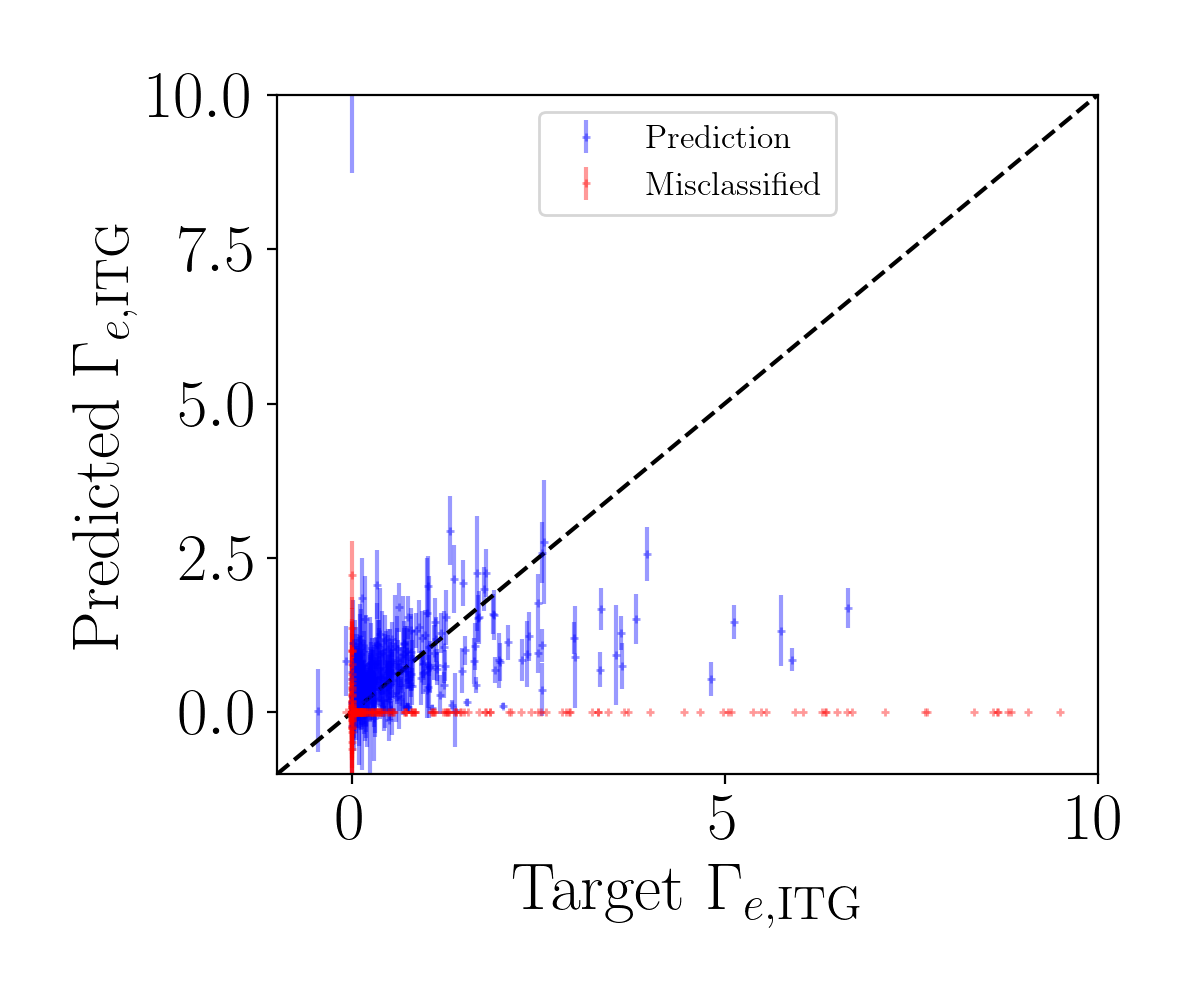}%
	\includegraphics[scale=0.27]{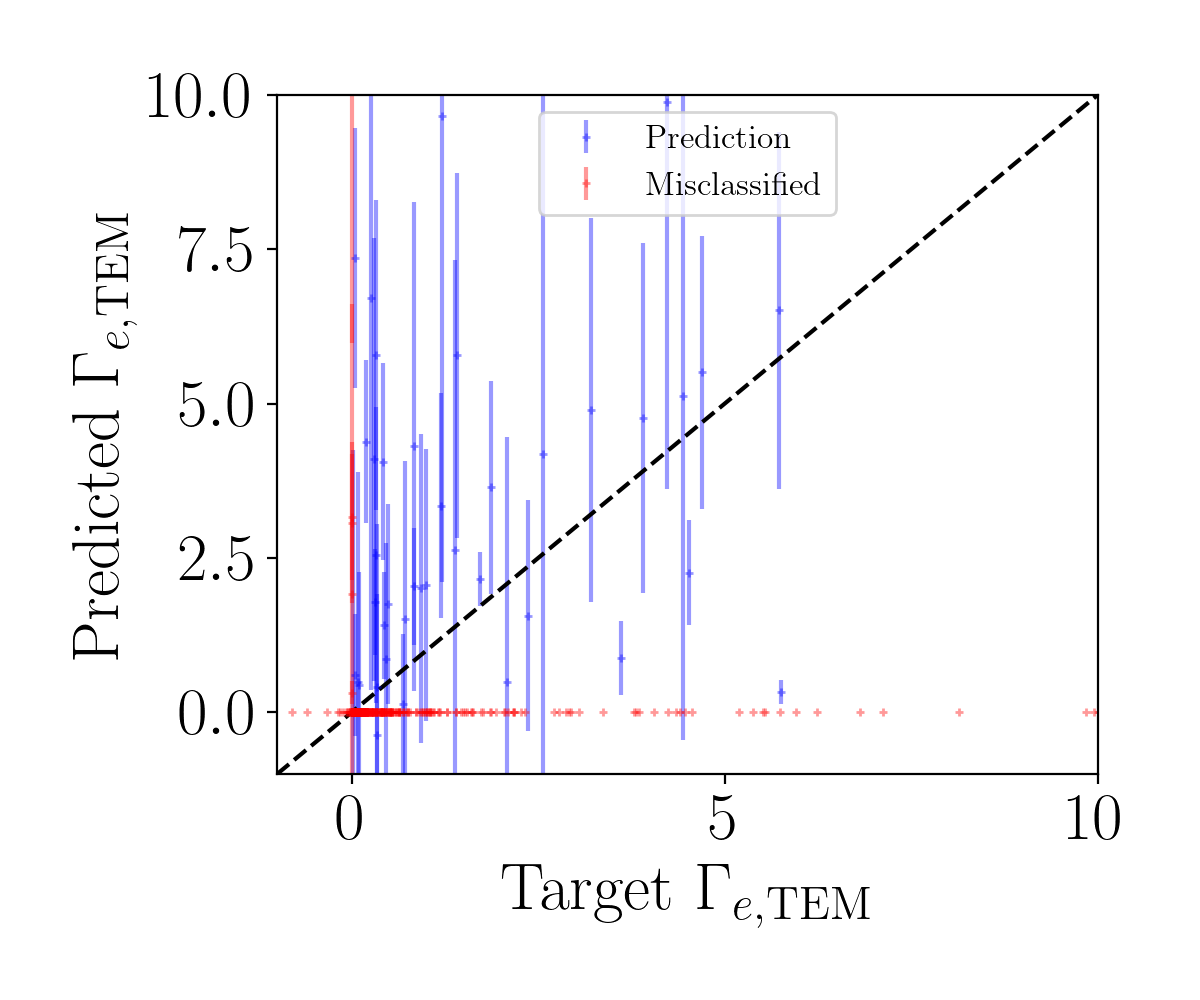}
	\caption{Comparison of the predicted electron particle flux, $\Gamma_e$, against the target flux as evaluated by QuaLiKiz for the ITG (left) and TEM (right) turbulent modes at the initial (top) and the final iteration using the proposed AL pipeline and acquisition function (middle) and using random sampling (bottom).}
	\label{fig:ElectronParticleFluxPerformance}
\end{figure}

\begin{figure}
	\centering
	\includegraphics[scale=0.27]{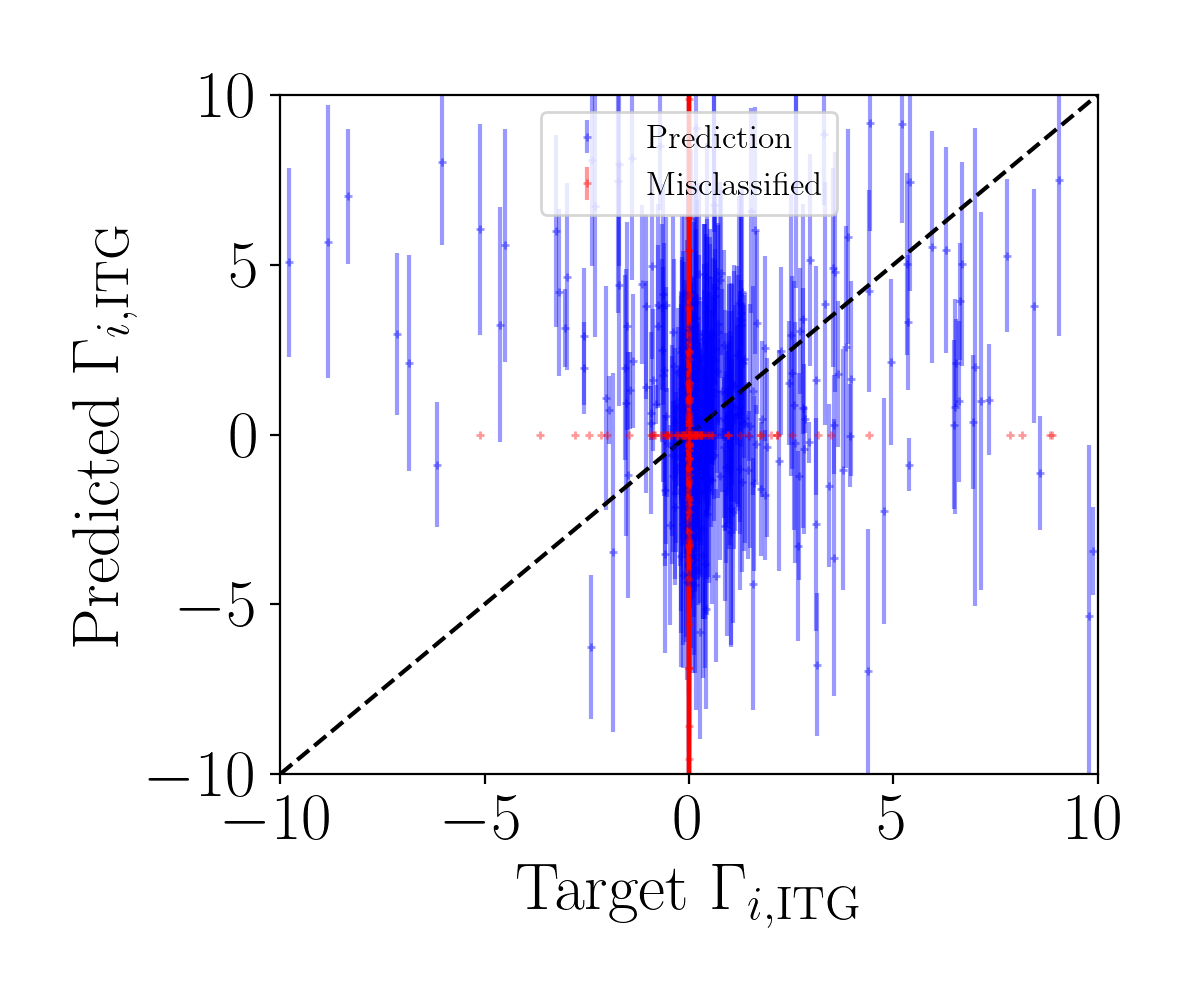}%
	\includegraphics[scale=0.27]{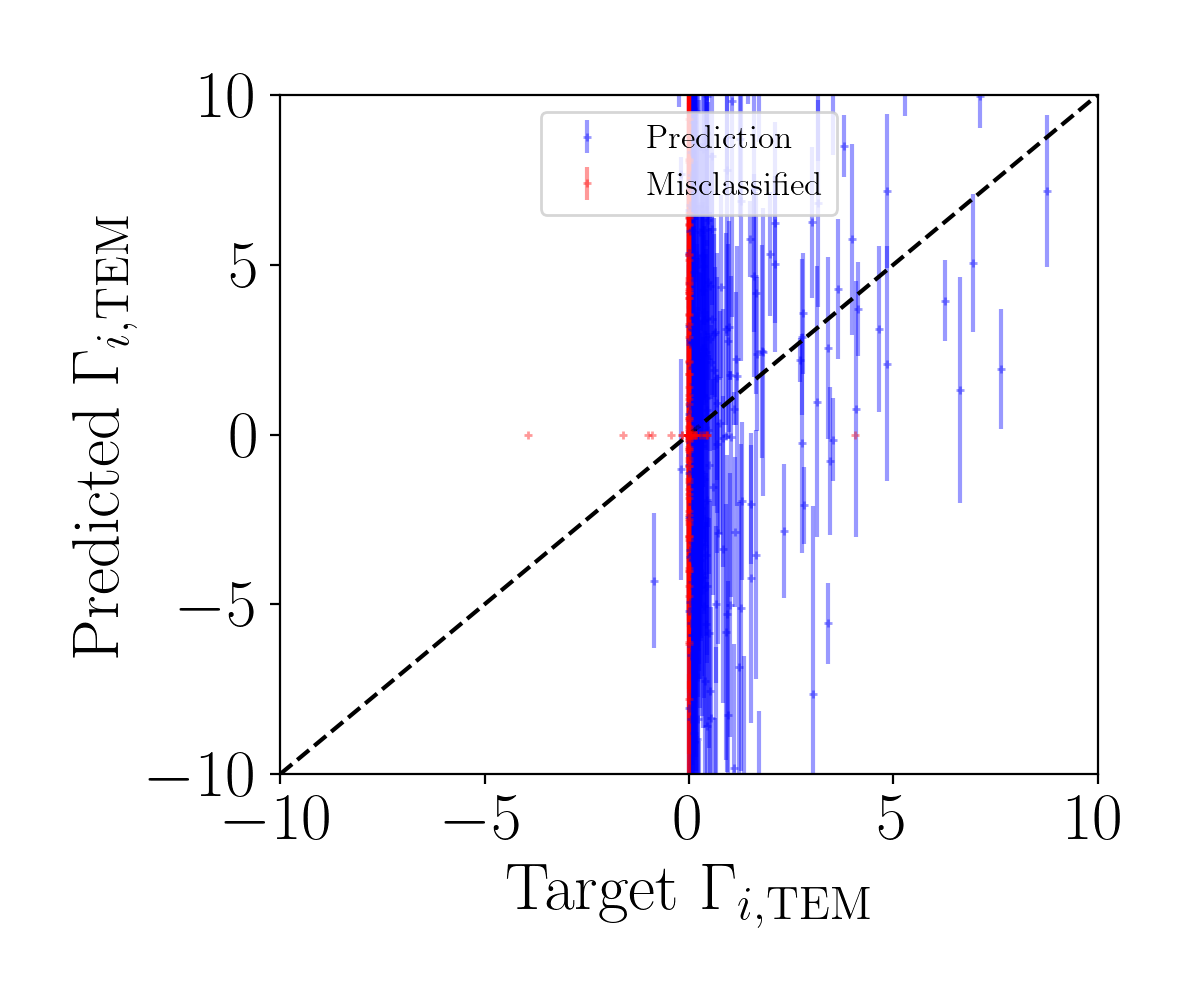}\\
	\includegraphics[scale=0.27]{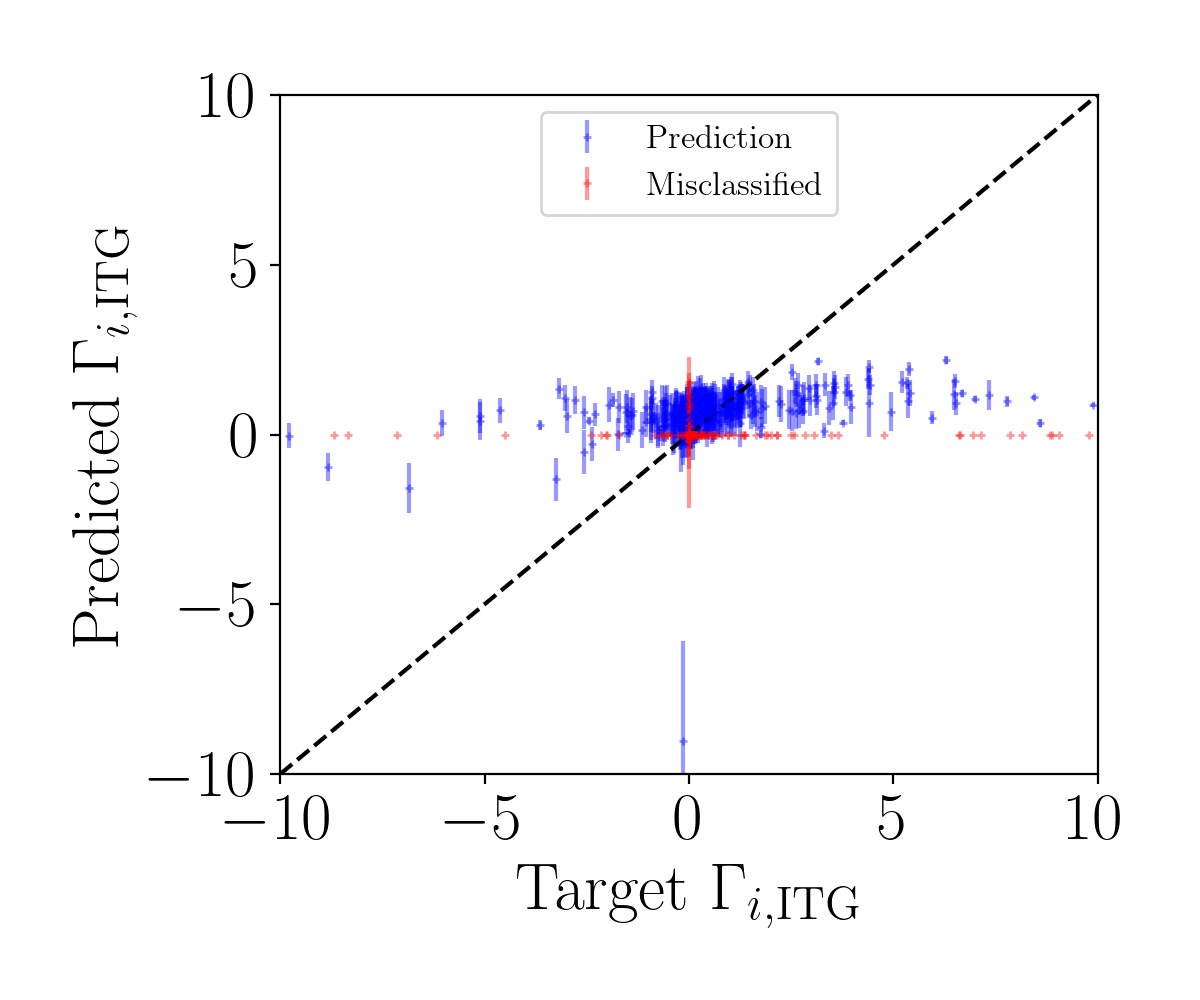}%
	\includegraphics[scale=0.27]{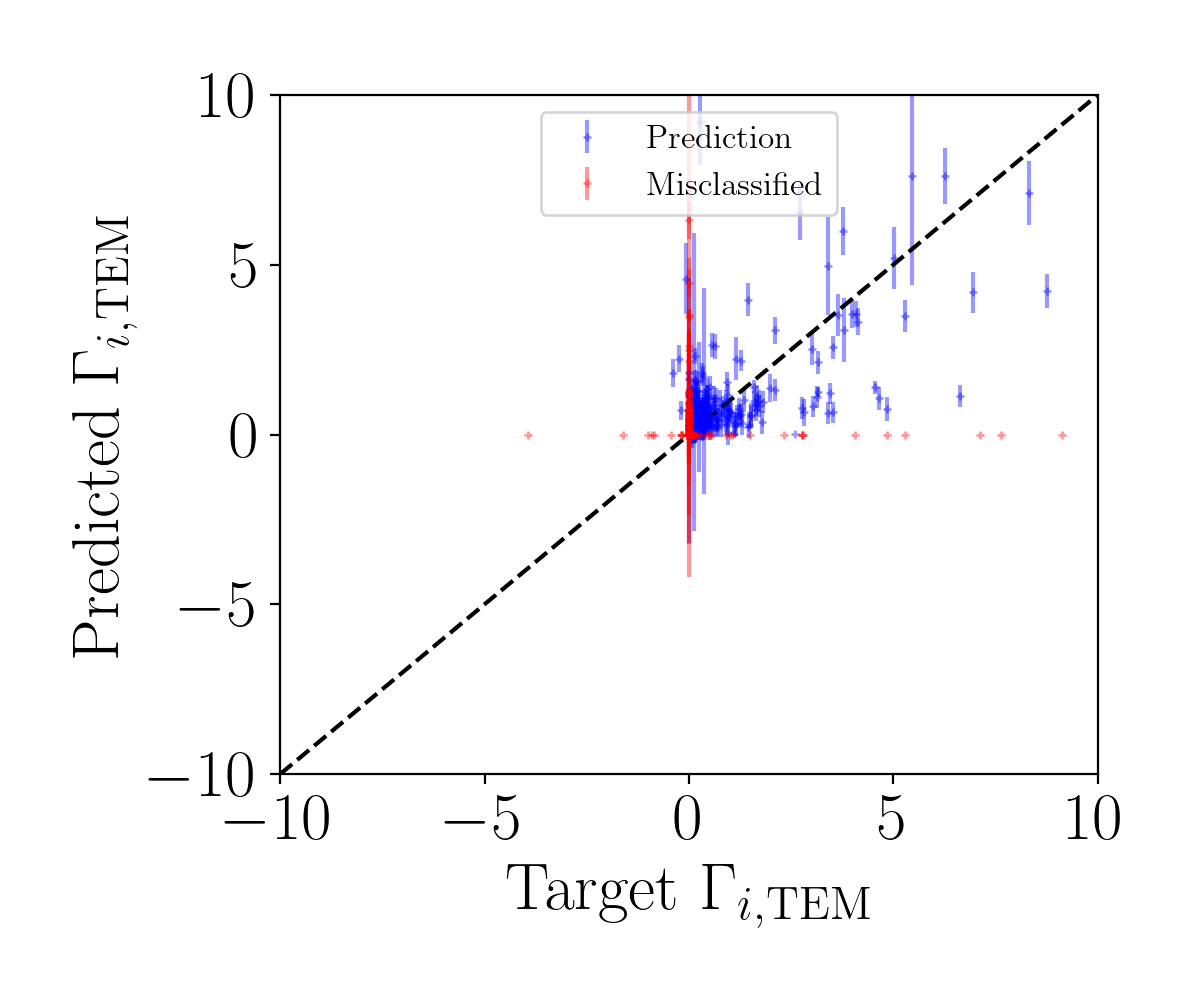}\\
	\includegraphics[scale=0.27]{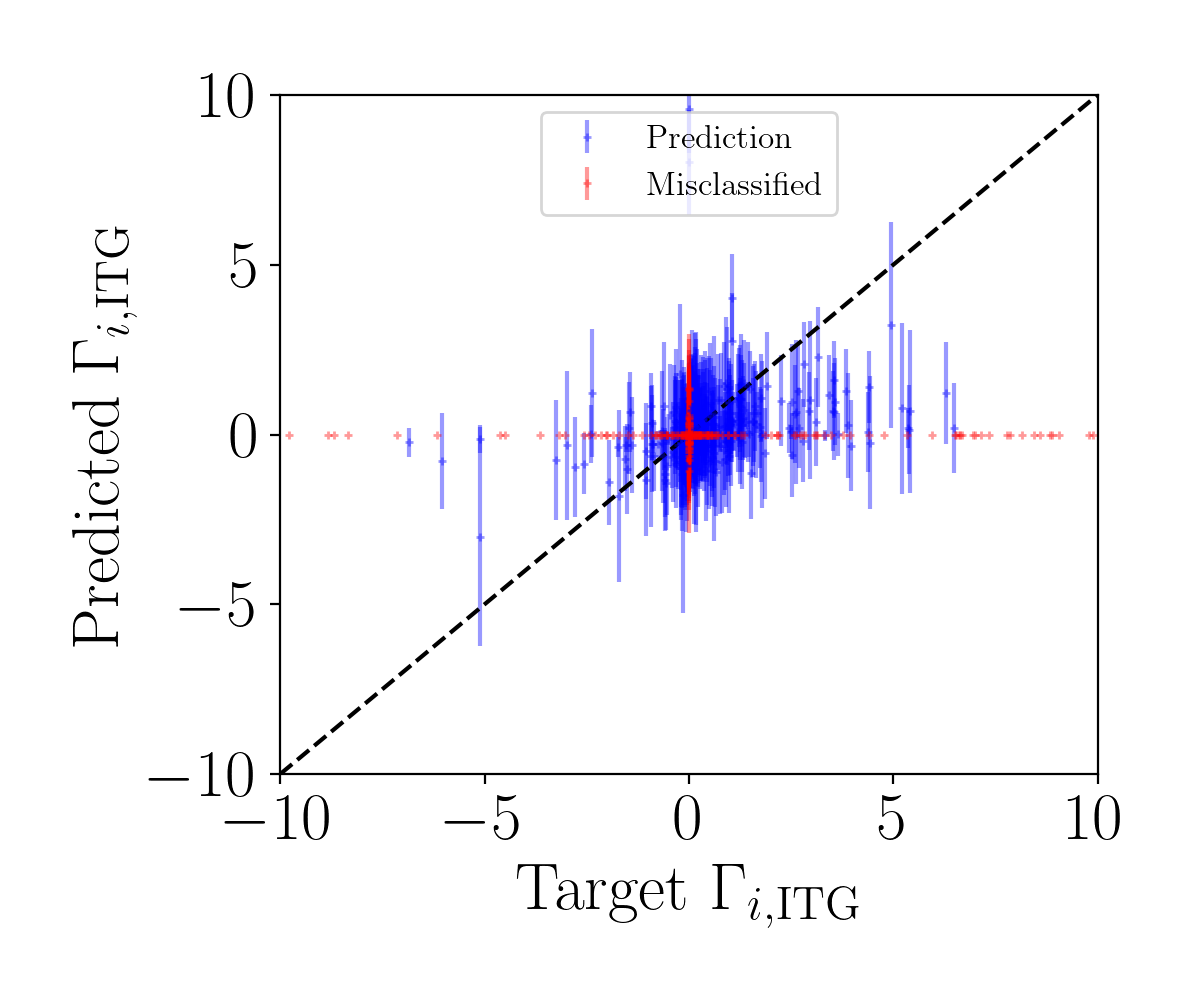}%
	\includegraphics[scale=0.27]{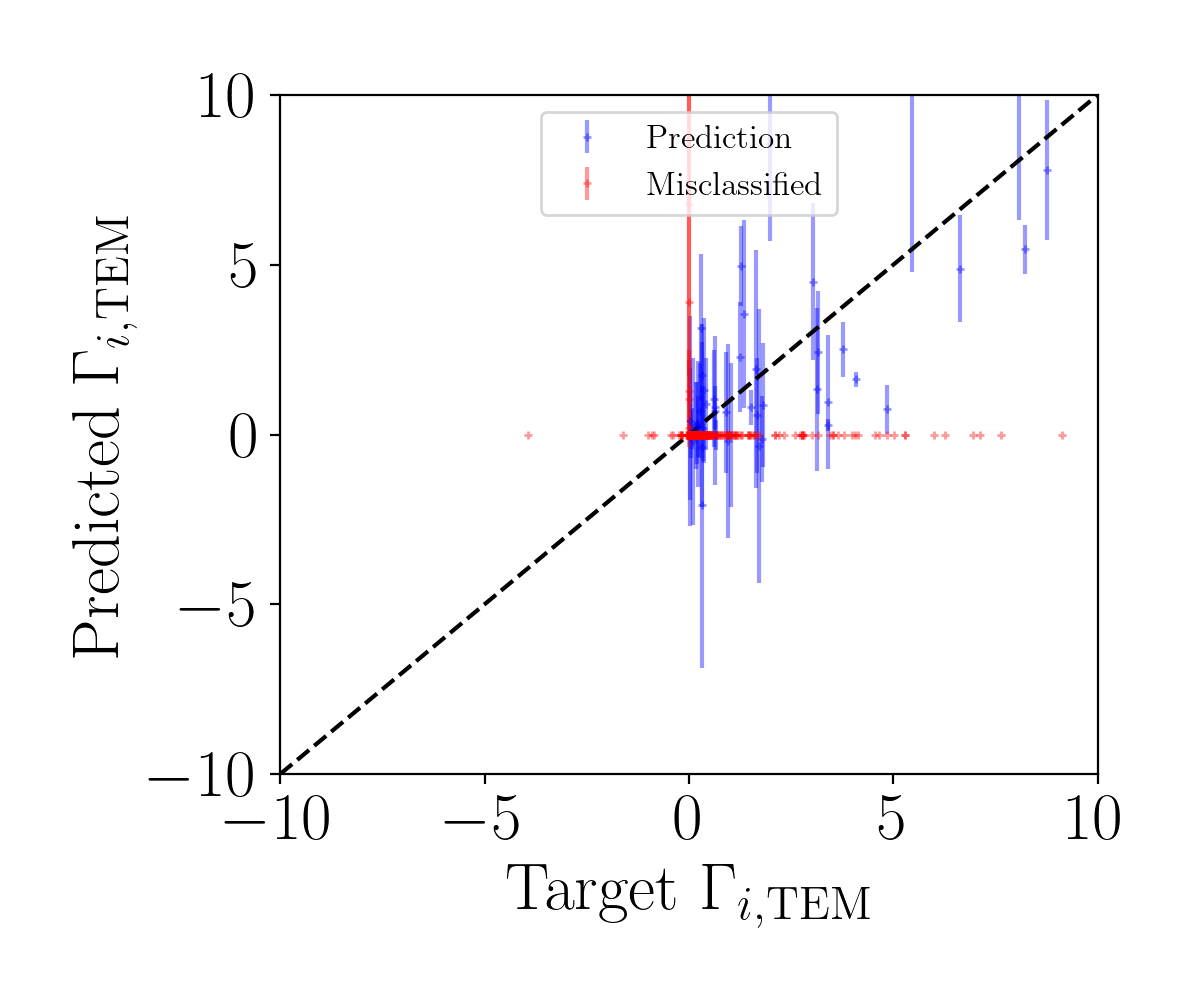}
	\caption{Comparison of the predicted ion particle flux, $\Gamma_i$, against the target flux as evaluated by QuaLiKiz for the ITG (left) and TEM (right) turbulent modes at the initial (top) and the final iteration using the proposed AL pipeline and acquisition function (middle) and using random sampling (bottom).}
	\label{fig:IonParticleFluxPerformance}
\end{figure}

\begin{figure}
	\centering
	\includegraphics[scale=0.27]{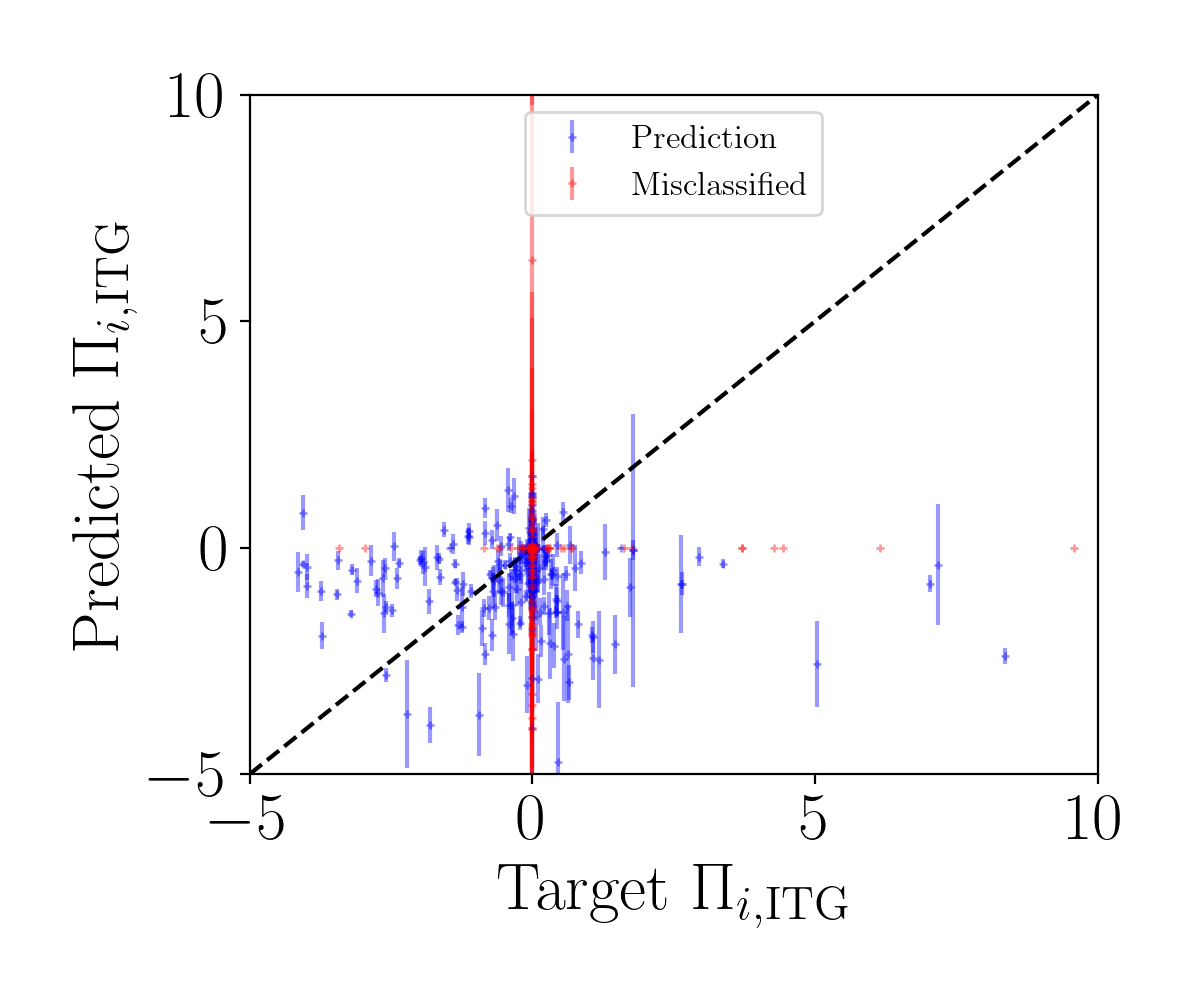}%
	\includegraphics[scale=0.27]{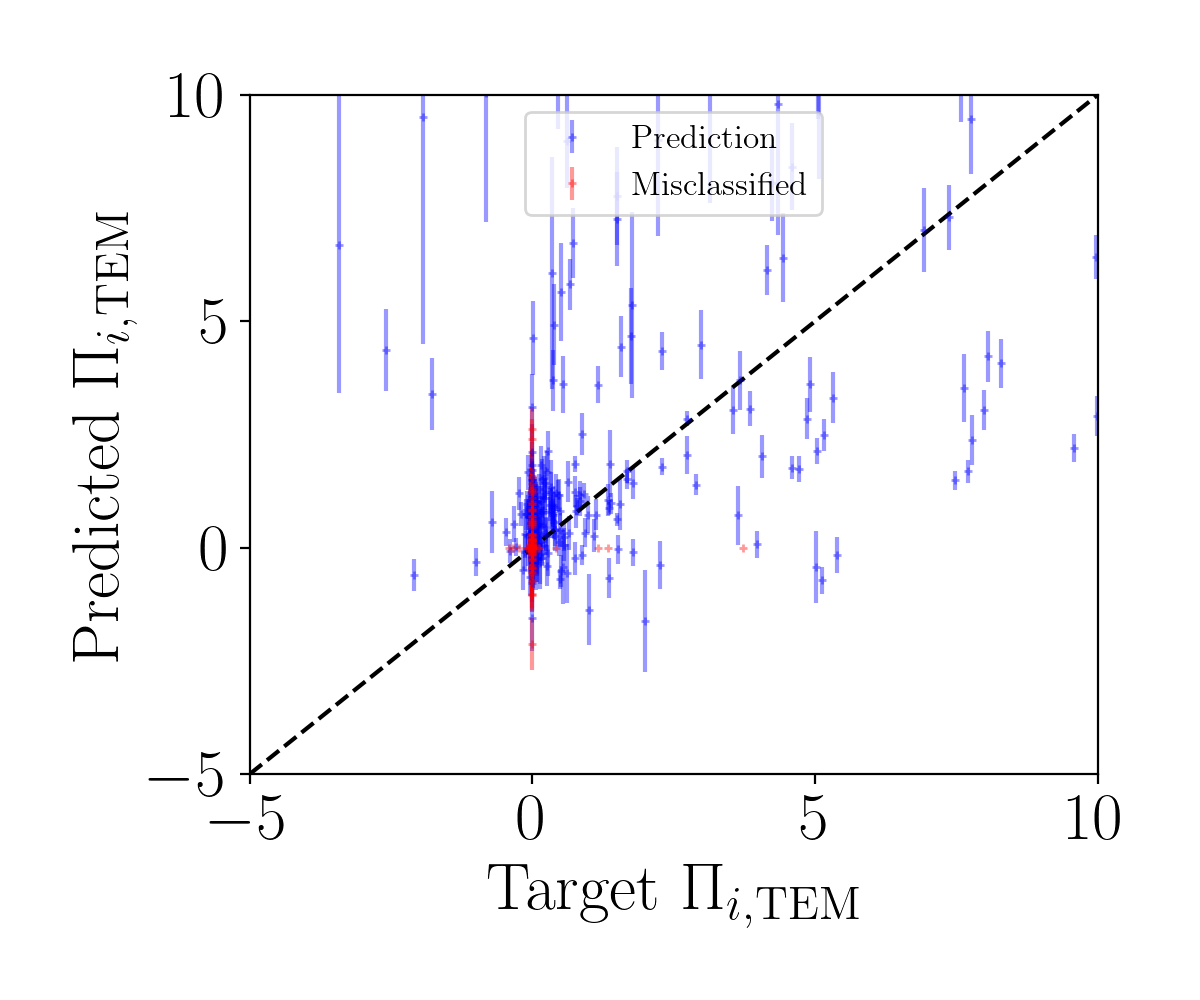}\\
	\includegraphics[scale=0.27]{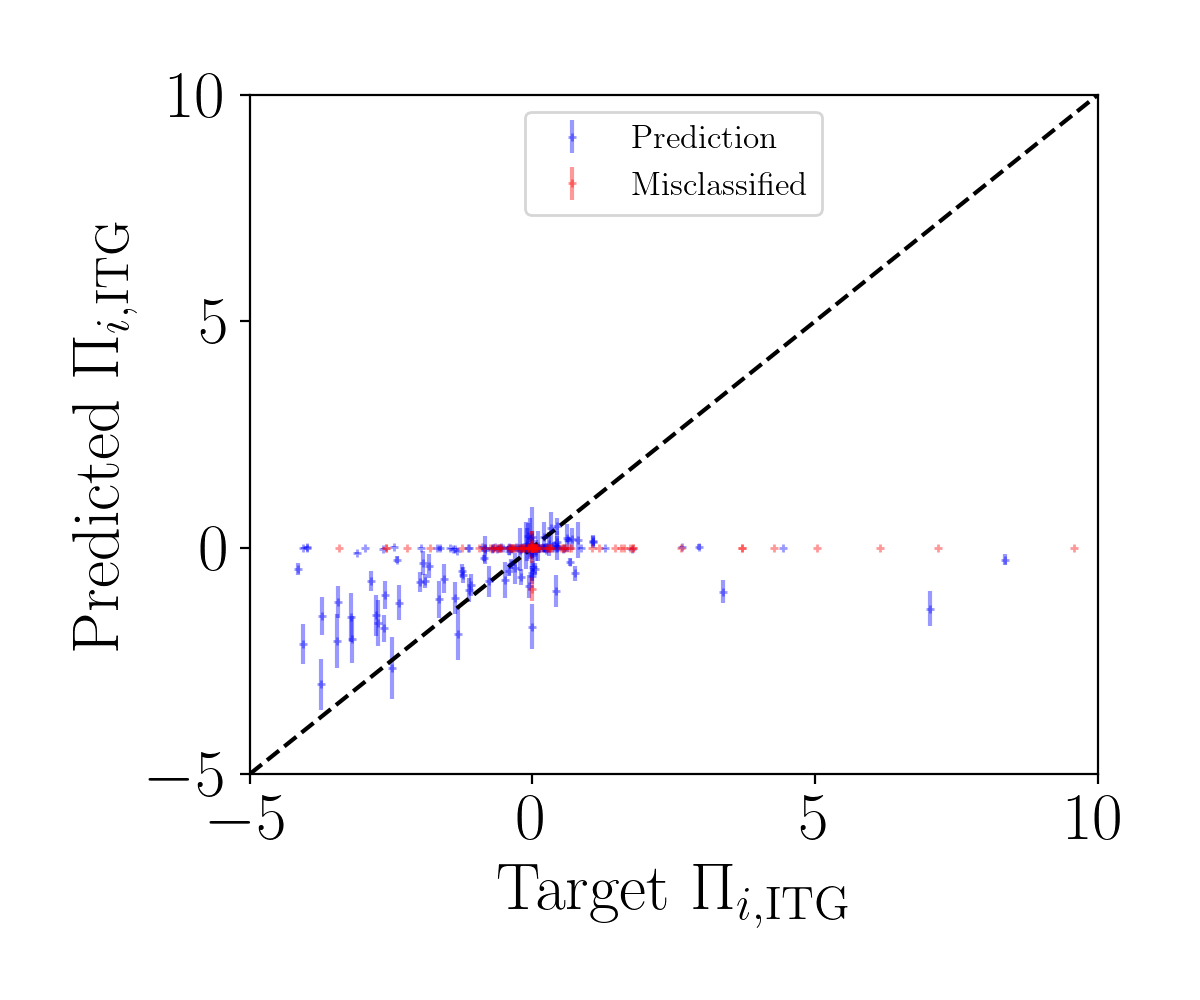}%
	\includegraphics[scale=0.27]{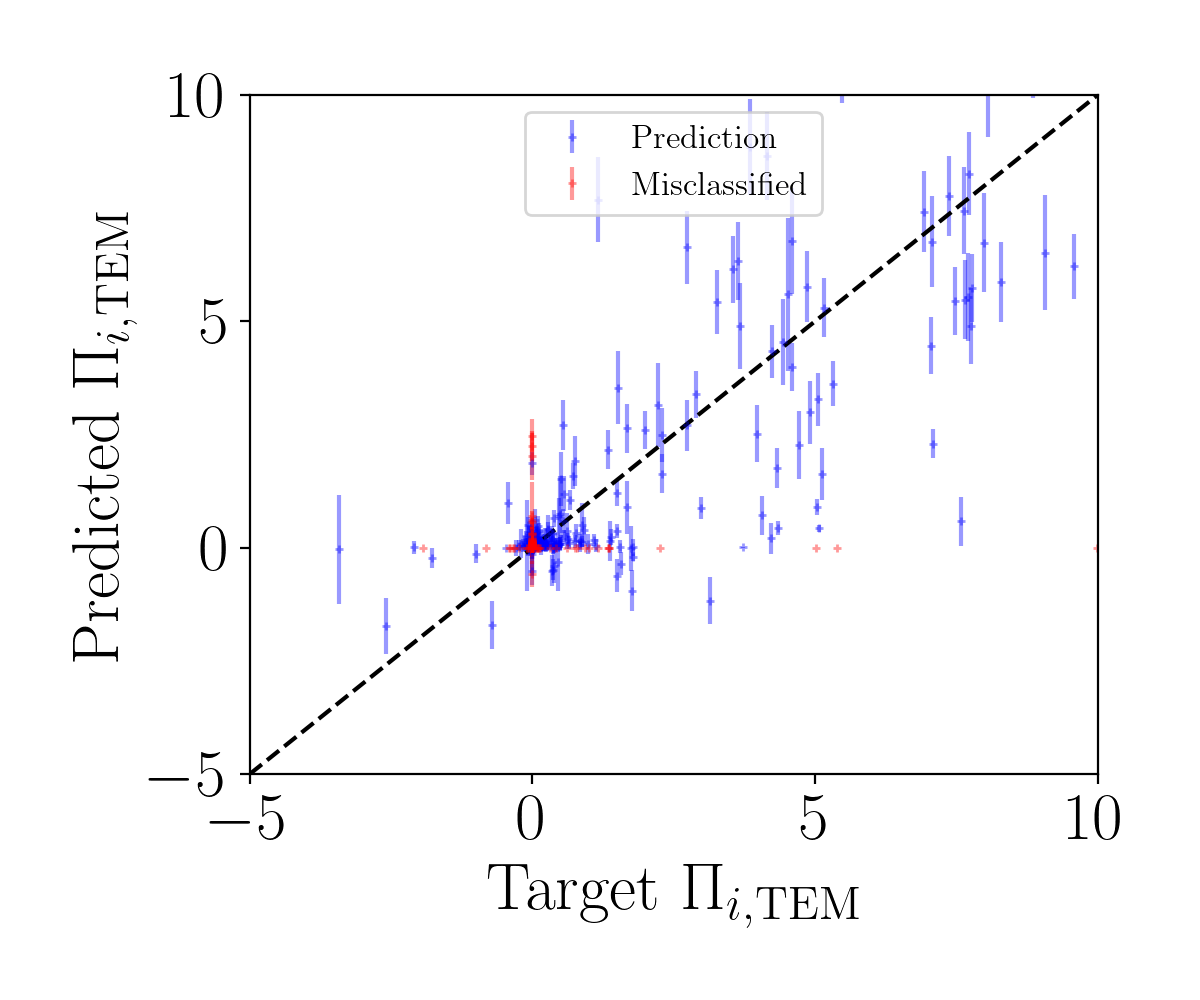}\\
	\includegraphics[scale=0.27]{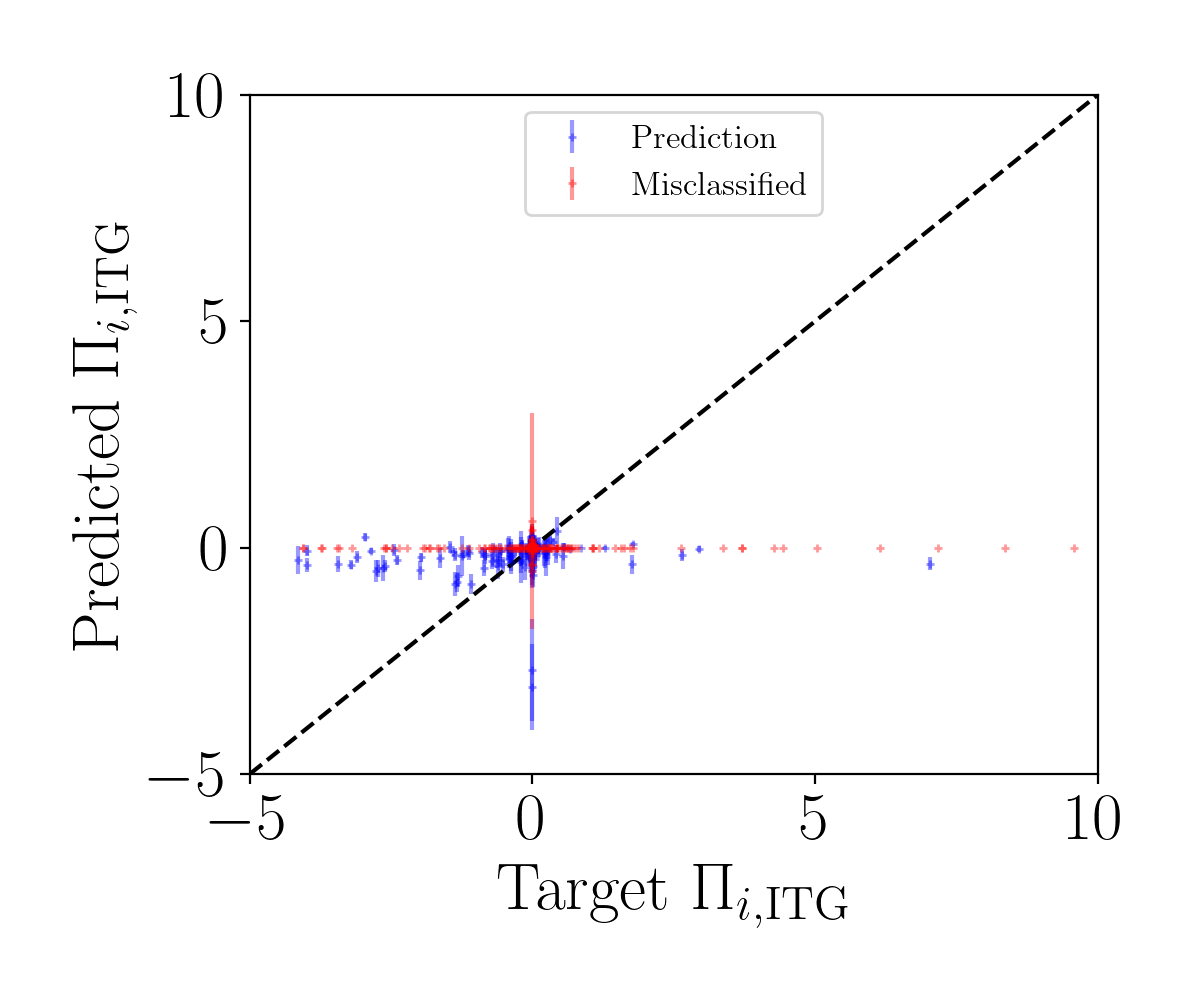}%
	\includegraphics[scale=0.27]{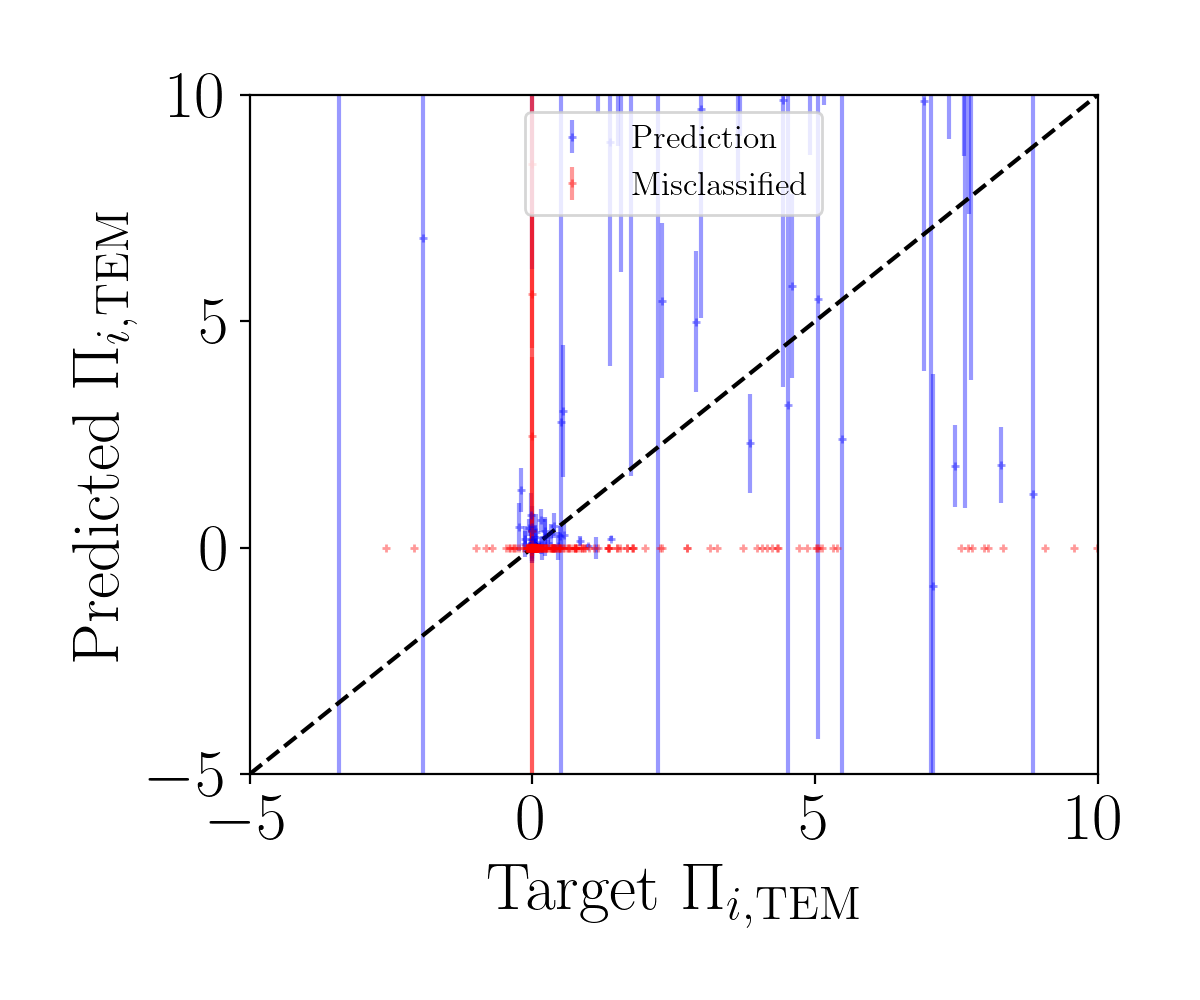}
	\caption{Comparison of the predicted ion heat flux, $\Pi_i$, against the target flux as evaluated by QuaLiKiz for the ITG (left) and TEM (right) turbulent modes at the initial (top) and the final iteration using the proposed AL pipeline and acquisition function (middle) and using random sampling (bottom).}
	\label{fig:IonMomentumFluxPerformance}
\end{figure}

This section provides the regressor performance plots for the remaining output variables not shown in the main study, in Figures~\ref{fig:IonHeatFluxPerformance} -- \ref{fig:IonMomentumFluxPerformance}, to improving readability. It was deemed important to include these both for transparency regarding the generality of the method and for future reference.

\section{Additional dataset comparison plots}
\label{app:DatasetDistributionComparison}

\begin{figure*}[p]
	\centering
	\includegraphics[scale=0.3]{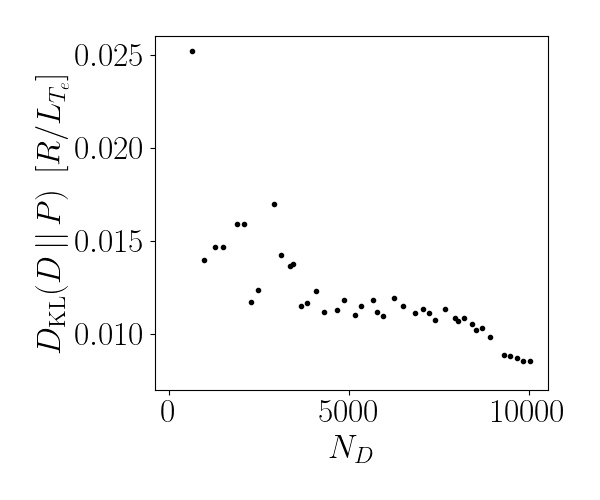}%
	\includegraphics[scale=0.3]{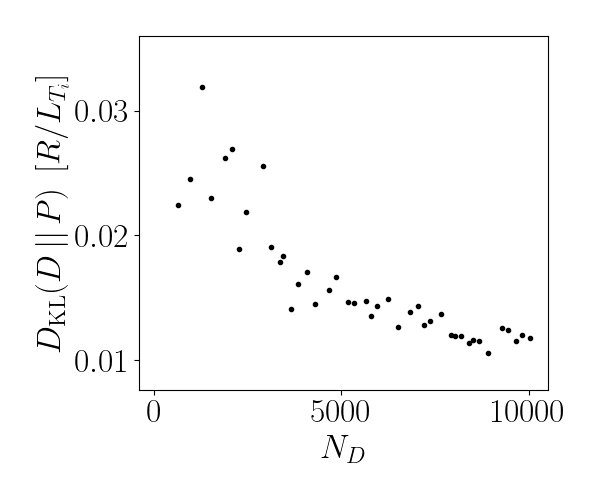}%
	\includegraphics[scale=0.3]{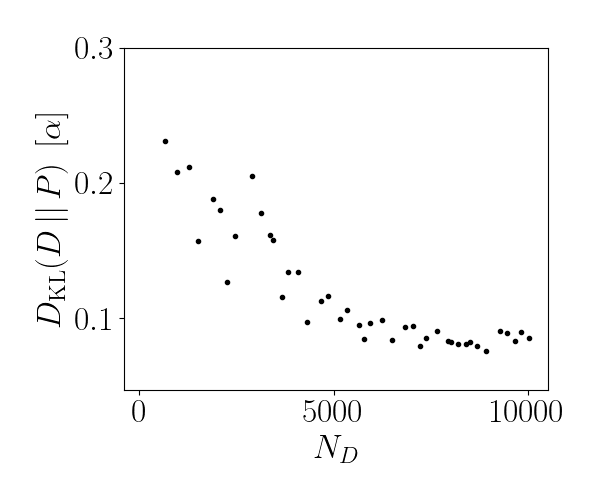}\\
	\includegraphics[scale=0.3]{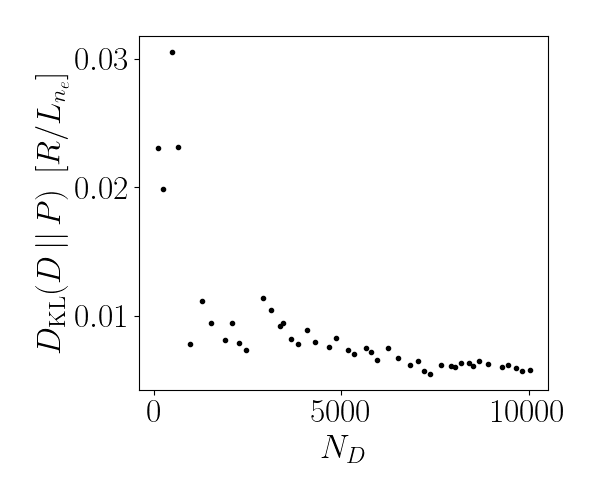}%
	\includegraphics[scale=0.3]{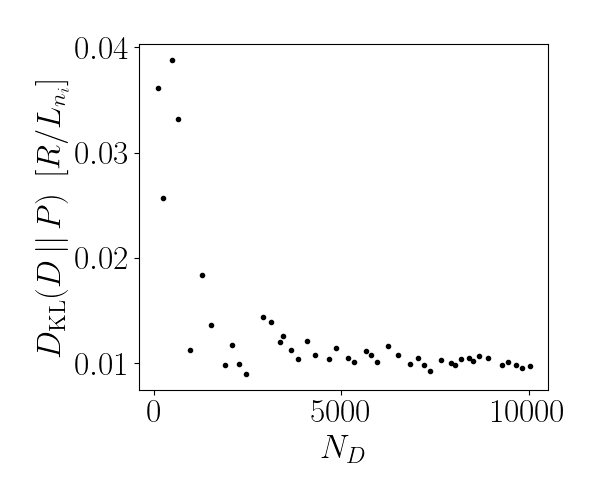}%
	\includegraphics[scale=0.3]{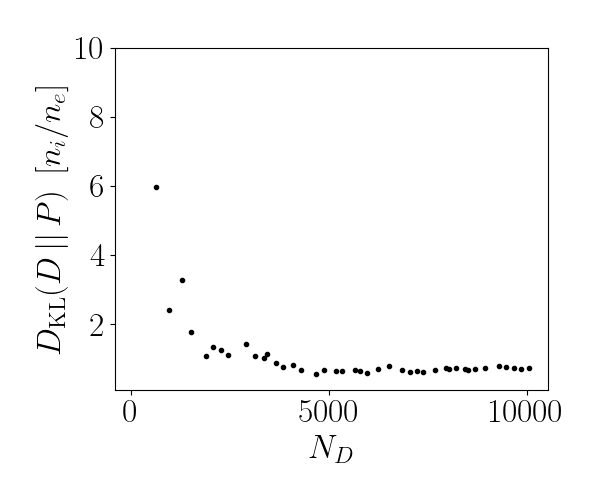}\\
	\includegraphics[scale=0.3]{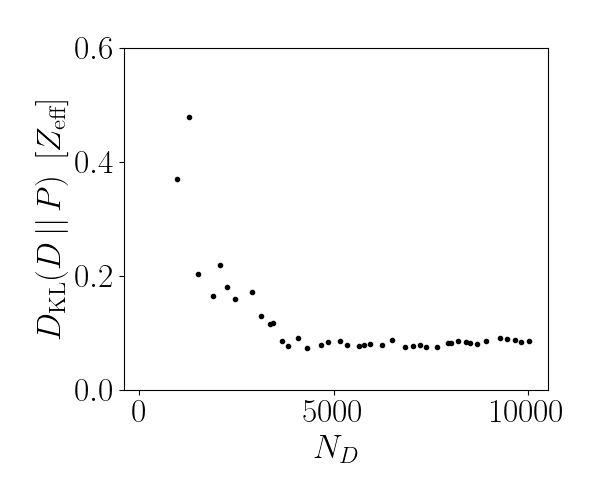}%
	\includegraphics[scale=0.3]{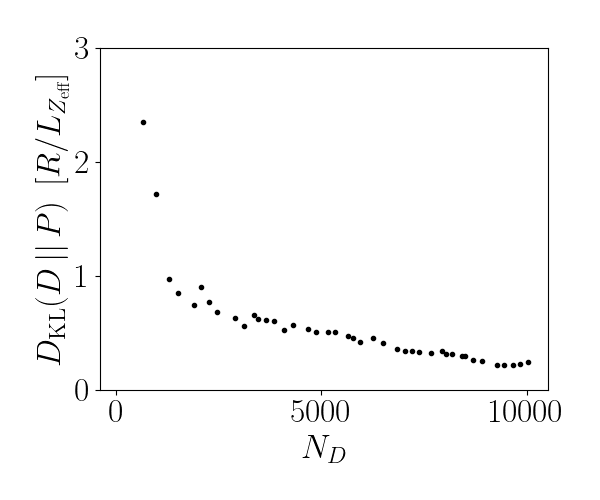}%
	\includegraphics[scale=0.3]{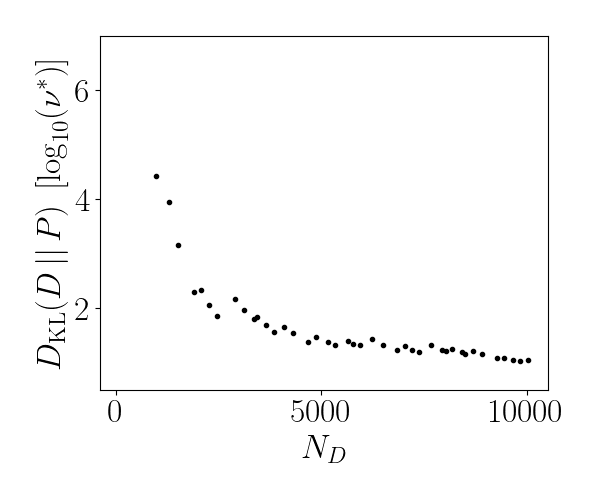}\\
	\includegraphics[scale=0.3]{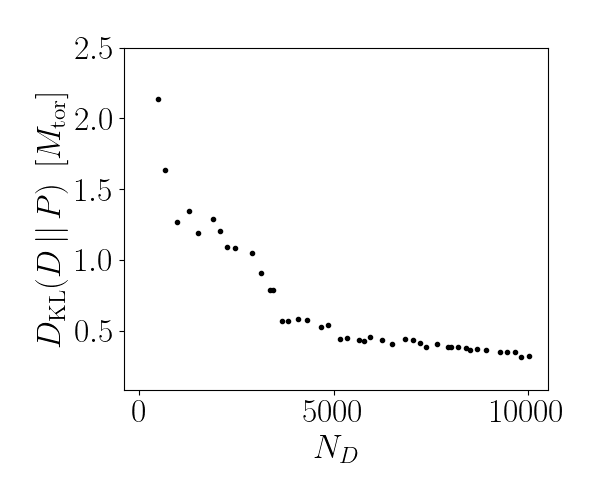}%
	\includegraphics[scale=0.3]{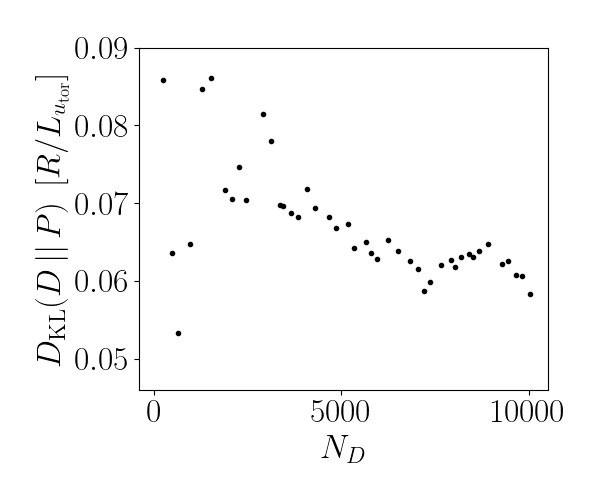}\\
	\includegraphics[scale=0.3]{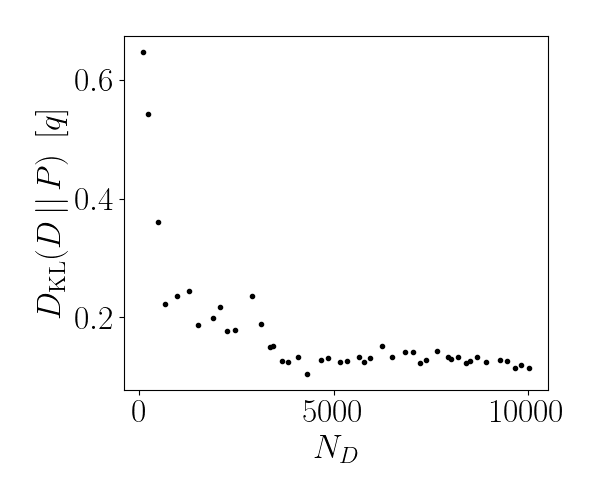}%
	\includegraphics[scale=0.3]{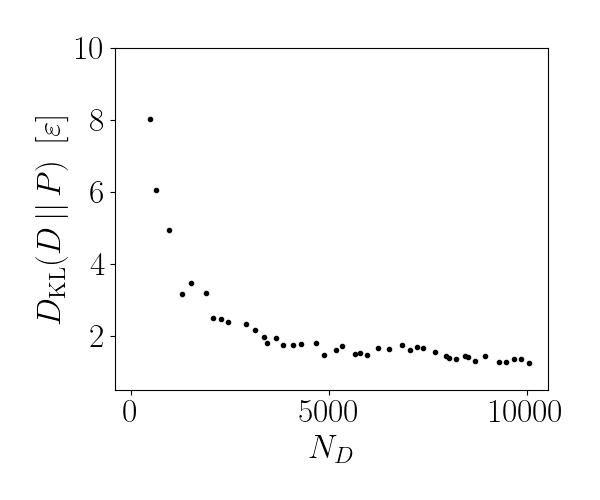}
	\caption{KL-divergence metric, $D_{\text{KL}}$, as a function of the dataset size, $N_D$, of the remaining 1D parameter distributions in Table~\ref{tbl:NeuralNetworkInputs} but not shown in Figure~\ref{fig:ConstructedDatasetKLDivergence}, computed between the constructed dataset, $D$, and the unlabelled pool, $P$, for which the latter acts as a proxy for the underlying data point sampling distribution due to the sheer number of points within.}
	\label{fig:ConstructedDatasetKLDivergenceRemaining}
\end{figure*}

This section provides the final dataset distribution comparison plots for the remaining input variables not shown in the main study, in Figure~\ref{fig:ConstructedDatasetKLDivergenceRemaining}, to improving readability. It was deemed important to include these both for transparency regarding the generality of the analysis and for future reference.

\end{document}